# An investigation of the photometric variability of confirmed and candidate Galactic Be stars using ASAS-3 data

Klaus Bernhard,[1,2]★ Sebastián Otero,[1] Stefan Hümmerich,[1,2] Nadejda Kaltcheva,[3] Ernst Paunzen,[4] Terry Bohlsen[5]
[1]*American Association of Variable Star Observers (AAVSO), 49 Bay State Rd, Cambridge, MA 02138, USA*
[2]*Bundesdeutsche Arbeitsgemeinschaft für Veränderliche Sterne e.V. (BAV), Munsterdamm 90, D-12169 Berlin, Germany*
[3]*Department of Physics and Astronomy, University of Wisconsin Oshkosh, 800 Algoma Boulevard, Oshkosh, WI 54901, USA*
[4]*Department of Theoretical Physics and Astrophysics, Masaryk University, Kotlářská 2, 611 37 Brno, Czech Republic*
[5]*Mirranook Observatory, Boorolong Rd, Armidale, NSW, 2350, Australia*

Accepted XXX. Received YYY; in original form ZZZ

**ABSTRACT**
We present an investigation of a large sample of confirmed ($N$=233) and candidate ($N$=54) Galactic classical Be stars (mean $V$ magnitude range of 6.4 to 12.6 mag), with the main aim of characterizing their photometric variability. Our sample stars were preselected among early-type variables using light curve morphology criteria. Spectroscopic information was gleaned from the literature, and archival and newly-acquired spectra. Photometric variability was analyzed using archival ASAS-3 time series data. To enable a comparison of results, we have largely adopted the methodology of Labadie-Bartz et al. (2017), who carried out a similar investigation based on KELT data. Complex photometric variations were established in most stars: outbursts on different time-scales (in 73±5 % of stars), long-term variations (36±6 %), periodic variations on intermediate time-scales (1±1 %) and short-term periodic variations (6±3 %). 24±6 % of the outbursting stars exhibit (semi)periodic outbursts. We close the apparent void of rare outbursters reported by Labadie-Bartz et al. (2017) and show that Be stars with infrequent outbursts are not rare. While we do not find a significant difference in the percentage of stars showing outbursts among early-type, mid-type and late-type Be stars, we show that early-type Be stars exhibit much more frequent outbursts. We have measured rising and falling times for well-covered and well-defined outbursts. Nearly all outburst events are characterized by falling times that exceed the rising times. No differences were found between early-, mid- and late-type stars; a single non-linear function adequately describes the ratio of falling time to rising time across all spectral subtypes, with the ratio being larger for short events.

## 1 INTRODUCTION

According to the still widely-employed definition by Jaschek et al. (1981), Be stars are non-supergiant B stars whose spectra show, or have shown at some time, emission in one or more of the hydrogen Balmer lines. While useful for initial classification, this definition is very broad and does not take into account the underlying mechanism responsible for the observed line emission, which is a general signature of circumstellar gas of a certain density. Thus, Balmer line emission may be observed in very different and not necessarily related objects like e.g. Herbig Ae/Be stars (Rivinius et al. 2013). Another, regularly-observed spectroscopic characteristic of Be stars is the presence of singly-ionized metal lines, like e.g. Fe II (Hanuschik 1994; Gray & Corbally 2009).

The focus of the present investigation is on the 'classical Be stars', a term that has recently been employed to exclude other emission-line objects like mass-transferring binary systems and Herbig Ae/Be stars (Porter & Rivinius 2003; Subramaniam et al. 2012).[1] These stars are rapidly-rotating main-sequence B objects notorious for forming gaseous, outwardly-diffusing Keplerian disks (Rivinius et al. 2013) that may develop and disperse on time-scales of months to years.[2] Cases are known in which a 'regular' B star, which has never shown any signs of emission, suddenly develops a disk, as has been observed in e.g. $\omega$ Ori and $\delta$ Sco (Guinan & Hayes 1984; Fabregat et al. 2000). Discovered more than 150 years ago (Secchi 1866), Be stars still present puzzles to astronomers, although recent decades have seen significant advances.

★ E-mail: klaus.bernhard@liwest.at

[1] If not indicated otherwise, here and throughout the paper, the term Be stars always refers to the classical Be stars.

Be stars exhibit complex variability on very different time-scales ranging from a few minutes to decades (Sterken et al. 1996; Labadie-Bartz et al. 2017, LB17 hereafter). Studying these variations is important to derive information on the (interplay of) astrophysical phenomena involved. Short-term variability in Be stars has been commonly observed in ground-based photometric studies, especially among earlier spectral types (Cuypers et al. 1989; Hubert & Floquet 1998). However, only with the advent of high-precision space photometry, it has become clear that, apparently, short-term variability is ubiquitous in these stars and rich frequency spectra have been observed in many objects (Gutiérrez-Soto et al. 2007; Emilio et al. 2010; Rivinius et al. 2017).

Periodic variations on intermediate time-scales (days to months) have also been reported in Be stars. For example, Mennickent et al. (1994) and Sterken et al. (1996) have established the existence of periodic and quasi-periodic variability in several Be stars on time-scales between days and months. Non-radial pulsation cannot be reconciled with these long periods, although the beating of closely-spaced non-radial pulsation frequencies has been postulated as a possible explanation (Sterken et al. 1996; LB17). Binarity and disk-related phenomena (e.g. the propagation of density waves in the disk) can also lead to this kind of variations (Rivinius et al. 2013), and outbursts may (re)occur on similar time-scales. Studies with the BRITE satellites have drawn a more differentiated picture and shown that several mechanisms might be at work in a single star (Baade et al. 2017a).

Photometric variability on long time-scales (months to decades) are generally attributed to (changes in) the circumstellar disk, most notably its development and dispersion. Disks are created through events referred to as outbursts, in which mass is elevated from the stellar surface and the development of, and mass-transfer to, the disk is initiated. Outbursts are accompanied by characteristic photometric variations. Depending on the inclination angle of the system, the object may get brighter or dimmer (Haubois et al. 2012). When seen pole-on, the resulting energy distribution will be that of the stellar continuum plus excess from the colder (and hence redder) disk ('Be phase'). At visual wavelengths, brightenings of up to $\sim$0.4 mag are observed (Haubois et al. 2012), while the disk excess may dominate the total flux at near-infrared and longer wavelengths. On the other hand, if the star-disk system is seen edge-on, the circumstellar disk absorbs and scatters part of the stellar flux ('shell phase'), which – depending on the size of the pseudo-photosphere (Harmanec 1983; Vieira et al. 2015) – may result in a dimming at visual wavelengths (up to about 0.2 mag; Haubois et al. 2012), as is observed in the so-called shell stars (Rivinius et al. 2013, and references therein). In a few Be stars (most notably Pleione; cf. Tanaka et al. 2007), the position angle of the disk changes due to precession (Martin et al. 2011), which leads to transitions between the Be and shell phases. Other mechanisms, for instance binarity-induced phenomena or periodic density oscillations in the disk, may contribute to the very-long period variations seen in Be stars.

[2] The lower limit can be much shorter. For example, Peters (1986) reported the development of H$\alpha$ emission in the transient Be star $\mu$ Cen in only two days. See also Baade et al. (1988).

The development of Be star disks, which have become known as 'decretion disks' (Pringle 1992), is rather well understood; once matter has been ejected, it is governed by viscous processes. However, the mechanism(s) behind the formation and maintaining of the circumstellar disks in Be stars have remained elusive. Be stars are fast rotators (rotation rate of about 75 % of critical or above). However, the majority of them likely does not achieve critical rotation rates (critical rotation in Be stars has been initially suggested by Struve 1931), and a mechanism besides rotation is needed to trigger the 'Be phenomenon'. Be star disks are known to form and dissipate over relatively short time scales, which are too short to be related to stellar evolutionary effects. Whatever mechanism is operating must therefore be capable of switching on and off (Rivinius et al. 2013).

The most promising mechanism in this respect is pulsation. While pulsation was suggested as a potential trigger of mass loss in Be stars at an early stage (Baade 1988), it has remained open for a long time whether the short-period variability ($P < 2.0$ d) observed in Be stars is due to rotational modulation (Balona 1990) or pulsation (Baade 1987, see e.g. the discussion in Porter & Rivinius 2003). Recent evidence strongly favours the scenario that all Be stars are in fact non-radially pulsating stars (Semaan et al. 2011; Rivinius et al. 2013; Baade et al. 2017a).

While single non-radial pulsation modes are not suited to trigger mass loss, beating effects may produce higher amplitudes (Neiner et al. 2002; Rivinius et al. 2013; Labadie-Bartz et al. 2017) and a connection between pulsational amplitude and circumstellar activity has been established (Carciofi et al. 2008; Neiner et al. 2013). Interestingly, so-called 'difference frequencies' ($\Delta$ frequencies) may show amplitudes in excess of the amplitude sum of their associated pulsational base frequencies, influencing the mass transfer to the circumstellar environment (Baade et al. 2016, 2017a,b). In summary, compelling evidence now exists that pulsation is at the root of the mass ejection events observed in Be stars.

Many reviews on Be stars have been published, and we do not attempt to give an exhaustive overview. For a summary of the earlier literature, the reader is referred to Underhill & Doazan (1982), while Rivinius et al. (2013) provide an excellent survey of the current knowledge.

The present work presents an investigation of a large sample of confirmed ($N = 233$) and candidate ($N = 54$) Galactic Be stars, using archival photometric and spectroscopic observations as well as newly-acquired spectra, with the main aim of describing their photometric variability. Our methodological approach is outlined in Section 2. Results are presented in Section 3 and discussed in Section 4.

## 2 METHODOLOGY

### 2.1 Target selection

Our sample was initially recruited from a list of early-type variable stars compiled by one of us (S.O.), which was assembled by a systematic investigation of photometric time series data from the All Sky Automated Survey (Pojmański 2002, ASAS hereafter) archive. To this end, the light curves of bright objects ($V_T \lesssim 10.5$ mag) with suitable Tycho-2 colours (($B_T - V_T$) $\lesssim 0.6$ mag; Høg et al. 2000) were visually inspected using a semi-automated approach. The emphasis was on discovering new variables; therefore, objects with well-defined variability types contained in catalogues like the General Catalogue of Variable Stars (GCVS; Samus et al. 2017) and the International Variable Star Index (VSX) of the American Association of Variable Star Observers (Watson 2006) were rejected.

Suspected or misclassified variables and variables of undetermined or doubtful type were kept.

To build up an initial sample of classical Be stars for the present study, the resulting list of objects was searched for objects exhibiting a variability pattern in agreement with a Be star classification. To guide us in this endeavor, we have employed the variability types developed by LB17 to describe the diverse manifestations of photometric variability in these stars. In summary, the light curves were searched for the presence of the following variations, which are discussed in more detail in Section 3.2 (the corresponding LB17 types are provided in parentheses):

a) long-term changes in mean brightness on the order of years to decades (type LTV),
b) outburst variation, i.e. a sudden change in flux that is followed by a (generally) more gradual decline to baseline brightness (type ObV),
c) periodic variations on intermediate time-scales of days to months (type IP),
d) short-period variability (defined as periodic variability with $P \leq 2$ d by LB17; type NRP).

Be stars often exhibit several or all of the above-listed types of variation in their light curves (LB17). It is therefore reasonable to expect complex light changes during the nearly 10 years of ASAS-3 coverage. (As can be seen from the set of light curves provided in Fig. B1, the results of our study have subsequently shown that this is indeed a reasonable assumption.) Similar complex light variations are not expected in most other variable stars in the outlined spectral range, such as $\beta$ Cep stars, slowly-pulsating B (SPB) stars or $\alpha^2$ Canum Venaticorum variables, which especially holds true for the large-amplitude outbursts and long-term mean magnitude changes observed in Be stars. Therefore, attention was paid in particular to items a) and b).

In this way, the light curves of all objects exhibiting a variability amplitude of at least 0.05 mag were subjected to a careful visual inspection, in agreement with the approach of LB17. A specific variability type was only registered if it could be clearly identified in the light curve.

301 objects were preselected in this way and further investigated using literature information gleaned from the SIMBAD (Wenger et al. 2000) and VizieR (Ochsenbein et al. 2000) databases. 239 stars could be confirmed as emission-line stars by their spectral type(s) in the literature (cf. the spectral types given in Table A1). We note that although Be stars are non-supergiant objects by definition, we a priori decided to only exclude stars of luminosity type Ia. This decision has been made in order to take into account the great uncertainty in (luminosity) classification that exists among a significant part of our sample stars. Indeed, as discussed in Section 2.3, the variable nature of Be star spectra results in a great range of listed spectral types for some objects.

As the focus of the present investigation is on the classical Be stars (cf. Section 1), care was taken to exclude B[e] stars and Herbig Ae/Be stars. To this end, the literature was searched for classificatory information, and a very few B[e] stars that had been selected for the initial sample were thus subsequently removed. In the case of the Herbig Ae/Be stars, a two-fold approach was taken that relied on literature information and colour-colour plots. It has been shown that Herbig Ae/Be stars can be efficiently distinguished from classical Be stars at infrared wavelengths (Rivinius et al. 2013), as they generally exhibit significant infrared excesses due to the presence of dust in the circumstellar environment. The disks of classical Be stars, on the other hand, contain no dust; the observed infrared excess in these objects is due to free-free radiation of hydrogen.

We have therefore investigated our sample stars using infrared observations from the 2MASS (Skrutskie et al. 2006) and WISE (Wright et al. 2010) catalogues and investigated all stars with significant near-infared ($(J-K)-(B-V) \gtrsim 0.8$ mag) and mid-infrared ($V-[22] \gtrsim 5$ mag) excesses. Most of these objects were found to show peculiar light curves and have been classified at least once as young stellar objects in the literature. Furthermore, we checked for the presence of diffuse nebulae by a visual inspection of the corresponding WISE images using the ALADIN visualization tool (Bonnarel et al. 2000). The presence of diffuse nebulae was revealed in all cases. Consequently, these objects were assumed to be Herbig Ae/Be stars and removed from the sample.

Only two eclipsing binary systems are present in our sample, which – judging from their light curves – might be detached systems harbouring classical Be stars. We have chosen to keep them in the sample, in accordance with the approach of LB17. We also searched for the possible presence of cataclysmic variables, which are easily revealed by their infrared colours (contribution of the donor star) and X-ray properties, but no such objects were found.

In addition to the classifications in the literature, some further objects could be confirmed by LAMOST spectra, our own spectra and *uvbyβ* photometry. All in all, 233 classical Be stars were selected in this way. The high detection rate indicates that our selection criteria based on light curve morphology are a viable and efficient method of identifying classical Be stars among early-type stars in large photometric time-series databases. We therefore felt justified in including the remaining 54 stars as candidate Be stars into the final sample. These stars exhibit a variability pattern in agreement with a classical Be star classification but have never been identified as emission-line objects in the literature. As it is not uncommon for Be star disks to (re)appear and disperse on time-scales that may reach years or decades, a Be star need not necessarily show emission at all epochs. It is thus possible that Be stars have been missed in spectroscopic surveys if they did not show emission at the corresponding epoch of observation.

In summary, the final sample encompasses 287 stars – 233 spectroscopically-confirmed Be stars and 54 Be star candidates in the mean $V$ magnitude range of 6.4 to 12.6 mag that were selected by light curve morphology criteria. For convenience, all stars were numbered in order of increasing right ascension (No. 1 – No. 287). Throughout this study, in the discussion of stars of interest, the internal identification number is always listed in parentheses, in order to provide an easy identification in the corresponding tables.

The distribution of our sample stars in Galactic coordinates is shown in Fig. 1, together with the sample of LB17. As expected, Be stars are mostly confined to the Galactic disk. There is an obvious gap in the distribution of our sample stars (from $l \approx 45°$ to $l \approx 180°$). This is due to the fact that in the outlined RA range, the Galactic disk reaches declinations northerly of +28°, which is outside the coverage of our photometric data source (cf. Section 2.4).

### 2.2 The Labadie-Bartz et al. study

During the preparatory stages of our investigation, LB17 published a thorough analysis of the photometric variability of Be stars employing observations from the Kilodegree Extremely Little Telescope (KELT; Pepper et al. 2007) transit survey. To this end, well-known Be stars from the Be Star Spectra (BeSS) database (Neiner et al. 2011) were chosen and cross-matched with entries in the KELT database, which resulted in a sample of 610 stars. Because of sig-nificant saturation effects in the light curves, 510 objects were finally included into the analysis. These objects are situated in the mean $V$ magnitude range of 6.0 to 12.8 mag. KELT data boast coverage up to ~10 yr (although only 217 objects of the LB17 study exhibit a time baseline exceeding four years; cf. also Section 4) and have a typical sampling cadence of 30 min. The typical photometric error for the KELT data is 7 mmag (LB17).

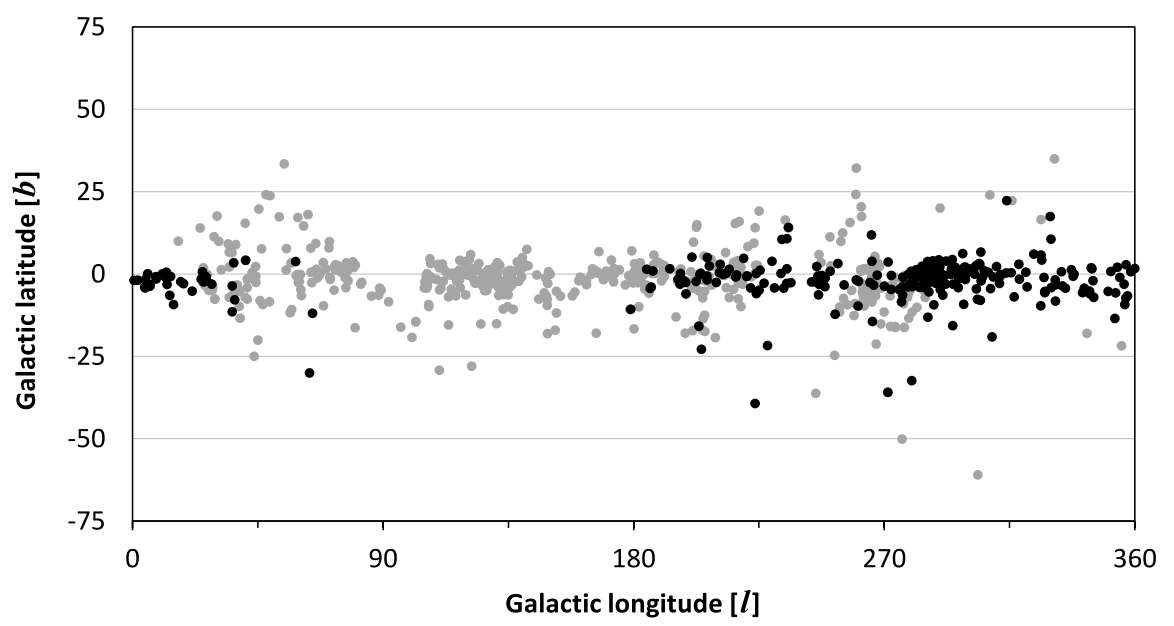

**Figure 1.** Distribution in Galactic coordinates of our sample (black dots) and the sample of LB17 (grey dots). The obvious gap in the distribution of our sample stars in the right-ascension (RA) interval from $l \approx 45°$ to $l \approx 180°$ is due to the coverage of our photometric data source (cf. Section 2.4). Both samples complement each other well; together, they cover approximately the whole Galactic disk.

LB17 investigated the presence and occurrence rate of outbursts, long-term variability related to the circumstellar disk and non-radial pulsations. Their work therefore shares the same aim as our study, viz. the investigation of the photometric variability of classical Be stars. In order to enable a comparison of results, we have chosen to adopt the aforementioned authors' methodology wherever appropriate. Thus, in the following chapters dealing with the photometric analysis, relevant results from LB17 are mentioned, employed for comparison and included into our analysis. This approach was chosen because we think that the resulting enlarged and homogeneous sample of Be stars with known properties will facilitate further research on these objects and thereby serve the Be star community best. We stress, however, that the validity of this comparison is severely compromised by the different data sources employed and the different ways the samples were compiled.

### 2.3 Spectroscopic classification

When investigating Be stars, it is useful to subdivide the sample into early-type, mid-type and late-type Be stars (e.g. Gutiérrez-Soto et al. 2007). We have divided our sample according to the scheme employed by LB17 and consider Be stars with spectral types earlier than B4 as *early-type Be stars*, objects with spectral types of B4, B5 and B6 as *mid-type Be stars*, and stars with spectral type of B7 and later as *late-type Be stars*. Objects without a specific spectral type in the literature are listed as *unclassified Be stars*.

The spectroscopic classification of our target stars was gleaned from various literature sources, which are listed in the presentation of results in Table 2. Spectra from the BeSS database were also secured and primarily used to check the literature types and to search for the presence of emission lines. For some of our stars, spectra from the LAMOST DR2 archive (Cui et al. 2012; Luo et al. 2016) are available, which were also taken into consideration (Figure C1). Additional spectra of several of our target stars (unclassified or ambiguously classified objects with few or no spectroscopic observations in the BeSS database or the literature) were obtained at Mirranook Observatory using a LISA spectrograph on a C11 279/2800 mm Schmidt-Cassegrain telescope. The LISA is a commercially available classic spectrograph optimised for 400-700 $\mu$m and was used with a 23 $\mu$m slit.[3] The employed camera is an Atik 314+ cooled CCD camera with 6.45 $\mu$m pixels giving well sampled images with a 23 $\mu$m slit.[4] The spectra taken had a S/N≈60 and R≈1500 and were instrument-corrected using a Miles standard star (Sánchez-Blázquez et al. 2006) taken at similar airmass and processed with the ISIS software package (Anderson et al. 2013). They were employed to search for the presence of Balmer emission in these stars and are presented in the Appendix (Figure C1). We note that only six stars of the present sample remain without spectroscopic classification.

The variable and unusual nature of Be star spectra introduces ambiguities into the spectral classification (cf. the discussion in Harmanec 1983). At lower dispersions, problems can e.g. arise due to contamination from shell/emission components of the lines or Be star envelopes can be mistakenly interpreted as the photospheres of stars of later spectral (sub)types. Gray & Corbally (2009) caution that, depending on the strength of the Be phenomenon, spectral classification becomes increasingly difficult, becoming highly uncertain for the more extreme Be stars. The well-known Be star KX And may serve as an example and a warning. The literature spectral types for this star range from B0 IV-IIIe to A5p, with all kinds of intermediate types and luminosity classifications from V to Ia (Harmanec 1983). Similarly-diverging luminosity classifications have been reported for several stars of our sample. For instance, HD 29557 has been classified as B5Ib/IIp: shell? (Houk & Smith-Moore 1988a) and B3Ve (Levenhagen & Leister 2006). We therefore caution that the spectral types given, and the subdivision into early-, mid- and late-types based thereon, will have been affected by this uncertainty in classification.

### 2.4 The ASAS-3 photometric archive

ASAS is a photometric survey which aims at the detection and investigation of all kinds of photometric variability. ASAS constantly monitored the entire southern sky and part of the northern sky up to a declination of about $\delta < +28°$. Most data were acquired during the third phase of the project, ASAS-3, which lasted from 2000 until 2009 (Pojmański 2002). The ASAS instruments were situated at the 10-inch astrograph dome of the Las Campanas Observatory in Chile and consisted of two wide-field telescopes equipped with f/2.8 200 mm Minolta lenses and 2048 x 2048 AP 10 Apogee detectors. A sky coverage of 8°.8x8°.8 was achieved, with a CCD resolution of about 14″.8 / pixel, which led to a positional accuracy of around 3 – 5″ for bright stars and up to 15.5″ for fainter objects. Therefore, blending issues arise and photometry is rather uncertain in crowded fields such as star clusters.

ASAS monitored about $10^7$ sources in the magnitude range between the saturation limit[5] $V = 7$ (up to $V \approx 8.5$ for few frames; cf. David et al. 2014) and the detection threshold at $V = 14$. During the third project phase, observations were acquired in the Johnson $V$ passband. The typical scatter of an observation in the magnitude range $8 \lesssim V \lesssim 10$ is about 0.01 mag (e.g. Pigulski 2014). Because of the long time baseline of the project, the detection of periodic signals with very small amplitudes is possible. Periodic signals with amplitudes as low as ~3 mmag have been detected in ASAS-3 data (Pigulski 2014; Hümmerich et al. 2016), in good agreement with Pigulski & Pojmański (2008), who find that the detection threshold (defined as four times the average amplitude in the Fourier spectra) in the frequency range of 0-40 $\mathrm{d}^{-1}$ typically amounts to 3-5 mmag (cf. in particular their Fig. 16).

[3] https://www.shelyak.com/produit/pf0021vis-lisa-slit-visible/?lang=en
[4] https://www.atik-cameras.com/product/atik-314l-plus/
[5] To our knowledge, the literature does not provide any information on whether detector saturation or A/D saturation is involved. Consulting the manual of the employed CCD cameras, we strongly suggest that the former holds true.

ASAS maintained a rather strict observing cadence, which results in strong daily aliasing. A field was typically observed one to three times per day (Pigulski 2014), although for several stars, observations up to five times per day are available. Therefore, care has to be taken in the interpretation of the resulting Fourier amplitude spectra. Pigulski & Pojmański (2008) used ASAS-3 data to investigate a sample of $\beta$ Cep stars and identified periodic variability down to periods of the order of ~0.07 d. ASAS-3 data should therefore be well suited to the detection of the short period pulsations in Be stars.

### 2.5 Data processing and period analysis

Data of our target stars were downloaded from the ASAS-3 website.[6] Data points with a quality flag of 'D' (='worst data, probably useless') were removed and all light curves were inspected visually. Obvious outliers and data points associated to exceedingly large error bars were deleted. As the removal of only a few datapoints may have a significant impact on the frequency analysis, care was taken in this process.

While the majority part of ASAS-3 datasets is homogeneous and of good quality, several issues exist that may render ASAS-3 datasets unreliable (cf. the discussion in David et al. 2014). Significant additional scatter might be present due to flux contribution from one or more near-by objects ('blending'). Furthermore, a dataset may contain exposures suffering from saturation effects, and scatter may be introduced by a star's position close to the edge of the CCD. Unfortunately, concerning the latter issue, no information is provided in ASAS data, so its impact cannot be estimated. Some ASAS datasets are affected by instrumental long-term trends of low amplitude, which might introduce spurious signals and mimic the long-term variations seen in Be stars. These, however, are generally of very low amplitude, unlike the long-term variability observed in most of our sample stars.

For objects brighter than $V = 8.5$ mag, all datasets were checked for saturation effects. These can be rather straightforwardly identified and distinguished from intrinsic variability by a consistency check of the magnitudes in the five different extraction apertures indicated by the ASAS system. Saturation is known to result in significantly (and randomly) discrepant values between the apertures (David et al. 2014). Saturation was assumed to occur if the magnitude difference between the smallest aperture (2 px; ASAS-denomination 'MAG_0') and the largest aperture (6 px; ASAS-denomination 'MAG_4') amounted to at least 0.05 mag. This limit is an experiential value based on our own experience in dealing with the ASAS-3 data. In consequence, datasets exhibiting a discrepancy well beyond 0.01 mag that is suspected of being attributable to saturation were rejected.

[6] http://www.astrouw.edu.pl/asas/

Blending issues have been the most frequent problem we encountered while working with the data of our programme stars. As this is not relevant to the goals of our investigation, we have not corrected the ASAS-3 light curves presented in Fig. B1 for the influence of close neighbouring stars. However, in order to get a clearer idea of the real amplitude of the photometric variations and to provide a correct $V$ magnitude range for cataloging purposes, we include a corrected $V$ range in Table 2. Depending on the brightness of the objects, light contamination from stars as far away as 30″ to 45″ may affect the results. The light contribution of all close neighbouring stars that could be identified was removed by subtracting the intensities derived from $V$ magnitudes of catalogues with superior resolution. The respective $V$ magnitudes were taken from the General Catalogue of Photometric Data (GCPD; Mermilliod et al. 1997), the AAVSO Photometric All-Sky Survey (APASS; Henden & Munari 2014) or the Yale/San Juan Southern Proper Motion (SPM 4.0; Girard et al. 2011) catalogues when available. When no entry in those catalogues existed, the corresponding magnitudes were transformed from the Carlsberg Meridian Catalog 15 (CMC15; cf. Dymock & Miles 2009) or UCAC3 catalogues (UCAC3; Zacharias et al. 2010; Pavlov 2009) using 2MASS colours.

The ASAS-3 magnitudes were also corrected for known zero-point offsets affecting some fields, especially in the Southern hemisphere. In order to do that, the closest GCPD constant field star with a published $V$ magnitude was selected and its magnitude compared with the ASAS-3 value. The differences we found range from 0.00 mag in Northern fields up to 0.05 mag in far Southern fields. Column six in Table 2 lists the magnitude ranges we obtained after applying these corrections.

Column seven indicates the total $V$ range of the star as gleaned from its recorded photometric history according to data from sources such as ASAS-3, GCPD, *Hipparcos* (van Leeuwen et al. 1997), APASS and SPM 4.0. HIPPARCOS data have been transformed to Johnson $V$ using the table in Bessell (2000). SPM 4.0 magnitudes have been used only when the catalogue flags state that the magnitudes are derived from CCD $V$ photometry. Comparisons between APASS and GCPD data indicate that APASS is well-tied to the standard system. We note that the ranges derived in this way should be treated with some caution. However, the magnitudes from the employed sources have shown consistency with the ASAS-3 light curves spanning several years. We therefore have no reason to suspect erroneous data in the few cases where the derived total ranges state brighter or fainter magnitudes than has actually been observed in ASAS-3.

The search for periodic signals was done using the software package PERIOD04, which has been based on a discrete Fourier transform algorithm and allows least-squares fitting of multiple frequencies to the data (Lenz & Breger 2005). In order to look for periodic signals, the whole datasets were searched in the frequency realm of 0-50 $\mathrm{d}^{-1}$ and consecutively prewhitened with the most significant frequency, until no significant frequencies remained. In case periodic signals were identified, the data were folded with the resulting frequencies and visually inspected. Only frequencies producing convincing phase plots were kept. Aperiodic signals like long-term variations and outbursts were identified by visual inspection.

**Table 1.** Statistical information on the composition of the final sample. We consider Be stars with spectral types earlier than B4 as *early-type Be stars*, objects with spectral types of B4, B5 and B6 as *mid-type Be stars*, and stars with spectral type of B7 and later as *late-type Be stars*.

| Type | number | (candidates thereof) |
|---|---|---|
| early-type Be stars | 101 | (14) |
| mid-type Be stars | 50 | (12) |
| late-type Be stars | 49 | (14) |
| unclassified Be stars | 87 | (14) |
| total number | 287 | (54) |

## 3 RESULTS

### 3.1 Presentation of results

Table 1 gives statistical information on the composition of the final sample. As described in Section 2.3, the division into early-type, mid-type and late-type Be stars has been based on available spectral types. Objects without a specific spectral type in the literature are listed as unclassified Be stars.

Table 2 presents essential data for our sample stars. We have chosen to employ GSC1.2 numbers as primary identifiers throughout the paper because they are available for all stars in our sample. Where available, more commonly used identifiers, such as HD numbers, are given in column three of Table 2. The light curves of all objects are presented in the Appendix (Figure B1).

We have also investigated the possibilities of colour excess and intrinsic colour determination with our sample stars. As this is not the main aim of the present work, these results have not been employed for the discussion of our results (cf. Section 4) and are presented in the Appendix (Section D).

### 3.2 Photometric variability on different time-scales

#### *3.2.1 Long-term variability*

Be stars are known to exhibit long-term changes in mean brightness on the order of years to decades. This kind of variability is linked with the presence or absence of the circumstellar disk, respectively its formation and dissipation. Emission from the disk is correlated with the brightness of the star. Photometric long-term variations may also originate in cyclic changes of the relative intensity of the violet (V) and red (R) peaks of the Balmer emission lines (termed V/R variations), accompanied by radial velocity changes of the emissions wings and the absorption core (Harmanec 1983). The presence of relatively long-lived one-armed density wave patterns in the disk may lead to cycle lengths, and hence photometric variability, on a time-scale of typically years (Okazaki 1991; Rivinius et al. 2013).

Exemplary light curves illustrating long-term variability are shown in Fig. 2. In most cases, the base-line of the ASAS-3 survey (∼10 yr) is insufficient to make a statement about the presence of periodicity in these variations. They appear mostly aperiodic, although wave-like oscillations are present in some objects, cf. e.g. GSC 06275-00943 (#272). Some stars only show slow brightenings or fadings during the observed time-span, as seen e.g. in GSC 09008-04083 (#197). Quite often, the long-term variations are interspersed with other forms of variability; note e.g. the occurrence of short-term variations in GSC 08877-00138 (#3). We have identified long-term variability in 102 stars (36±6 %) of our sample. However, it has to be noted that the differentiation between long-term variations and outbursts of very long duration (cf. Section 3.2.2) is somewhat arbitrary and, because of the limited time base of the ASAS-3 data, not always possible.

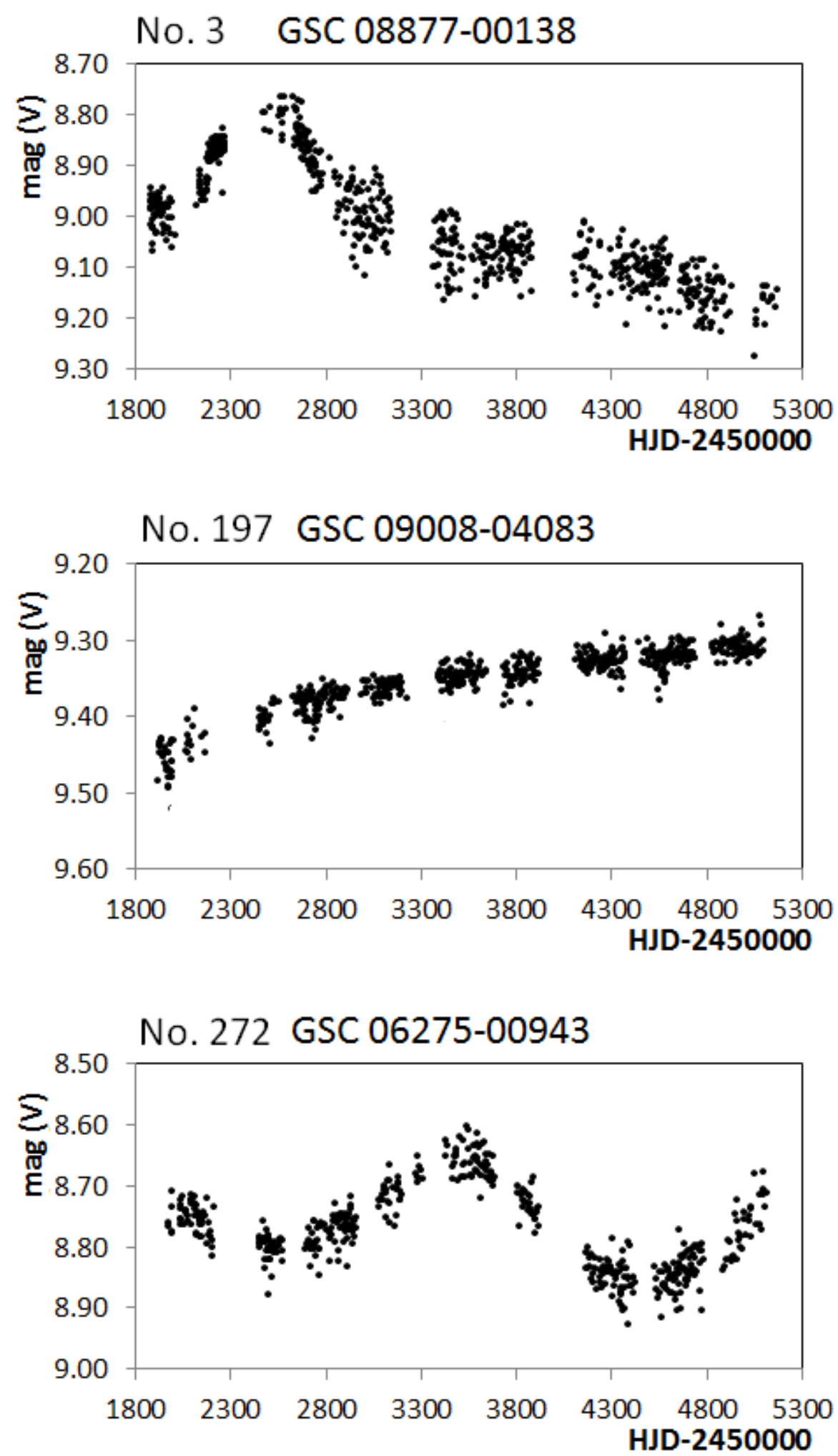

**Figure 2.** Exemplary ASAS-3 light curves of several Be stars of the present sample, illustrating long-term variations. The 'increased scatter' in the light curve of GSC 08877-00138 (#3), which is especially visible after ∼HJD 2452800, is due to short-term variations with $P = 1.10487(9)$ d.

#### *3.2.2 Outbursts*

LB17 have defined an outburst as a feature in the light curve delineated by a sharp departure from baseline brightness (either positive or negative) followed by a (generally) gradual return towards the baseline (cf. also Sigut & Patel 2013). While not always true, this is a reasonable working hypothesis, as has been shown e.g. by Labadie-Bartz et al. (2018), and will be enlarged on below in this section.

Following the approach of LB17, we have searched for the presence of outbursts in the light curves of all programme stars by visual inspection and noted their frequency of occurrence. Several circumstances, however, have complicated the counting process. Be star outbursts exhibit a bewildering diversity in duration, evolution and amplitude and it is sometimes hard to differentiate whether a feature is an outburst or some other form of variability. In addition, the superposition of several types of variability and continuously changing baseline flux have rendered the identification difficult. If an outburst has not ended before another one begins (i.e. the star rebrightens before attaining baseline brightness), both outbursts were taken into account (LB17). Generally, only stars showing unambiguous outbursts according to the definition given above have been considered. In agreement with LB17, we note that the outburst rates determined in this way are rather approximate and should be considered a lower limit, as outbursts with amplitudes below the detection threshold of the ASAS-3 data will likely be present. The occurrence of outbursts with amplitudes on the order of ∼10 mmag has been established using ultra-precise space-based photometry (Balona et al. 2015). Furthermore, at an inclination angle of ∼70°, a balance between excess emission and absorption by the disk has been predicted by the calculations of Haubois et al. (2012) and no net change in optical flux is expected, although the cancellation need not necessarily be perfect. Still, it is easily conceivable that outbursts will have been missed in such constellations.

**Table 2.** Essential data for our sample stars, sorted by increasing right ascension. The columns denote: (1) Internal identification number. Stars were numbered in order of increasing right ascension. (2) Identification from the Guide Star Catalog (GSC), version 1.2. (3) Identification from the Henry Draper (HD) catalogue or other conventional identifier. (4) Right ascension (J2000; Høg et al. 2000). (5) Declination (J2000; Høg et al. 2000). (6) $V$ magnitude range, as derived from ASAS-3 data. Ranges that have been corrected for light contamination from close neighbouring stars (cf. Section 2.5) are marked by an asterisk. (7) $V$ magnitude range, as gleaned from a star's recorded photometric history (cf. Section 2.5). (8) Spectral type from the literature. (9) Subtype (E=early; M=mid; L=late; U=unclassified). (10) Emission flag (emission confirmed: l=in literature/BeSS spectra; s=in own spectra; la=in LAMOST spectra; u=from *uvbyβ* photometry). (11) variability type, following LB17: ObV=outbursts present; SRO=semi-regular outbursts; LTV=long-term variability present; NRP=non-radial pulsator candidate (periodic variability with $P \leq 2$ d); IP=intermediate periodicity (periodic variability with $2 < P \leq 200$ d); EB=eclipsing binary. (12) variability type, following the GCVS and VSX classifications. (13) variability period(s), as derived from analysis of ASAS-3 data. Only part of the table is printed here for guidance regarding its form and content. The complete table is given in the appendix (Table A1).

| (1) No. | (2) ID GSC | (3) ID alt | (4) $\alpha$ (J2000) | (5) $\delta$ (J2000) | (6) Range($V$) [mag] | (7) Range($V$) lit. [mag] | (8) Spec.type lit. | (9) subtype [E/M/L/U] | (10) emission flag | (11) Var.type [LB17] | (12) Var.type [GCVS/VSX] | (13) Period(s) [d] |
|---|---|---|---|---|---|---|---|---|---|---|---|---|
| 1 | GSC 06464-00405 | HD 29557, NSV 16132 | 04 38 16.174 | -24 39 30.77 | 8.48-8.63 | 8.45-8.64 | B5Ib/IIp: shell? (Houk & Smith-Moore 1988b), B3Ve (Levenhagen & Leister 2006) | M | l | ObV | GCAS | |
| 2 | GSC 01845-02192 | HD 32318, NSV 16230 | 05 03 17.288 | +23 49 17.46 | 8.39-8.63 | 8.38-8.63 | B3IV (Straižys et al. 2003), A2e (McCarthy & Treanor 1965) | U | l | ObV | GCAS | |
| 3 | GSC 08877-00138 | HD 33599, NSV 16255 | 05 07 12.946 | -61 48 18.31 | 8.79-9.20 | 8.79-9.20 | B2Vep (Levenhagen & Leister 2006), B5p shell (Houk & Cowley 1975) | M | l, u | ObV, NRP | GCAS+LERI | 1.10487(9) |
| 4 | GSC 04755-00818 | HD 293881, ASAS J051449-0310.0 | 05 14 49.040 | -03 09 59.06 | 11.30-11.58* | 11.30-11.58 | B9 (Cannon & Mayall 1949), em (Wiramihardja et al. 1991) | L | l | LTV | GCAS | |
| 5 | GSC 09162-00751 | HD 269649, ASAS J053022-6919.7 | 05 30 22.553 | -69 19 38.95 | 10.41-10.81* | 10.41-10.81 | em (Howarth 2012), B2.5: (Rousseau et al. 1978) | E | l | SRO | GCAS | |
| 6 | GSC 00115-01423 | HD 37330, NSV 2478 | 05 37 53.455 | +00 58 06.98 | 7.30-7.47 | 7.30-7.49 | B6Vne (Warren & Hesser 1978) | M | l, u | LTV, NRP | GCAS+LERI | 0.77143(5) |
| 7 | GSC 01310-01587 | HD 37901, ASAS J054240+2134.7 | 05 42 39.812 | +21 34 43.29 | 8.95-9.08 | 8.95-9.08 | A0II/III (Hardorp et al. 1965) | L | | LTV | GCAS | |
| 8 | GSC 01311-01238 | HD 38191, ASAS J054456+2127.6 | 05 44 56.235 | +21 27 38.48 | 8.42-8.74 | 8.42-8.74 | B1:V:ne: (Morgan et al. 1955) | E | l | ObV, LTV | GCAS | |
| 9 | GSC 06491-00717 | HD 40193, SAO 171041 | 05 56 22.484 | -22 39 00.05 | 9.15-9.30* | 9.14-9.30 | B8Vne: (Houk & Smith-Moore 1988b), Be (Bidelman & MacConnell 1973) | L | l | ObV | GCAS | |
| 10 | GSC 01868-01264 | BD+25 1081, ASAS J060149+2537.9 | 06 01 49.500 | +25 37 52.73 | 10.34-10.44 | 10.23-10.44 | OB- (McCuskey 1967) | U | | LTV | GCAS | |

In this manner, outbursts were identified in 208 stars (73±5 %) of our sample. Outbursts were detected in 78 (77 %) of early-type Be stars (mean outburst rate of $N_{\rm ob} = 2.3 \pm 3.1\,yr^{-1}$), 33 (66 %) of mid-type Be stars (mean outburst rate of $N_{\rm ob} = 0.7 \pm 1.8\,yr^{-1}$), 28 (57 %) of late-type Be stars (mean outburst rate of $N_{\rm ob} = 0.2 \pm 0.4\,yr^{-1}$), and 69 (79 %) of unclassified Be stars (mean outburst rate of $N_{\rm ob} = 1.3 \pm 2.1\,yr^{-1}$). A surprisingly large number of these stars (24±6 %) exhibits outbursts with semi-regular or even distinct periodicity.

Some examples of outbursts of different amplitudes, durations, shapes and time-scales of (re)occurrence are provided in the following. Outbursts with very long durations are shown in Fig. 3. Repetitive outbursts are illustrated in Fig. 4, while outbursts of short duration are shown in Fig. 5. As can be clearly seen, outbursts occur either sporadically, as e.g. in GSC 08974-00327 (#174; Fig. 4), or with some regularity. The latter phenomenon is observed in particular among shorter outbursts. In some objects, the light curves are totally dominated by outbursts, and no quiescent phases of roughly constant brightness are present.

All objects exhibiting 'negative' outbursts according to the definition above have been considered as shell stars. These objects are seen roughly pole-on and the formation of the disk results in a reduced flux output of the system at visual wavelengths: instead of a brightening event, a drop in brightness is observed (cf. also Section 1). Fig. 6 illustrates some characteristic examples of shell star light curves. Note the unusual light curve of GSC 06853-02519 (#250; Fig. 6), which shows a drop in brightness exceeding 0.7 mag ($V$).

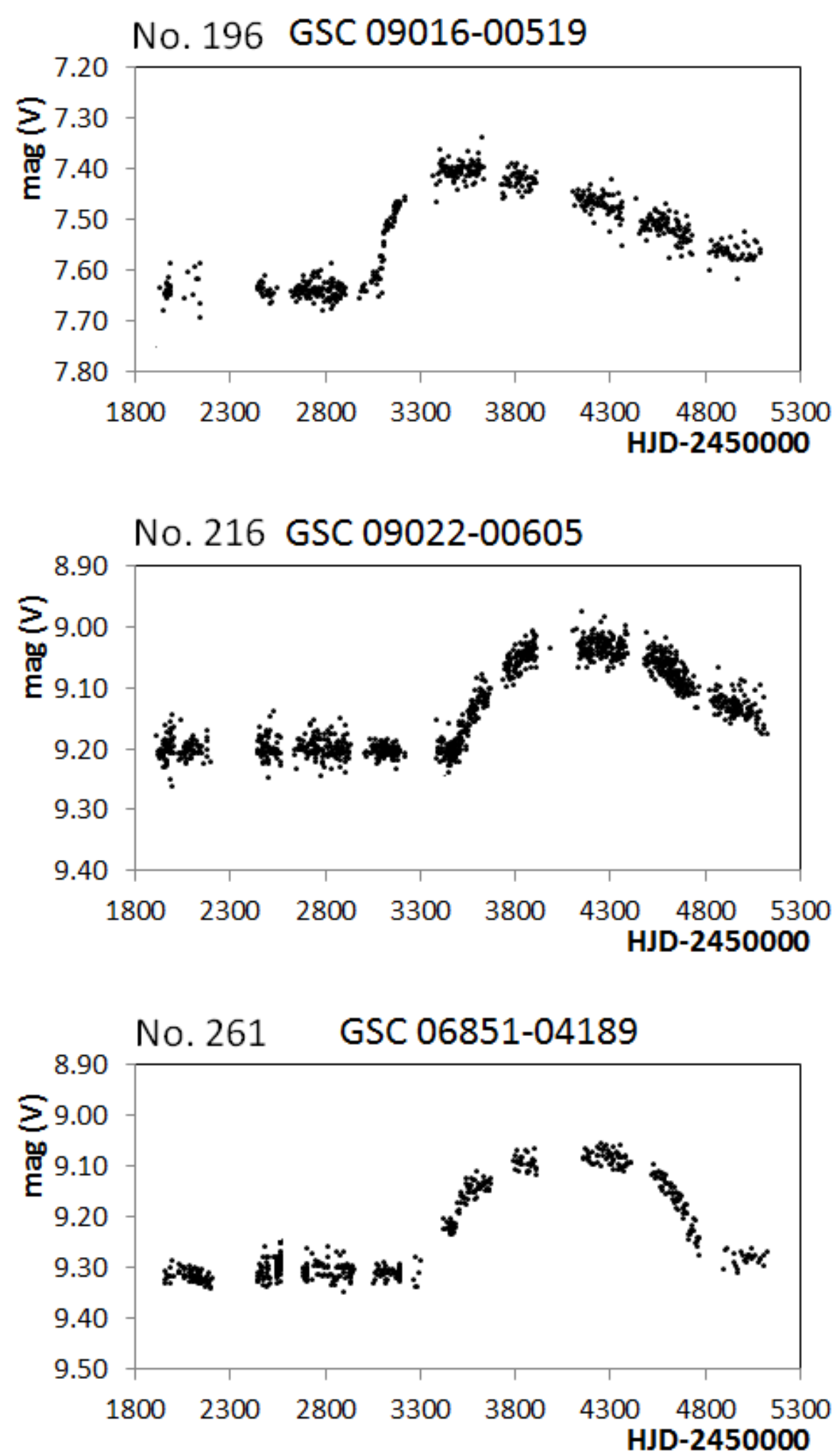

**Figure 3.** Exemplary ASAS-3 light curves of several Be stars of the present sample, illustrating outbursts of very long duration.

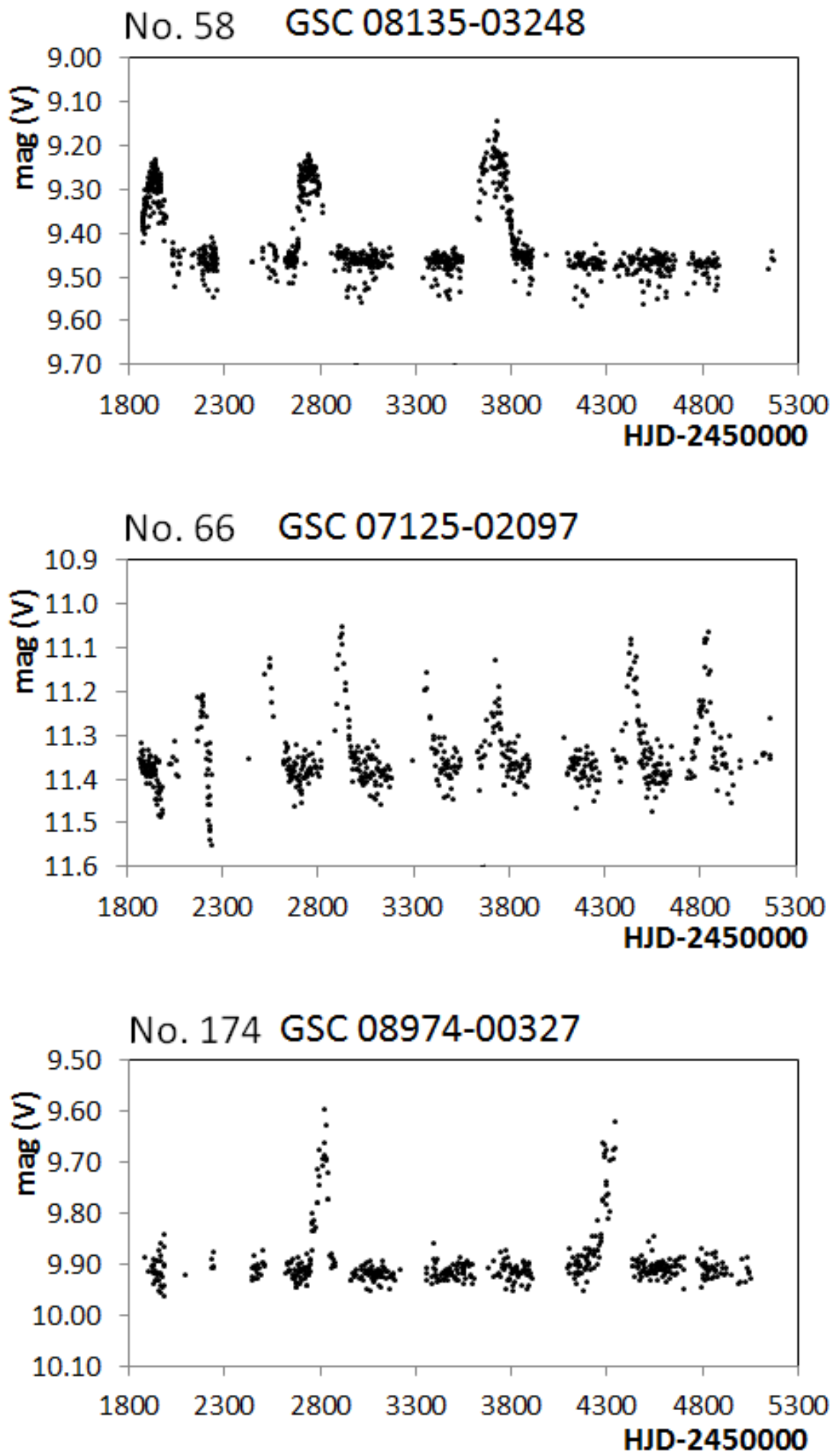

**Figure 4.** Exemplary ASAS-3 light curves of several Be stars of the present sample, illustrating repetitive outbursts. Note that GSC 08135-03248 (#58) is also an eclipsing binary with an orbital period of $P = 1.28212(1)$ d.

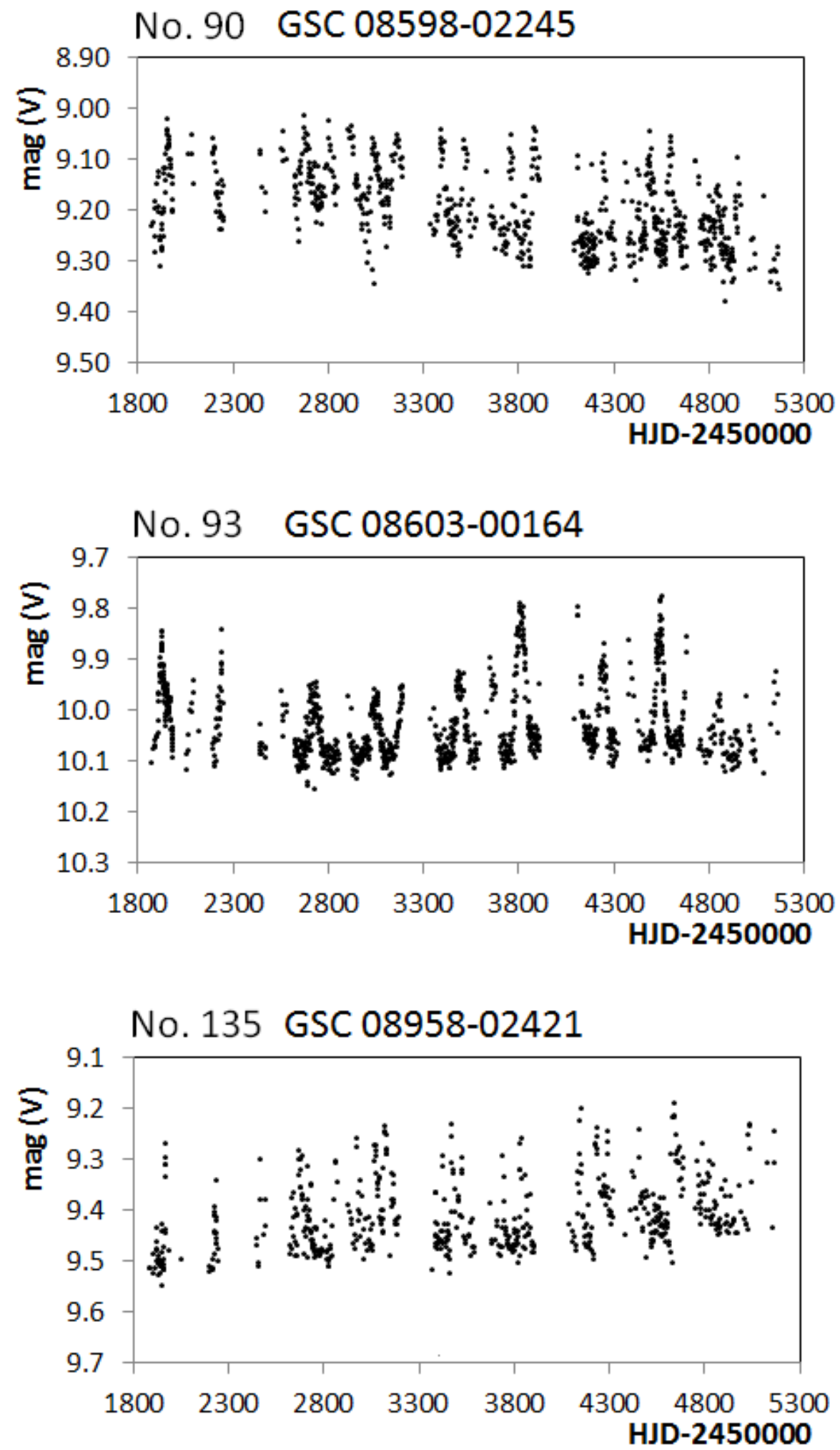

**Figure 5.** Exemplary ASAS-3 light curves of several Be stars of the present sample, illustrating outbursts of short duration.

We have measured peak-to-peak amplitudes for all objects exhibiting outburst events and calculated corresponding mean values (Fig. 7). Although the errors are considerable, our results indicate that the mean amplitude of outbursts decreases from early-type through late-type Be stars.

As a next step of analysis, the time spent in outburst among the stars of the different sub-types was measured. The results are presented in Fig. 8. There is an apparent tendency for earlier stars to spent more time in outburst which decreases towards later spectral types. However, the uncertainties are large, and the trend is not statistically significant.

Using KELT data, Labadie-Bartz et al. (2018) have correlated rising and falling times for 70 outbursts observed in 24 stars (18 early-, 4 mid-, and 2 late-type Be stars). They find that the vast majority of outburst events are indeed characterized by falling times exceeding the rising times, with a median rising time of 8.3 d and a median falling time of 16.0 d. Interestingly, for late-type Be stars, their results indicate that the ratio of falling time to rising time is significantly higher than for early- and mid-type objects (slope of the lines of best fit to the early-type, mid-type, and late-type stars are 1.97, 1.88, and 6.54, respectively). The authors conclude that their results suggest that, relative to the rising time, the inner disk dissipates quickly in hotter stars, and more slowly in cooler objects. They do, however, caution that their results suffer from a small sample size and significant scatter.

In order to investigate this matter with a larger sample size, we have measured rising and falling times for well-covered and well-defined outbursts in our sample stars. In order to render the results comparable, we followed the methodology outlined by Labadie-Bartz et al. (2018), who measured the duration of the rising and subsequent falling phases by a close visual inspection of the light curve. In this way, rising and falling times were correlated for 233 outbursts observed in 100 stars (41 early-, 14 mid-, 9 late-type, and 36 unclassified Be stars). In accordance with Labadie-Bartz et al. (2018), we confirm that nearly all outburst events are characterized by falling times that exceed the rising times. We find a median rising time of 28.5 d and a median falling time of 51.0 d, and slopes of 2.48, 3.35, and 2.72 for early-type, mid-type, and late-type stars, respectively.

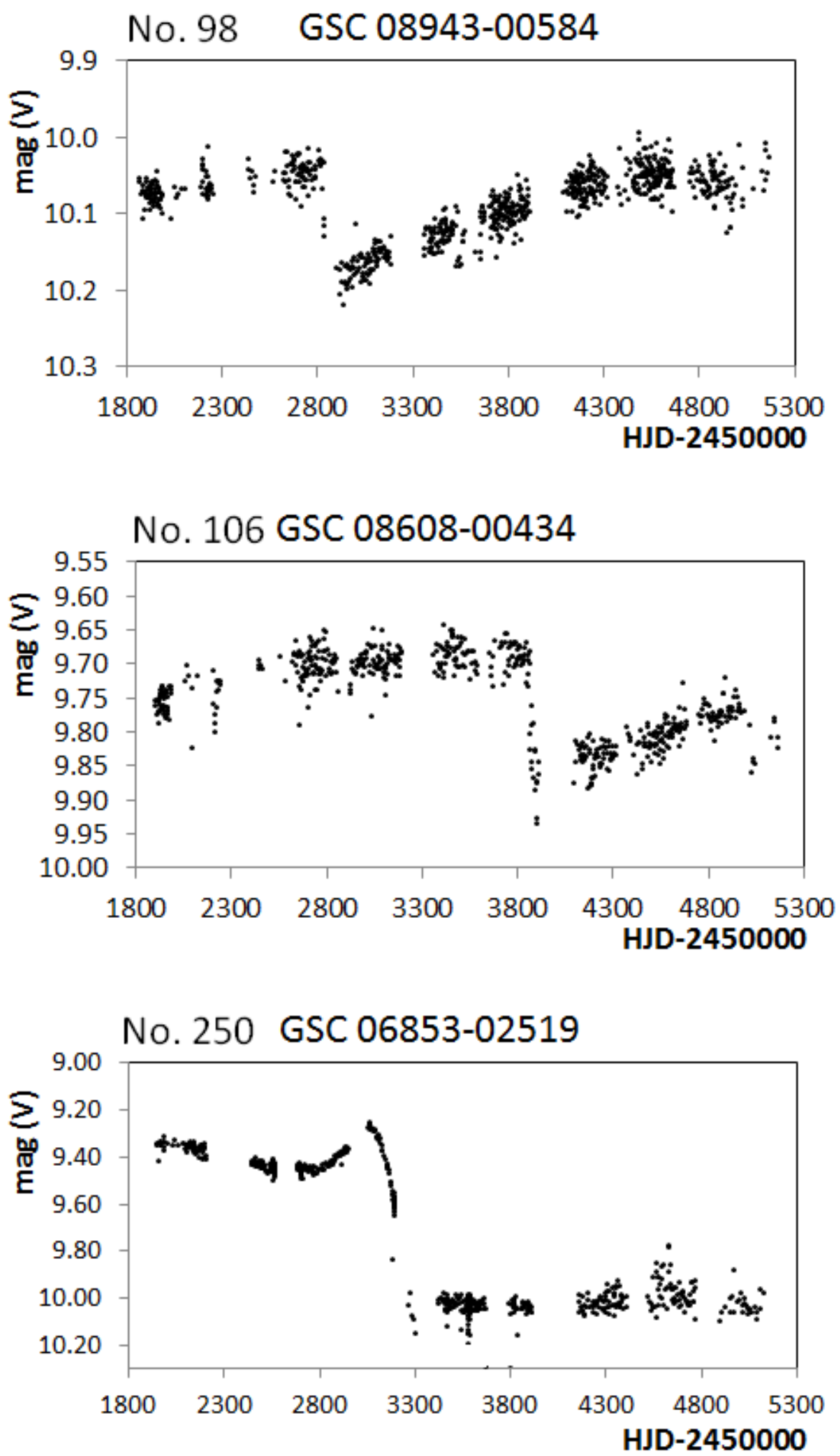

**Figure 6.** Exemplary ASAS-3 light curves of several shell stars of the present sample, illustrating the 'negative' outbursts typically observed in these objects.

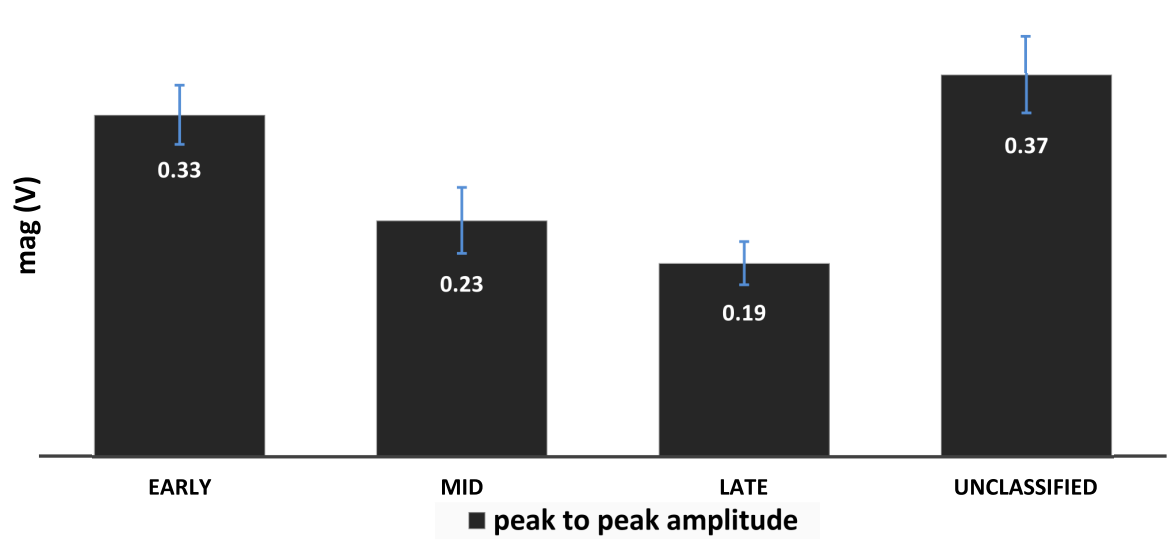

**Figure 7.** Mean peak-to-peak amplitudes of objects exhibiting outbursts among the different sub-types of Be stars, as derived from ASAS-3 data.

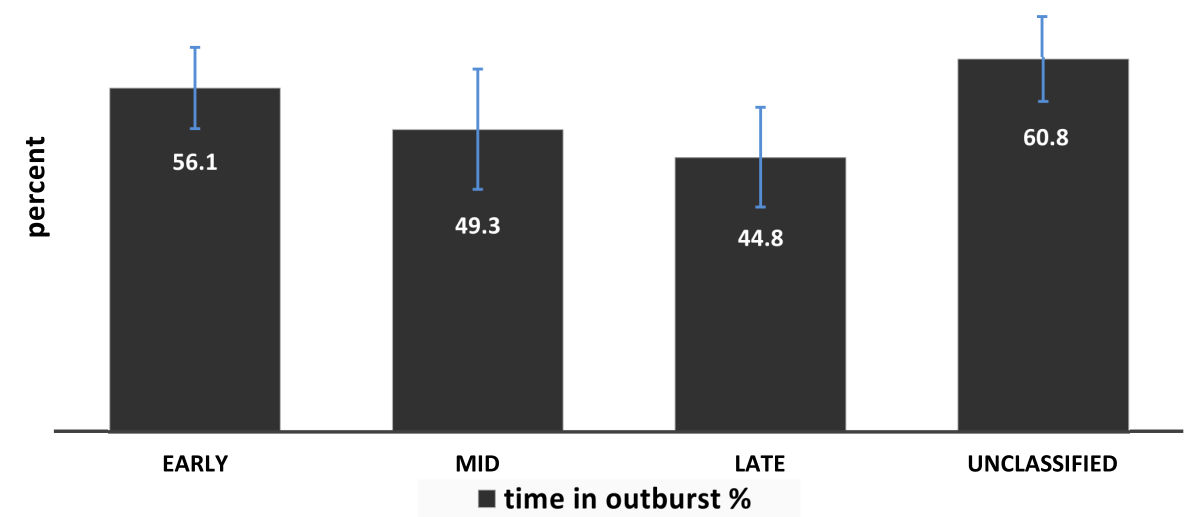

**Figure 8.** Time spent in outburst among stars of the different sub-types.

Interestingly, and in contrast to the results of Labadie-Bartz et al. (2018), we find that a single non-linear function adequately describes the ratio of falling time to rising time across all spectral subtypes, with the ratio being larger for short events (Fig. 9). No differences were found between early-, mid- and late-type stars, which suggests that the results of Labadie-Bartz et al. (2018) are influenced by their small sample size, as indeed cautioned by the authors. We note that in outburst events with short rising times up to about 65 days, the corresponding falling times seem to be only weakly related with the rising times.

Generally, the onset of an outburst is more sharply defined than its end; therefore, rise times are easier to determine than falling times. In consequence, the errors in the estimations of the falling times will be larger and may, in unfavourable cases, reach $\sim$10 %. However, the employed logarithmic representation in Fig. 9 is very robust against measurement uncertainties, and the derived relation still holds if we assume unrealistically high measurement errors on the order of $\sim$30 %. We are therefore confident that the observed deviation from linearity is not caused by increased relative uncertainties towards longer outburst events.

### *3.2.3 Periodic variations on intermediate time-scales*

Periodic variations on intermediate time-scales (days to months) are also frequently seen in Be stars. Diverse mechanisms have been deemed responsible for these variations, the most promising and widely-accepted theories being binarity-induced variability and disk-related phenomena (cf. Section 1). In particular, it has been shown that the decretion disk interacts both radiatively and gravitationally with the companion star (Panoglou et al. 2016, 2017).

In agreement with LB17, we define variability on intermediate time-scales (denoted as 'IP' in Table 2) as periodic variability with $2 < P \leq 200$ d (cf. in particular their Table 3). Outbursts which, of course, may also (re)occur on similar time-scales, are mostly semi-regular and have therefore been excluded from the 'IP' class.

We have identified only four stars (1±1 % of our sample) that clearly show variability on intermediate time-scales (Fig. 10), which is in striking contrast to the results of LB17 who have identified this type of variations in 39 % of their sample stars. However, as the amplitudes of IP variations are generally of very low amplitude (mostly ∼10 mmag according to the examples illustrated in LB17), this result is not surprising. KELT observations are of higher accuracy (typical photometric error of about 7 mmag for a single measurement, as opposed to about 10 mmag for ASAS-3 data; cf. Section 2.4) and boast superior temporal resolution (typical sampling cadence of 30 min, as opposed to mostly one to three measurements per day for ASAS-3 data).

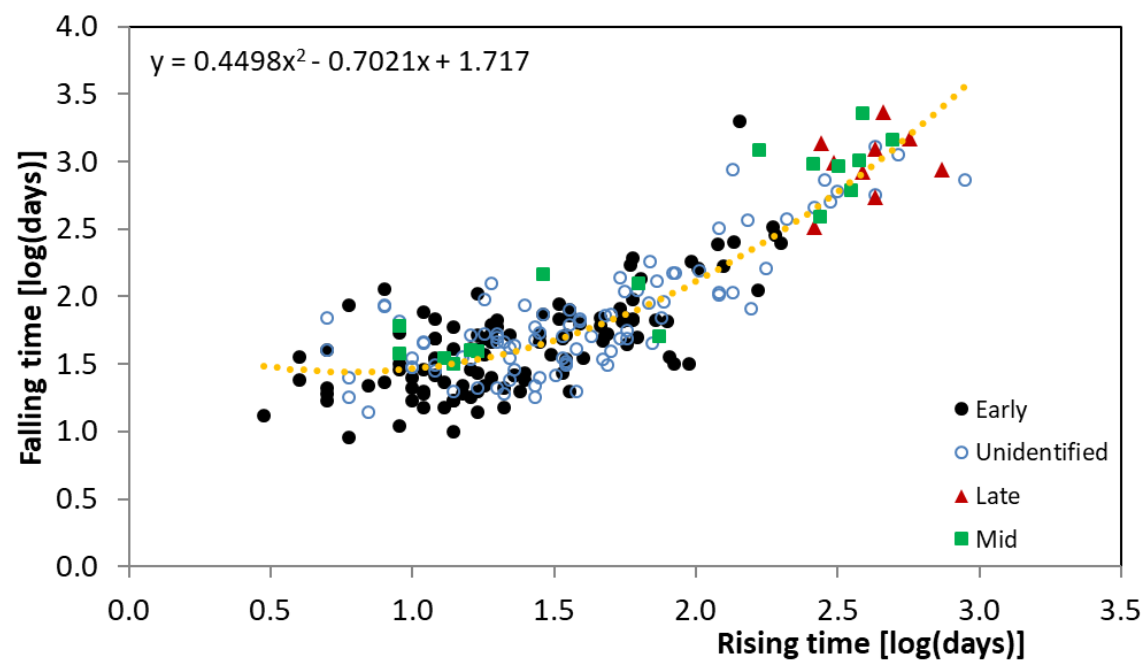

**Figure 9.** The duration of the rising and falling times for 233 outbursts observed in 100 stars (41 early-, 14 mid-, 9 late-type, and 36 unclassified Be stars, as indicated in the legend). The dotted line indicates a polynomial fit of the second order, which adequately describes the ratio of falling time to rising time across all spectral subtypes.

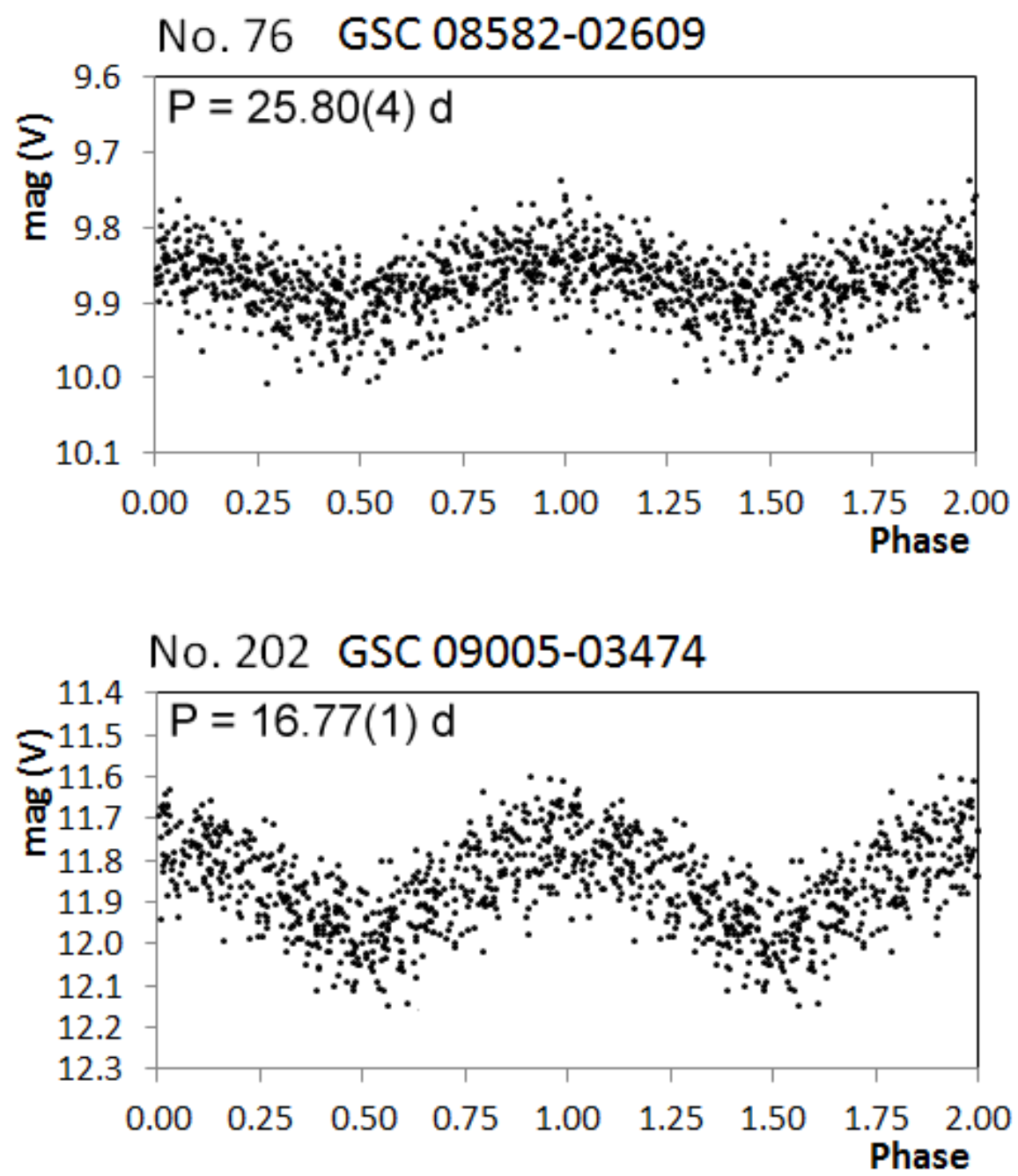

**Figure 10.** ASAS-3 light curves of two Be stars of the present sample that exhibit variations on intermediate time-scales. The light curves of GSC 08582-02609 (#76) and GSC 09005-03474 (#202) have been folded with, respectively, $P = 25.80(4)$ d and $P = 16.77(1)$ d. Unless indicated otherwise, here and throughout the paper, phase plots have been based on the whole available range of ASAS-3 data.

While under favourable circumstances, periodic signals with amplitudes as low as 3 mmag are detectable in ASAS-3 data (cf. Section 2.4), the complex and often irregular variability displayed by most of our sample stars is of considerably higher amplitude, which renders the detection of the low-amplitude IP variations very difficult - in particular in the low-cadence ASAS-3 data, which are not favourable for identifying signals at the shorter end of the IP periods. Consequently, the peak-to-peak amplitudes of the IP variations identified in this study are on the order of 20 mmag, while IP variations with a peak-to-peak amplitude as low as 3.2 mmag have been identified in KELT data (LB17). In summary, we put the observed discrepancy in numbers down to the different characteristics of the employed data sources.

### 3.2.4 *Periodic short-term variations*

As has been pointed out in the introduction, there is growing evidence that all Be stars are in fact non-radial pulsators. Some are suspected of showing transient modes lasting for days to months, which often precede outburst events (Gutiérrez-Soto et al. 2007; Huat et al. 2009). In agreement with LB17, short period variability is here defined as periodic variability with $P < 2$ d (denoted as 'NRP' in Table 2). Evidence for the ubiquitous presence of short-period pulsations comes mostly from ultra-precise space photometry. Pulsations with amplitude less than 1 mmag are commonly observed, which will be missed in ground-based surveys like ASAS or KELT.

In low-inclination Be stars, the variability is often dominated by variations in the disk, which may occult the stellar variability (Baade et al. 2016) and render the detection of low-amplitude variations impossible. The short period variations are also often superimposed by complex variations with much larger amplitude, which renders the detection of the periodic signals difficult. Furthermore, the strict observing cadence of the ASAS survey (typically one to three, and in favourable cases up to five, observations per day; cf. Pigulski 2014) is not favourable to resolving these high-frequency signals. This may explain why we could establish short period variability in only 6±3 % (17 objects) of our sample stars, which is in contrast to the findings of LB17 that detected NRP in 25 % of their sample. Among the 17 stars showing NRP variability, there are eight early-type Be stars, four mid-type Be stars, two late-type Be stars, and three unclassified Be stars. Some exemplary phase plots of stars exhibiting short-term variations are shown in Fig. 11.

### 3.2.5 *Eclipsing binary systems*

Two eclipsing binary systems are present in our sample. GSC 08135-03248 (#58) shows eclipses with a period of 1.28212(1) d and an amplitude of 0.08 mag. It also exhibits three outbursts in its light curve that reach amplitudes of up to 0.25 mag. Unfortunately, no recent spectroscopic classification is available for this object, which has been classified as B2V by Houk (1978). JD light curve and phase plot are shown in Fig. 12.

GSC 08975-00799 (#180) is an eclipsing binary system with an orbital period of 1.68445(5) d and an eclipse amplitude of only 0.06 mag. No outbursts are present in the light curve, which shows long-term variations typical of Be stars. The most recent spectroscopic classification is from Houk & Cowley (1975), who derived a spectral type of B3II. JD light curve and phase plot are illustrated in Fig. 13. We note that the eclipse amplitude is small and the observational scatter rather large. We therefore cannot definitely exclude an orbital period of 0.84223(3) d (half the given value). However, we are confident of our period solution, as the corresponding phase plots indicates the presence of minima of different depth.

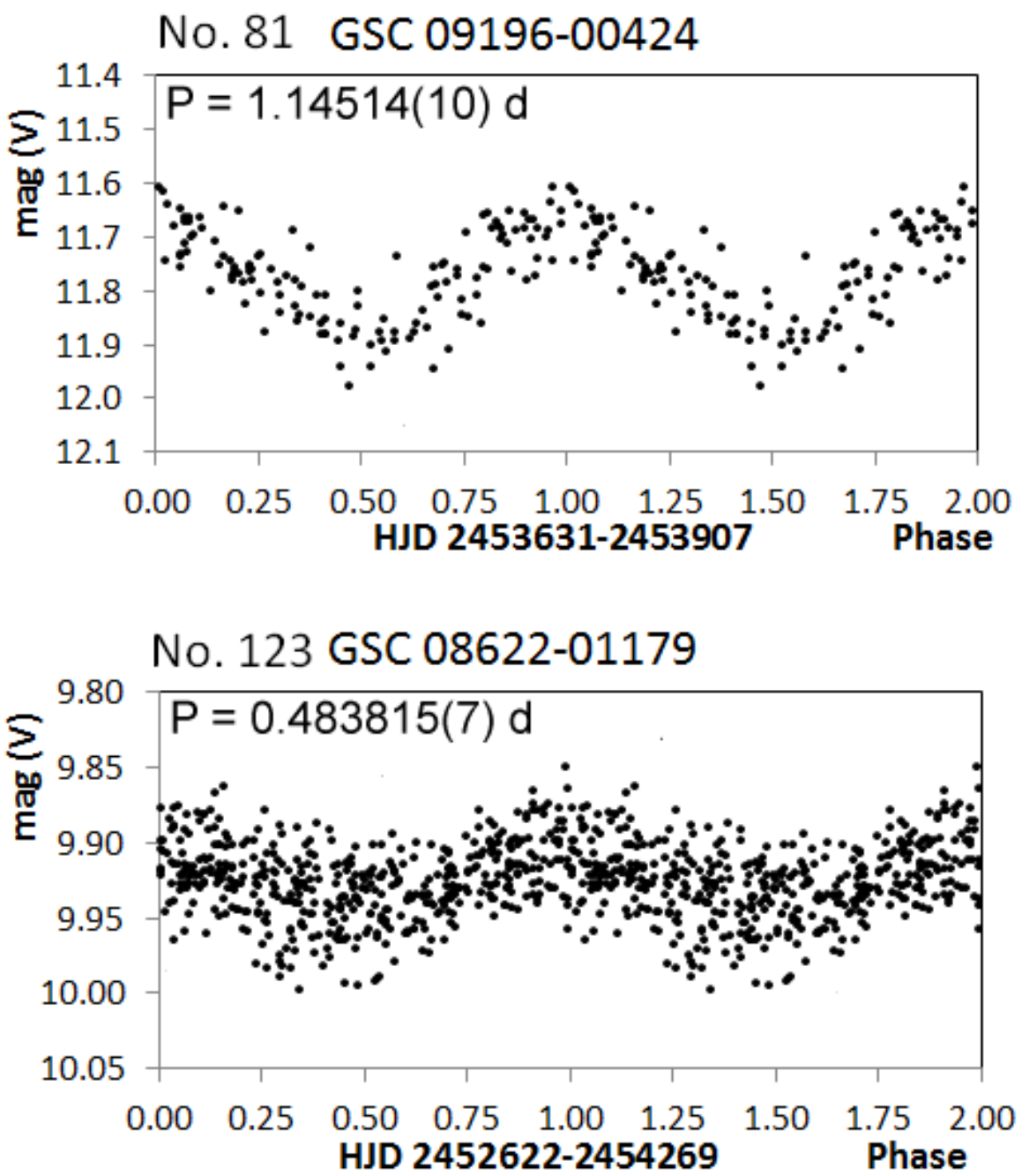

**Figure 11.** Exemplary ASAS-3 light curves of two Be stars of the present sample, illustrating variations on short time-scales. The light curves of GSC 09196-00424 (#81) and GSC 08622-01179 (#123) have been folded with, respectively, $P = 1.14514(10)$ d and $P = 0.483815(7)$ d. In order to bring out the short-term variations more clearly, the phase plots have been based on only part of the ASAS-3 data, as indicated below the abscissae.

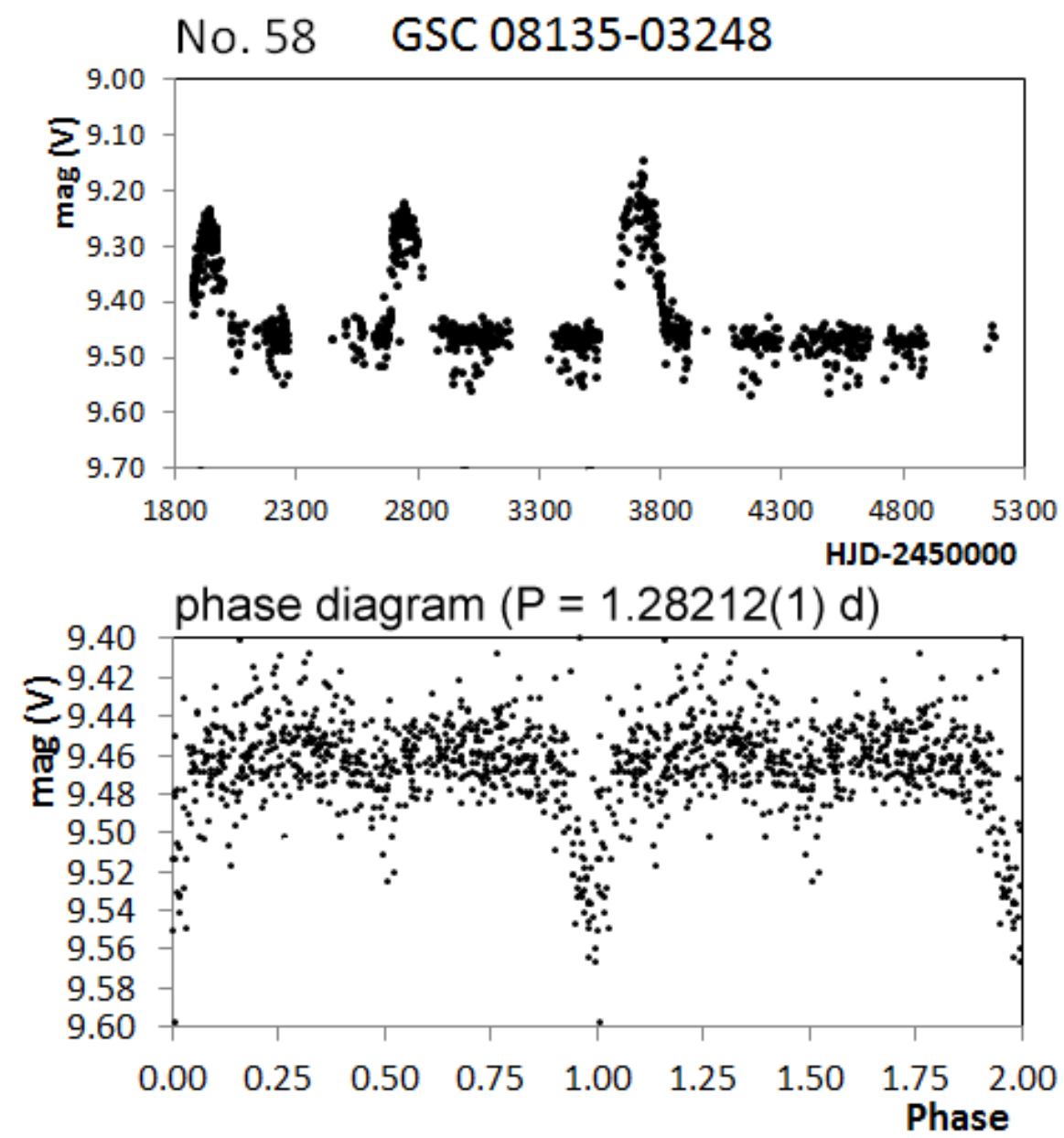

**Figure 12.** ASAS-3 light curve of the eclipsing binary GSC 08135-03248 (#58). Upper and lower panel illustrate the JD light curve and the phase plot ($P = 1.28212(1)$ d), respectively. Outbursts have been clipped in the lower panel.

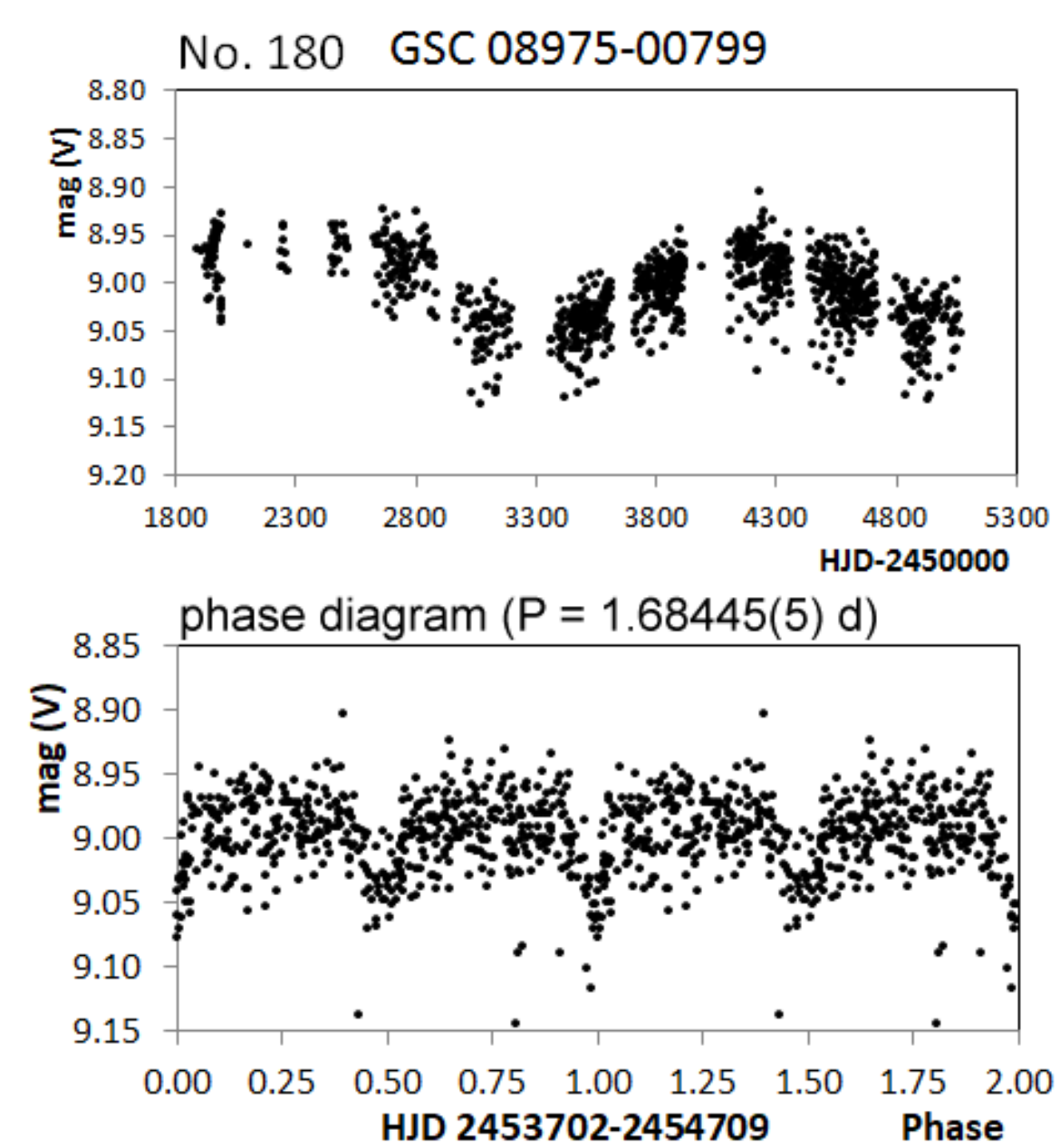

**Figure 13.** ASAS-3 light curve of the eclipsing binary GSC 08975-00799 (#180). Upper and lower panel illustrate the JD light curve and the phase plot ($P = 1.68445(5)$ d), respectively. In order to bring out the eclipses more clearly, the phase plot has been based on only part of the ASAS-3 data, as indicated below the abscissa.

Both objects show light curve morphology in agreement with a Be star classification and, judging from their eclipse profiles and periods, might be detached systems harbouring classical Be stars. In accordance with the approach of LB17, we have therefore chosen to keep them in the sample. While we have no evidence for the presence of emission features in GSC 08135-03248, *uvbyβ* photometry indicates that GSC 08975-00799 (#180) is indeed an emission-line star. However, we caution that only further spectroscopic investigations will be able to settle this question conclusively. If proven to indeed harbour classical Be stars, these binary systems would be of considerable interest, as Be stars with short orbital periods are rare objects and it is not clear how decretion disks can develop around a single star in such a system.

## 4 DISCUSSION

LB17 have correlated the observed frequency of outbursts with spectral sub-type. In agreement with the literature (Rivinius et al. 2013), they have found that early-type Be stars are generally more active than mid-type and late-type objects. However, recent results based on space photometry (Baade et al. 2016) suggest that (part of) this phenomenon may be rather due to differences in amplitude than outburst activity. Shokry et al. (2018) confirm that late-type Be stars are generally less variable than their early-type counterparts.

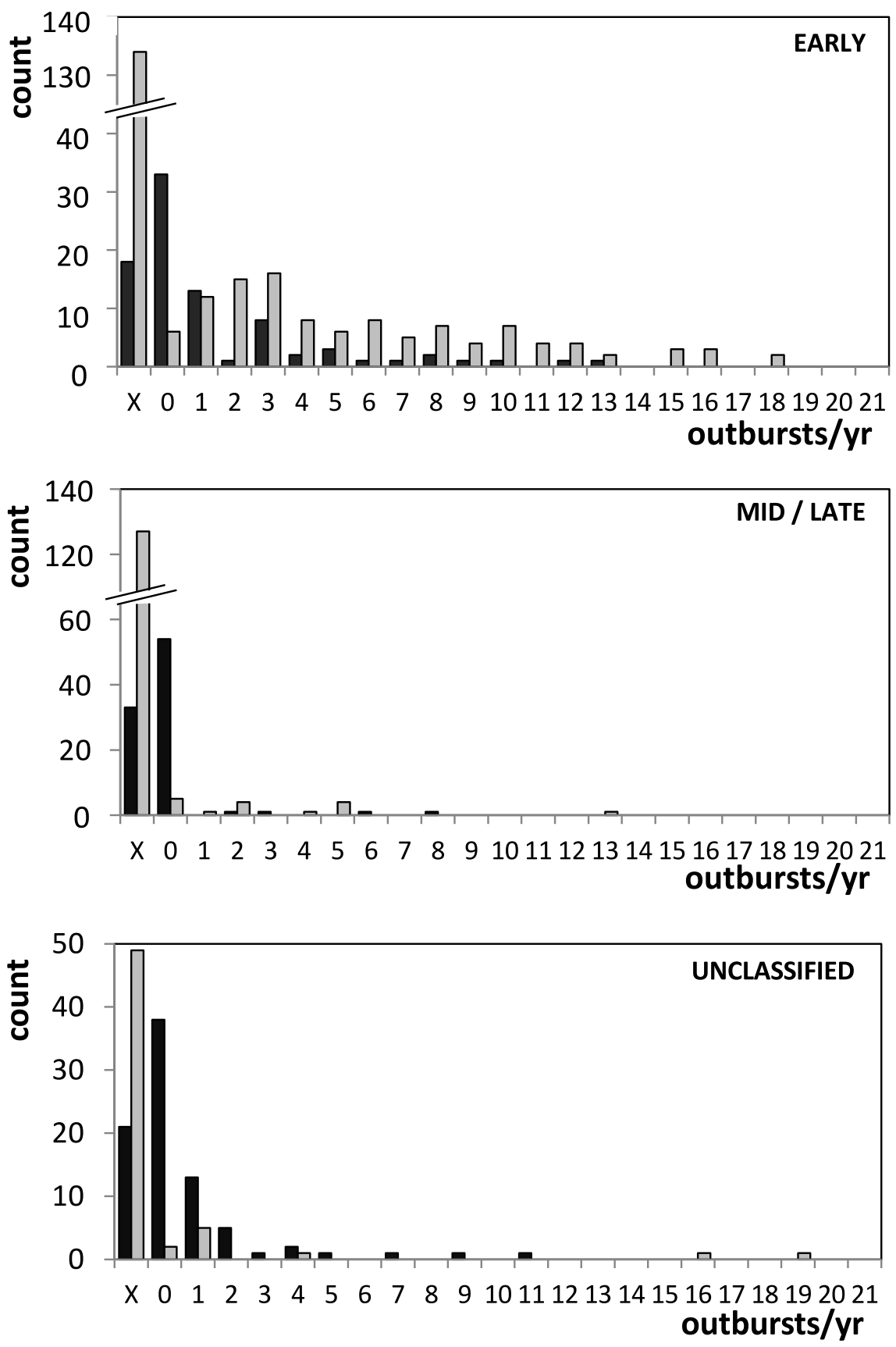

**Figure 14.** Outbursts per year observed among the different sub-types of Be stars. Dark and light bars indicate, respectively, the results of this investigation and LB17. The columns denoted as 'X' give the number of stars that do not show definite outburst signatures in the covered time span. Note the broken ordinates in the upper and middle panels, and the high fraction of stars with outbursts among the unclassified Be stars, which is due to a bias in the preselection process of Be star candidates via light curve morphology criteria (cf. Section 2.1).

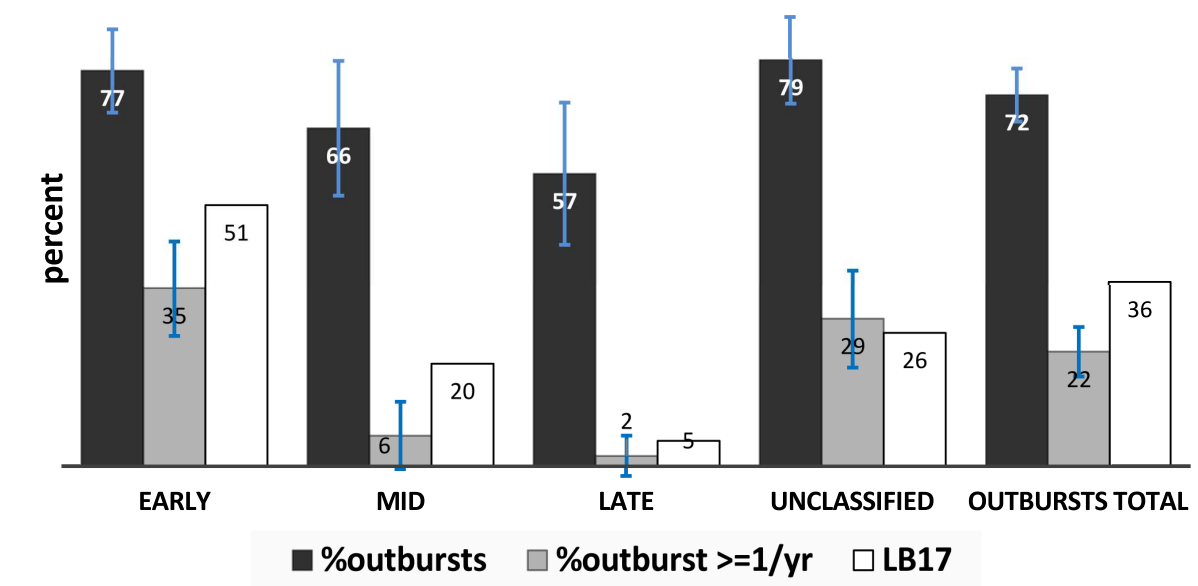

**Figure 15.** Percentage of stars exhibiting outbursts, as derived in the present study and LB17. The black bars denote the percentage of stars showing outbursts in this study, while the grey bars indicate the percentage of stars with ≥1 outburst $\mathrm{yr}^{-1}$ in this study. The open bars represent the percentage of stars with outbursts in LB17. Please note that for the results from LB17, no uncertainties are available. Results are indicated for the analysis of (from left to right) early-type Be stars, mid-type Be stars, late-type Be stars, unclassified Be stars and the total sample of stars.

However, their results indicate that the Be/B star fraction may not strongly depend on spectral subtype and that late-type Be stars might be more common than is generally agreed upon.

A breakdown of the number of outbursts per year observed among the different sub-types of Be stars is provided in Fig. 14. Outburst rates have been determined as number of outbursts per monitoring interval and subsequently normalized to frequency per year. Also included is information on the number of stars that do not exhibit outbursts in the covered time. It is obvious that a significant amount of early-type stars shows numerous outbursts per year, while mid-type and late-type objects are obviously less active in that respect. We thus confirm the general trend that, in the accuracy limit of the employed data, early-type Be stars show more frequent outbursts (mean outburst rate of $N_{\mathrm{ob}} = 2.3{\pm}3.1\,yr^{-1}$) than mid- ($N_{\mathrm{ob}} = 0.7{\pm}1.8\,yr^{-1}$) and late-type objects ($N_{\mathrm{ob}} = 0.2{\pm}0.4\,yr^{-1}$).

From this figure, it is apparent that the fraction of stars with outbursts is highest in the unclassified Be stars. This is due to a bias in the preselection process of Be star candidates via light curve morphology criteria (cf. Section 2.1), which favours objects exhibiting clear Be star variability signatures, and the most dominant and readily recognizable feature in this respect is the presence of outbursts in the light curve.

However, in terms of numbers, our results differ significantly from those of LB17, in particular in regard to the outburst rates of mid-type and late-type Be stars. The discrepancy in numbers is most pronounced for objects with rare outbursts (<1 outburst $\mathrm{yr}^{-1}$; Fig. 14). For example, among mid-type and late-type Be stars, LB17 have identified only five such objects (cf. their Figure 1), which is in contrast to the 54 objects identified in the present study.

The considerable uncertainties inherent to the counting process (cf. Section 3.2.2) might go some way in explaining the observed discrepancies. However, the most important aspect is surely the difference in time baseline between the employed data sources. While both ASAS and KELT boast coverage up to ∼10 yr, the number of stars actually attaining maximum temporal coverage is vastly dissimilar. LB17 find that only 217 (out of the 510 investigated) objects (∼43 %) have KELT data with a time baseline exceeding four years. In contrast, data with a time baseline of 10 yr and near-continuous sampling exist for practically all of our sample stars. Longer datasets are much more suited to the identification of Be stars with rare outbursts. In fact, most of our sample stars exhibit <1 outburst $\mathrm{yr}^{-1}$ (Fig. 14), with many objects showing only a single outburst during the ∼10 yr of ASAS-3 coverage. These stars may easily have been missed in surveys with shorter time baseline. The impact of the time baseline on the detection of outbursts is also illustrated by a comparison of the vastly discrepant number of stars without any outbursts signatures in both studies (Fig. 14).

In summary, with our work, which is based on a data source boasting a near-continuous (albeit sparsely-sampled) coverage of ∼10 yr for almost all sample stars, we are able to close the apparent void of 'rare outbursters' proposed by LB17 and show that Be stars with infrequent outbursts are not rare.

An overview over the percentage of stars that show outbursts among the different sub-types of Be stars is provided in Fig. 15. Also illustrated are the results of LB17 (open bars), who find that outbursts have a higher occurrence rate in early-type stars than in mid- and late-type objects. However, no statistically significant trend is seen in the results for the ASAS sample (dark bars). While we do find a decreasing incidence of stars showing outbursts from early-type to late-type objects, the uncertainties are large, and the distribution is essentially flat. Outbursts, therefore, are a common characteristic of Be stars of all subtypes.

As the longer time baseline of ASAS renders these data more suitable to identifying objects with rare outbursts, we have recalculated our results taking into account only objects with ≥1 outburst $yr^{-1}$ (grey bars), as these frequent outbursters would have been detected more easily in the shorter time-baseline KELT data. Interestingly, the resulting distribution indeed agrees with the results of LB17, and the difference in percentage between outbursting early-type and later-type Be stars is statistically significant. It therefore seems likely that the results of LB17 have been influenced by their data source's bias against objects with rare outbursts.

24±6 % of our sample stars exhibit semi-regular or even distinctly periodic outbursts. The observed periods are widely different, ranging from e.g. $P \sim 120.5(8)$ d for GSC 08598-02245 (#90) to $P \sim 1560(50)$ d for GSC 05978-00030 (#48). This is in agreement with the results of LB17, who identified semi-regular outbursts in 21 % of their sample stars. This rate is surprisingly high, and theory needs to account for these (semi)regular outbursts, which imply that the mechanism responsible for triggering mass loss must be capable of switching on and off on short time scales – often with some periodicity. In our opinion, this is another point in favour of non-radial pulsation as the dominant mechanism involved, as discussed in Baade et al. (2017b).

The fraction of stars showing long-term variations is similar in both studies (LB17: 37 %; this study: 36±6 %). However, there are striking differences in number for stars showing variations on intermediate and short time-scales (cf. Section 3.2). As has been pointed out, these differences are likely not real but due to the characteristics of the employed data source and the accompanying difficulties in data analysis.

Depending on the inclination angle of the star-disk system, brightenings of up to about 0.4 mag and fadings of about 0.2 mag in the $V$-band have been predicted (Haubois et al. 2012). While these predictions seem to be accurate for the brightening events, there are several shell stars in our sample that exhibit drops in brightness significantly exceeding the predicted limit, most notably GSC 04826-01079 (#39; drop in brightness of about 0.4 mag), GSC 08338-02080 (#230; 0.5 mag) and GSC 06853-02519 (#250; ~0.7 mag). An explanation for this difference is still pending. In GSC 08338-02080 (#230), the drops in brightness are rather symmetric and seem to occur with a period of roughly 600 d. Perhaps, then, some kind of eclipses are at the root of the observed light changes in this star. It will be interesting to investigate what (if any) other mechanisms contribute to the observed drops in brightness, which might provide valuable constraints for theoretical modeling efforts.

## 5 CONCLUSION

With our study, we significantly enlarge the sample of Galactic Be stars with a detailed description of their photometric variability properties, building on the work of LB17. Wherever appropriate, we have employed the methodological approach of the aforementioned investigators in order to render the results comparable. Together with the results of LB17, our sample will greatly facilitate future research on these objects. The main findings of the present study are summarized in the following:

- Complex photometric variations were established in most of our sample stars: outbursts on different time-scales (in 73±5 % of stars), long-term variations (36±6 %), variations on intermediate time-scales (1±1 %) and short-term variations indicative of non-radial pulsation (6±3 %).
- A surprisingly large number (24±6 %) of outbursting stars exhibit semi-regular or even distinctly periodic outbursts.
- GSC 08135-03248 (#58) and GSC 08975-00799 (#180) have been identified as eclipsing binary systems that might harbour a classical Be star component. These systems are potentially of great interest and merit further attention.
- Using a data source that boasts a near-continuous coverage of ~10 yr for almost all sample stars, we are able to close the apparent void of 'rare outbursters' proposed by LB17 and show that Be stars with infrequent outbursts are not rare.
- We confirm the general trend that, in the accuracy limit of the employed data, early-type Be stars show more frequent outbursts (mean outburst rate of $N_{\rm ob} = 2.3 \pm 3.1\,yr^{-1}$) than mid- ($N_{\rm ob} = 0.7 \pm 1.8\,yr^{-1}$) and late-type objects ($N_{\rm ob} = 0.2 \pm 0.4\,yr^{-1}$).
- On the other hand, we cannot confirm the finding of LB17 that outbursts have a higher occurrence rate in early-type stars than in mid- and late-type objects. This is likely a result of their data source's bias against objects with rare outbursts. No statistically significant trend is seen in the results for the ASAS sample, and the resulting distribution is essentially flat. Outbursts, therefore, are a common characteristic of Be stars of all subtypes.
- Our results indicate that the mean amplitude of outbursts decreases from early-type through late-type Be stars.
- We have measured rising and falling times for well-covered and well-defined outbursts in our sample stars and confirm that nearly all outburst events are characterized by falling times that exceed the rising times (median rising time of 28.5 d; median falling time of 51.0 d). No differences were found between early-, mid- and late-type stars. Instead, we find that a single non-linear function adequately describes the ratio of falling time to rising time across all spectral subtypes, with the ratio being larger for short events.

## ACKNOWLEDGEMENTS

We thank our referee, Dr. Dietrich Baade, for his thoughtful and detailed report which helped to greatly improve the paper. Guoshoujing Telescope (the Large Sky Area Multi-Object Fiber Spectroscopic Telescope LAMOST) is a National Major Scientific Project built by the Chinese Academy of Sciences. Funding for the project has been provided by the National Development and Reform Commission. LAMOST is operated and managed by the National Astronomical Observatories, Chinese Academy of Sciences. This research has made use of the SIMBAD database and the VizieR catalogue access tool, operated at CDS, Strasbourg, France, and the AAVSO Photometric All-Sky Survey (APASS), funded by the Robert Martin Ayers Sciences Fund.

## APPENDIX A: ESSENTIAL DATA

Table A1 Essential data for our sample stars, sorted by increasing right ascension. The columns denote: (1) Internal identification number. Stars were numbered in order of increasing right ascension. (2) Identification from the Guide Star Catalog (GSC), version 1.2. (3) Identification from the Henry Draper (HD) catalogue or other conventional identifier. (4) Right ascension (J2000; Høg et al. 2000). (5) Declination (J2000; Høg et al. 2000). (6) *V* magnitude range, as derived from ASAS-3 data. Ranges that have been corrected for light contamination from close neighbouring stars (cf. Section 2.5) are marked by an asterisk. (7) *V* magnitude range, as gleaned from a star's recorded photometric history (cf. Section 2.5). (8) Spectral type from the literature. (9) Subtype (E=early; M=mid; L=late; U=unclassified). (10) Emission flag (emission confirmed: l=in literature/BeSS spectra; s=in own spectra; la=in LAMOST spectra; u=from *uvbyβ* photometry). (11) variability type, following LB17: ObV=outbursts present; SRO=semi-regular outbursts; LTV=long-term variability present; NRP=non-radial pulsator candidate (periodic variability with $P \leq 2$ d); IP=intermediate periodicity (periodic variability with $2 < P \leq 200$ d); EB=eclipsing binary. (12) variability type, following the GCVS and VSX classifications. (13) variability period(s), as derived from analysis of ASAS-3 data.

| (1) No. | (2) ID GSC | (3) ID alt | (4) $\alpha$ (J2000) | (5) $\delta$ (J2000) | (6) Range(*V*) [mag] | (7) Range(*V*) lit. [mag] | (8) Spec.type lit. | (9) subtype [E/M/L/U] | (10) emission flag | (11) Var.type [LB17] | (12) Var.type [GCVS/VSX] | (13) Period(s) [d] |
|---|---|---|---|---|---|---|---|---|---|---|---|---|
| 1 | GSC 06464-00405 | HD 29557, NSV 16132 | 04 38 16.174 | -24 39 30.77 | 8.48-8.63 | 8.45-8.64 | B5Ib/IIp: shell? (Houk & Smith-Moore 1988b), B3Ve (Levenhagen & Leister 2006) | M | l | ObV | GCAS | |
| 2 | GSC 01845-02192 | HD 32318, NSV 16230 | 05 03 17.288 | +23 49 17.46 | 8.39-8.63 | 8.38-8.63 | B3IV (Straižys et al. 2003), A2e (McCarthy & Treanor 1965) | U | l | ObV | GCAS | |
| 3 | GSC 08877-00138 | HD 33599, NSV 16255 | 05 07 12.946 | -61 48 18.31 | 8.79-9.20 | 8.79-9.20 | B2Vep (Levenhagen & Leister 2006), B5p shell (Houk & Cowley 1975) | M | l, u | ObV, NRP | GCAS+LERI | 1.10487(9) |
| 4 | GSC 04755-00818 | HD 293881, ASAS J051449-0310.0 | 05 14 49.040 | -03 09 59.06 | 11.30-11.58* | 11.30-11.58 | B9 (Cannon & Mayall 1949), em (Wiramihardja et al. 1991) | L | l | LTV | GCAS | |
| 5 | GSC 09162-00751 | HD 269649, ASAS J053022-6919.7 | 05 30 22.553 | -69 19 38.95 | 10.41-10.81* | 10.41-10.81 | em (Howarth 2012), B2.5: (Rousseau et al. 1978) | E | l | SRO | GCAS | |
| 6 | GSC 00115-01423 | HD 37330, NSV 2478 | 05 37 53.455 | +00 58 06.98 | 7.30-7.47 | 7.30-7.49 | B6Vne (Warren & Hesser 1978) | M | l, u | LTV, NRP | GCAS+LERI | 0.77143(5) |
| 7 | GSC 01310-01587 | HD 37901, ASAS J054240+2134.7 | 05 42 39.812 | +21 34 43.29 | 8.95-9.08 | 8.95-9.08 | A0II/III (Hardorp et al. 1965) | L | | LTV | GCAS | |
| 8 | GSC 01311-01238 | HD 38191, ASAS J054456+2127.6 | 05 44 56.235 | +21 27 38.48 | 8.42-8.74 | 8.42-8.74 | B1:V:ne: (Morgan et al. 1955) | E | l | ObV, LTV | GCAS | |
| 9 | GSC 06491-00717 | HD 40193, SAO 171041 | 05 56 22.484 | -22 39 00.05 | 9.15-9.30* | 9.14-9.30 | B8Vne: (Houk & Smith-Moore 1988b), Be (Bidelman & MacConnell 1973) | L | l | ObV | GCAS | |
| 10 | GSC 01868-01264 | BD+25 1081, ASAS J060149+2537.9 | 06 01 49.500 | +25 37 52.73 | 10.34-10.44 | 10.23-10.44 | OB- (McCuskey 1967) | U | | LTV | GCAS | |
| 11 | GSC 00721-02056 | HD 250980, ASAS J060334+0939.9 | 06 03 34.246 | +09 39 54.65 | 9.45-9.58 | 9.45-9.58 | Be (Bidelman & MacConnell 1973), B2ne (Popper 1950) | E | l | LTV | GCAS | |
| 12 | GSC 01864-00314 | ALS 8684, ASAS J060445+2324.1 | 06 04 45.296 | +23 24 07.25 | 11.25-11.57 | 11.25-11.57 | OB (McCuskey 1967) | U | la | ObV | GCAS | |
| 13 | GSC 00738-01213 | ASAS J060954+1300.3 | 06 09 53.686 | +13 00 12.93 | 11.20-11.72 | 11.20-11.72 | OB:e (Stephenson & Sanduleak 1977a), binary | U | l, la | SRO | GCAS | 360(5) |
| 14 | GSC 00742-01475 | ASAS J061353+1418.0 | 06 13 52.522 | +14 18 02.15 | 10.89-11.24 | 10.89-11.24 | pec (McCuskey 1959) | U | | ObV | GCAS | |
| 15 | GSC 00739-01143 | BD+11 1084, ASAS 061535+1125.9 | 06 15 35.038 | +11 25 52.86 | 10.27-10.50 | 10.27-10.54 | OBe (Nassau et al. 1965) | U | l, la | ObV | GCAS | |
| 16 | GSC 00739-01342 | HD 254329, ASAS J061624+1223.8 | 06 16 24.004 | +12 23 50.06 | 9.20-9.50 | 9.20-9.50 | OBe (Nassau et al. 1965), B8e (Merrill & Burwell 1943) | L | l | ObV | GCAS | |
| 17 | GSC 01319-00734 | HD 255103, ASAS J061940+1822.3 | 06 19 39.964 | +18 22 17.30 | 10.49-10.84 | 10.48-10.84 | OB+e (Stephenson & Sanduleak 1971) | U | l | ObV | GCAS | |
| 18 | GSC 00743-02467 | HD 44351, ASAS J062159+1418.5 | 06 21 58.770 | +14 18 32.26 | 8.22-8.53 | 8.22-8.53 | B8Ib/II (Nassau et al. 1965), B3e shell (Merrill & Burwell 1949) | L | l | LTV | GCAS | |
| 19 | GSC 00732-02105 | HD 256577, ASAS J062400+0818.0 | 06 24 00.179 | +08 18 02.49 | 9.48-9.78 | 9.31-9.78 | B1V (Turner 1976), B2IV:e:p (Hiltner 1956) | E | l, u | SRO | GCAS | |
| 20 | GSC 00154-02436 | HD 259481, ASAS J063259+0456.4 | 06 32 59.376 | +04 56 22.50 | 9.41-9.68* | 9.35-9.68 | B0Ve (Vrancken et al. 1997) | E | l, la | ObV | GCAS | |
| 21 | GSC 00733-01509 | HD 259631, ASAS J063343+0802.1 | 06 33 43.417 | +08 02 10.00 | 9.51-9.57 | 9.51-9.59 | B1Vnne (Turner 1976) | E | l, u | LTV | BE | |
| 22 | GSC 00733-01932 | HD 260360, ASAS J063551+0756.4 | 06 35 51.182 | +07 56 24.56 | 10.85-11.07 | 10.77-11.07 | B3III (Voroshilov et al. 1985), em (Skiff 2014) | E | l | ObV | GCAS | |
| 23 | GSC 00146-01543 | HD 288847, ASAS J063627+0140.4 | 06 36 27.868 | +01 40 21.28 | 9.56-9.87 | 9.56-9.87 | B5 (Skiff 2014), em (Merrill & Burwell 1950) | M | l | ObV | GCAS | |
| 24 | GSC 00154-00165 | HD 260765, ASAS J063707+0535.0 | 06 37 06.595 | +05 34 57.99 | 10.27-10.49 | 10.16-10.49 | B8III (Voroshilov et al. 1985) | L | | LTV | GCAS | |
| 25 | GSC 00755-00857 | ASAS J064801+1300.2 | 06 48 01.247 | +13 00 13.03 | 11.59-11.97 | 11.59-11.99 | OB:e (Stephenson & Sanduleak 1977b) | U | l, la | ObV | GCAS | |
| 26 | GSC 00152-00780 | ASAS J064839+0205.3 | 06 48 39.065 | +02 05 20.06 | 11.53-12.04* | 11.53-12.05 | em (Robertson & Jordan 1989) | U | l | SRO | GCAS | |
| 27 | GSC 00148-02601 | HD 292379, ASAS J064926+0035.0 | 06 49 25.983 | +00 35 00.07 | 10.02-10.13* | 9.98-10.13 | em (MacConnell 1981), OB- (Nassau et al. 1965) | U | l | LTV | BE | |
| 28 | GSC 00160-01058 | HD 264600, ASAS J064937+0613.6 | 06 49 36.772 | +06 13 32.34 | 10.82-11.05 | 10.80-11.19 | B2Vnne (Vijapurkar & Drilling 1993) | E | l | SRO | GCAS | 259(5) |
| 29 | GSC 05387-01121 | HD 49888, NSV 3231 | 06 50 08.089 | -12 35 05.12 | 7.22-7.50 | 7.18-7.50 | em (Bidelman 1988), B5e (Neubauer 1943), B5Iab/Ib (Houk & Smith-Moore 1988b) | M | l, u | LTV | GCAS | |
| 30 | GSC 04801-00017 | ASAS J065244-0011.3 | 06 52 44.027 | -00 11 16.77 | 11.26-11.55 | 11.13-11.55 | B: (McCuskey 1956), em (Robertson & Jordan 1989) | U | l | ObV | GCAS | |
| 31 | GSC 05383-00187 | HD 50424, SAO 133821 | 06 52 53.052 | -10 00 26.93 | 8.79-8.92 | 8.79-8.98 | B8e (Stephenson & Sanduleak 1977a) | L | l, u | LTV | GCAS | |
| 32 | GSC 04805-00043 | HD 50891, ASAS J065459-0342.0 | 06 54 58.824 | -03 42 01.29 | 8.76-9.09 | 8.76-9.09 | B1IIIe (Negueruela et al. 2004), B0:ep (Morgan et al. 1955), B0.5Ve (Chojnowski et al. 2015) | E | l, s | SRO, LTV | GCAS | |
| 33 | GSC 05388-01118 | BD-12 1700, ASAS J065555-1300.0 | 06 55 55.052 | -13 00 01.82 | 10.47-10.93 | 10.47-10.93 | OBe (Stephenson & Sanduleak 1971), em (Merrill & Burwell 1950) | U | l | SRO | GCAS | |
| 34 | GSC 00748-01908 | HD 267003, ASAS J065729+0759.3 | 06 57 28.901 | +07 59 20.05 | 11.12-11.34 | 11.09-11.34 | Be (Stephenson & Sanduleak 1977b) | U | l, la | ObV | GCAS | |
| 35 | GSC 04809-00545 | HD 295852, ASAS 065752-0345.8 | 06 57 51.546 | -03 45 46.65 | 9.28-9.58 | 9.28-9.63 | Be (Bidelman & MacConnell 1973), B3e (Münch 1952) | E | l | ObV,LTV | GCAS | |
| 36 | GSC 00153-00891 | ASAS J065831+0259.1 | 06 58 31.540 | +02 59 05.46 | 12.35-12.78 | 12.35-12.78 | em (Stephenson & Sanduleak 1977b) | U | l | SRO | GCAS | |
| 37 | GSC 04801-01915 | ASAS J065910-0037.2 | 06 59 09.981 | -00 37 10.55 | 11.14-11.33 | 11.14-11.35 | em (Robertson & Jordan 1989), B5 (McCuskey 1956) | M | l | ObV | GCAS | |
| 38 | GSC 04826-00257 | ASAS J070241-0709.5 | 07 02 41.275 | -07 09 32.72 | 10.98-11.18* | 10.82-11.18 | em (MacConnell 1981) | U | l | LTV | GCAS | |
| 39 | GSC 04826-01079 | ASAS J070334-0712.4 | 07 03 34.270 | -07 12 27.79 | 11.17-11.41 | 11.17-11.41 | em (MacConnell 1981) | U | l | ObV | GCAS | |
| 40 | GSC 05393-02168 | BD-13 1825, ASAS J070658-1340.6 | 07 06 58.226 | -13 40 35.18 | 9.17-9.65 | 9.17-9.65 | Be (Bidelman & MacConnell 1973) | U | l, u | SRO | GCAS | |
| 41 | GSC 05968-03899 | HD 54576, SAO 152454 | 07 08 37.741 | -17 38 05.02 | 9.44-9.53* | 9.44-9.55 | B7/8 II, (Anderson & Francis 2012) | L | | ObV | GCAS | |
| 42 | GSC 05398-01016 | HD 55439, ASAS J071235-0950.7 | 07 12 34.812 | -09 50 42.09 | 8.29-8.52 | 8.29-8.52 | B2Ve (Claria 1974) | E | l, u, s | LTV, ObV | GCAS | |
| 43 | GSC 05973-00249 | NSV 17397, BD-20 1805 | 07 14 58.243 | -20 37 12.21 | 10.27-10.66* | 10.27-10.67 | B2IIIne (Vijapurkar & Drilling 1993) | E | l | ObV, LTV | GCAS | |
| 44 | GSC 05399-00962 | HD 56873, NSV 3525 | 07 18 30.832 | -10 35 20.15 | 10.52-10.74 | 10.51-10.78 | B8e: (Stephenson & Sanduleak 1977b) | L | l | ObV | GCAS | |
| 45 | GSC 07634-01561 | HD 57551, ASAS J071959-4029.7 | 07 19 59.383 | -40 29 41.60 | 9.68-9.90 | 9.68-9.90 | em (Henize 1976), B8III (Houk 1978) | L | l | ObV | GCAS | |
| 46 | GSC 04820-02947 | ASAS J072024-0337.7 | 07 20 24.300 | -03 37 38.72 | 10.26-10.40 | 10.26-10.41 | B7e (Stephenson & Sanduleak 1977a) | L | l | ObV | GCAS | |
| 47 | GSC 05978-01855 | BD-20 1891 | 07 22 36.249 | -20 54 34.82 | 9.45-9.66 | 9.45-9.66 | Be, (Stephenson & Sanduleak 1977a), B2.5Vne (Garrison et al. 1977) | E | l | ObV | GCAS | |
| 48 | GSC 05978-00030 | ASAS J072628-2205.2 | 07 26 28.027 | -22 05 13.78 | 11.78-12.31* | 11.78-12.31 | em (MacConnell 1981) | U | l | SRO | GCAS | 1560(50) |
| 49 | GSC 05983-00995 | HD 59479, GDS_J0729316-173620 | 07 29 31.668 | -17 36 20.68 | 9.89-10.06 | 9.83-10.06 | B5V (Münch 1952) | M | | ObV, LTV | GCAS | |
| 50 | GSC 07109-00828 | CPD-32 1553, ASAS J073300-3233.2 | 07 33 00.270 | -32 33 11.05 | 10.47-10.69 | 10.47-10.69 | Be (Bidelman & MacConnell 1973) | U | l | LTV | GCAS | |
| 51 | GSC 05405-00431 | HD 60993, NSV 17524 | 07 36 42.296 | -13 03 49.89 | 8.86-8.98 | 8.70-8.98 | B3n (Neubauer 1943) | E | u | LTV | GCAS | |
| 52 | GSC 05988-00265 | NSV 17539, BD-18 1948 | 07 39 24.704 | -18 48 32.93 | 10.33-10.46 | 10.30-10.69 | OB+e (Stephenson & Sanduleak 1971) | U | l | LTV | BE | |
| 53 | GSC 08552-00688 | HD 62483, NSV 17565 | 07 41 28.928 | -53 11 41.14 | 8.06-8.25 | 8.06-8.27 | B2III (Hill 1970), B2II (Houk 1978) | E | u | SRO | GCAS | |
| 54 | GSC 06552-00580 | ASAS J074314-2829.2 | 07 43 13.945 | -28 29 12.94 | 12.12-12.80 | 12.12-12.80 | OB+ (Orsatti 1992), em (Henize 1976) | U | l | SRO | GCAS: | 269(9) |
| 55 | GSC 06552-01189 | CD-29 4913, ASAS J074559-2933.2 | 07 45 59.001 | -29 33 16.13 | 11.05-11.46 | 11.05-11.53 | OB+e (Orsatti 1992) | U | l | LTV | GCAS | |
| 56 | GSC 07106-02534 | CD-30 5070, ASAS J074630-3115.6 | 07 46 30.000 | -31 15 32.05 | 11.13-11.43* | 11.13-11.54 | OBe (Orsatti 1992) | U | l | SRO | GCAS | 720(20) |
| 57 | GSC 07123-00519 | CD-33 4170, ASAS J074829-3324.2 | 07 48 29.103 | -33 24 10.14 | 11.74-12.05* | 11.69-12.06 | OB+e (Orsatti 1992) | U | l | ObV, SRO: | GCAS | |
| 58 | GSC 08135-03248 | HD 64596, ASAS J075233-4628.5 | 07 52 33.235 | -46 28 30.09 | 9.15-9.53 | 9.15-9.53 | B2V (Houk 1978) | E | | SRO, EB | EA+GCAS | 1.28212(1) |
| 59 | GSC 06565-01179 | NSV 17662, CPD-29 2176 | 07 55 42.491 | -29 33 53.56 | 10.37-10.95 | 10.22-10.98 | B0V(n)e (Vijapurkar & Drilling 1993), B0Ve (Fernie et al. 1966) | E | l | ObV | GCAS | |
| 60 | GSC 07124-01160 | ALS 913, ASAS J075930-3204.7 | 07 59 30.375 | -32 04 38.75 | 11.82-12.47* | 11.82-12.47 | OB+e (Orsatti 1992) | U | l | SRO | GCAS | 291(4) |
| 61 | GSC 07120-02077 | NSV 17698, CPD-30 2160 | 07 59 43.672 | -30 41 26.35 | 10.59-11.05* | 10.59-11.05 | B1IIInne (Vijapurkar & Drilling 1993) | E | l | SRO | GCAS | |
| 62 | GSC 06562-00688 | HD 67123, ASAS J080533-2740.5 | 08 05 32.896 | -27 40 20.40 | 9.99-10.26 | 9.99-10.26 | Be (Stephenson & Sanduleak 1977b), A3 (Cannon & Pickering 1919a) | U | l | ObV | GCAS | |
| 63 | GSC 05421-00568 | ASAS J080723-1247.1 | 08 07 23.294 | -12 47 06.71 | 11.22-11.62 | 11.22-11.62 | em (MacConnell 1981) | U | l | ObV | GCAS | |
| 64 | GSC 07659-01614 | CD-38 4113, ASAS J080845-3857.5 | 08 08 44.976 | -38 57 35.13 | 11.17-11.40* | 11.17-11.44 | em (MacConnell 1981) | U | l | ObV | GCAS | |

Table A1continued.

| (1) No. | (2) ID GSC | (3) ID alt | (4) $\alpha$ (J2000) | (5) $\delta$ (J2000) | (6) Range($V$) [mag] | (7) Range($V$) lit. [mag] | (8) Spec.type lit. | (9) subtype [E/M/L/U] | (10) emission flag | (11) Var.type [LB17] | (12) Var.type [GCVS/VSX] | (13) Period(s) [d] |
|---|---|---|---|---|---|---|---|---|---|---|---|---|
| 65 | GSC 05438-00850 | HD 68468, NSV 17783 | 08 12 00.394 | -14 10 08.35 | 8.67-8.82 | 8.67-8.82 | B shell, (Bidelman & MacConnell 1973), B3:e shell, (Guetter 1968) | E | l, s, u | ObV, LTV | GCAS | |
| 66 | GSC 07125-02097 | NSV 17791, CPD-32 2038 | 08 12 12.588 | -32 28 15.06 | 10.99-11.36* | 10.75-11.37 | B0IIIe (Bosch et al. 2003) | E | l | SRO | GCAS | |
| 67 | GSC 05435-00526 | HD 71072, NSV 17879 | 08 25 00.396 | -12 45 53.24 | 6.87-6.94 | 6.87-6.94 | B7IVe (Buscombe 1969) | L | l | LTV, NRP | BE | 1.09207(8) |
| 68 | GSC 07669-01154 | USNO-A2.0 0450-06879300 | 08 28 02.327 | -41 44 57.16 | 12.18-12.41 | 12.17-12.41 | | U | | ObV | GCAS | |
| 69 | GSC 07669-03714 | HD 72014, NSV 17900 | 08 28 52.061 | -42 35 14.91 | 6.23-6.68 | 6.20-6.68 | B1III (Cucchiaro et al. 1980), B0V:nne, (Garrison et al. 1977), B1/2V:nne (Houk 1978) | E | l, s, u | ObV | GCAS | |
| 70 | GSC 07139-02209 | HD 71964, SAO 199281 | 08 29 01.345 | -33 27 39.42 | 8.81-8.90* | 8.76-8-90 | A0III/IV (Houk 1982) | L | | LTV | GCAS: | |
| 71 | GSC 08151-01868 | CPD-45 2826, GDS_J0840509-460545 | 08 40 50.921 | -46 05 45.36 | 9.91-10.30* | 9.91-10.30 | B2Vne (Fitzgerald et al. 1979), B2:Vne (Feast et al. 1961) | E | l | ObV | GCAS | |
| 72 | GSC 08155-02404 | HD 74401, ASAS J084200-4739.9 | 08 42 00.380 | -47 39 53.92 | 8.90-8.99 | 8.90-9.02 | B1.5IIIne (Garrison et al. 1977) | E | l | LTV | BE | |
| 73 | GSC 08164-01530 | HD 76568, SAO 236395 | 08 54 57.808 | -50 51 50.36 | 9.39-9.51 | 9.39-9.51 | B1V:nne (Houk 1978), B7e (Stephenson & Sanduleak 1977a) | M | l | ObV | GCAS | |
| 74 | GSC 08594-00620 | HD 77032, ASAS J085715-5835.2 | 08 57 15.412 | -58 35 09.68 | 8.67-8.80 | 8.67-8.80 | B2:Vne (Garrison et al. 1977), B5Vne (Houk & Cowley 1975) | M | l, s, u | ObV | GCAS | |
| 75 | GSC 08165-00324 | CD-46 4821, ASAS J090014-4641.1 | 09 00 14.047 | -46 41 08.57 | 8.95-9.06 | 8.95-9.06 | B2Vn (Slawson & Reed 1988), Be (Stephenson & Sanduleak 1977a) | E | l | LTV | BE | |
| 76 | GSC 08582-02609 | HD 299794, ASAS J090036-5321.0 | 09 00 36.167 | -53 21 02.25 | 9.78-9.98 | 9.78-9.98 | em (Henize 1976), OB (Stephenson & Sanduleak 1971) | U | l | LTV, IP | BE | 25.80(4) |
| 77 | GSC 08591-00039 | HD 302105, SAO 236686 | 09 10 36.634 | -57 31 48.79 | 9.46-9.63* | 9.46-9.63 | B (Stephenson & Sanduleak 1977a), em (Henize 1976), B1IVn (Garrison et al. 1977) | E | l | LTV | BE | |
| 78 | GSC 07686-01898 | CPD-41 3411, ASAS J091049-4228.6 | 09 10 49.165 | -42 28 37.91 | 10.11-10.24* | 10.11-10.24 | B7e (Herbst 1975) | L | l | ObV | GCAS | |
| 79 | GSC 08587-02162 | HD 79778, ASAS J091351-5532.4 | 09 13 51.368 | -55 32 22.77 | 8.06-8.46* | 8.06-8.46 | B2IVne (Garrison et al. 1977) | E | l, s, u | SRO | GCAS | |
| 80 | GSC 08174-00235 | HD 298335, ASAS J092111-5030.8 | 09 21 10.586 | -50 30 44.92 | 10.15-10.53* | 10.15-10.53 | B7e (Stephenson & Sanduleak 1977a), B2/4 (Sundman et al. 1974), B5II/Ib (Kennedy 1996) | U | l | ObV | GCAS | |
| 81 | GSC 09196-00424 | NSV 4471 | 09 21 11.015 | -68 39 47.93 | 11.65-11.82 | 11.65-11.82 | | U | | LTV, NRP | LERI+BE | 1.14514(10) |
| 82 | GSC 08588-02569 | HD 81354, ASAS J092255-5544.2 | 09 22 54.755 | -55 44 11.97 | 9.14-9.37* | 9.14-9.37 | B3/5Ve (Houk & Cowley 1975) | M | l, u | ObV | GCAS | |
| 83 | GSC 08167-01520 | HD 82830, ASAS J093309-4645.9 | 09 33 09.092 | -46 45 54.48 | 9.21-9.39 | 9.20-9.39 | B0/2Ie (Houk 1978), Be (Bidelman & MacConnell 1973), B1III:ne (Garrison et al. 1977) | E | l, u | LTV, ObV | GCAS | |
| 84 | GSC 08593-01904 | HD 83060, NSV 18234 | 09 33 42.391 | -57 09 48.09 | 9.07-9.14* | 9.07-9.14 | Be (Stephenson & Sanduleak 1977a), B2Vnne (Houk & Cowley 1975), B2IV-Vnep (shell) (Garrison et al. 1977) | E | l | LTV, NRP | LERI | 0.73486(1) |
| 85 | GSC 08585-02212 | HD 83597, ASAS J093751-5340.9 | 09 37 50.894 | -53 40 49.23 | 9.08-9.34 | 9.08-9.34 | B1.5Ve (Garrison et al. 1977) | E | l, s, u | ObV, LTV | GCAS | |
| 86 | GSC 08593-00336 | HD 83781, SAO 237182 | 09 38 43.500 | -56 48 21.06 | 9.46-9.73 | 9.46-9.73 | B8/9V (Houk & Cowley 1975) | L | | ObV | GCAS | |
| 87 | GSC 07179-02573 | HD 83739 | 09 39 33.173 | -36 40 31.89 | 9.89-10.10 | 9.89-10.10 | em (Henize 1976), B8 (Cannon & Pickering 1919b) | L | l | LTV | GCAS | |
| 88 | GSC 08593-02269 | HD 84361 | 09 42 42.693 | -58 06 19.69 | 8.17-8.43* | 8.17-8.43 | em (Henize 1976), B2/3V (Houk & Cowley 1975) | E | l, u | ObV, LTV, IP | GCAS+LERI: | 4.62256(9) |
| 89 | GSC 08606-00053 | HD 85495, GDS_J0950146-580245 | 09 50 14.687 | -58 02 45.12 | 7.90-8.06 | 7.90-8.06 | em (Henize 1976), B4II/III (Houk & Cowley 1975),B5 (Loden et al. 1976) | M | l, s | LTV | GCAS | |
| 90 | GSC 08598-02245 | HD 85809, ASAS J095238-5419.1 | 09 52 38.445 | -54 19 07.59 | 9.07-9.42* | 9.07-9.47 | B2/3III/V (Houk & Cowley 1975) | E | | SRO | GCAS | 120.5(8) |
| 91 | GSC 08606-02020 | CPD-56 2606, ASAS J095308-5636.5 | 09 53 07.478 | -56 36 29.24 | 10.30-10.81 | 10.29-10.81 | em (Henize 1976), OB+e (Stephenson & Sanduleak 1971) | U | l | SRO | GCAS | |
| 92 | GSC 08610-03102 | HD 86657, ASAS J095758-5957.6 | 09 57 58.374 | -59 57 37.45 | 9.66-9.79* | 9.66-9.79 | B5 (Gieseking 1980) | M | | ObV, LTV | GCAS | |
| 93 | GSC 08603-00164 | HD 300425, ASAS J100139-5500.8 | 10 01 39.127 | -55 00 46.36 | 9.78-10.10 | 9.78-10.10 | em (MacConnell 1981), OBe (Stephenson & Sanduleak 1971) | U | l | SRO | GCAS | 152(4) |
| 94 | GSC 08599-02357 | HD 87203, ASAS J100152-5412.4 | 10 01 52.354 | -54 12 21.12 | 8.50-8.75 | 8.50-8.75 | Be (Stephenson & Sanduleak 1977a), B5Ve (Houk & Cowley 1975),O9.5/B1 (Nordström 1975) | M | l, s, u | ObV | GCAS | |
| 95 | GSC 08611-01377 | HD 87565 | 10 03 50.777 | -59 51 23.35 | 9.30-9.66* | 9.30-9.66 | B2Ve (González & Lapasset 2001), B2Vnne (Levato & Malaroda 1975) | E | l, u | ObV | GCAS | |
| 96 | GSC 08603-00677 | CPD-54 3090, ASAS 100446-5517.5 | 10 04 46.026 | -55 17 30.05 | 11.05-11.40 | 11.05-11.40 | B7e (Stephenson & Sanduleak 1977a), B2/4 (Nordström 1975) | M | l | ObV | GCAS | |
| 97 | GSC 08955-00160 | HD 87849 | 10 05 10.339 | -67 15 30.81 | 8.94-9.07* | 8.94-9.07 | B4V (Houk & Cowley 1975) | M | | ObV | GCAS | |
| 98 | GSC 08943-00584 | HD 304915 | 10 06 36.154 | -60 49 06.70 | 10.12-10.39* | 10.12-10.45 | em (MacConnell 1981), B5V (Loden et al. 1976) | M | l | ObV | GCAS | |
| 99 | GSC 08607-01004 | Hen 3-375, ASAS J101045-5620.0 | 10 10 45.371 | -56 20 01.71 | 12.09-12.57 | 12.09-12.57 | OB+:e (Stephenson & Sanduleak 1977a) | U | l | SRO | GCAS | |
| 100 | GSC 08607-00285 | HD 300584, NSV 18347 | 10 10 59.296 | -57 09 26.44 | 10.08-10.28 | 10.08-10.29 | O5/B0e (Loden et al. 1976), B1Ve (Graham 1970) | E | l | ObV | GCAS | |
| 101 | GSC 08604-00356 | HD 300540, Hen 3-388 | 10 13 31.708 | -54 53 59.40 | 9.81-9.93* | 9.81-9.93 | em (Henize 1976), B6/7 (Loden et al. 1976) | M | l | LTV, ObV | GCAS | |
| 102 | GSC 08947-00809 | NSV 18363, CPD-62 1480 | 10 13 35.005 | -63 00 11.72 | 11.11-11.27 | 10.82-11.27 | em (Henize 1976), OB+e (Stephenson & Sanduleak 1971) | U | l | ObV | GCAS | |
| 103 | GSC 08943-02244 | HD 305019, NSV 18365 | 10 15 02.400 | -60 55 21.23 | 10.17-10.55* | 10.17-10.57 | O em (Loden et al. 1976), B0ep (Humphreys 1973) | E | l | ObV | GCAS | |
| 104 | GSC 08943-00620 | ALS 1476, ASAS J101605-6032.8 | 10 16 04.411 | -60 32 52.25 | 11.32-11.85* | 11.32-11.85 | B2III (Vijapurkar & Drilling 1993), OB+ (Stephenson & Sanduleak 1971) | E | | LTV, ObV | GCAS | |
| 105 | GSC 08612-00975 | GDS_J1019327-582544 | 10 19 32.785 | -58 25 43.73 | 11.82-12.19* | 11.82-12.19 | B:e (Stephenson & Sanduleak 1977a) | U | l | ObV | GCAS | |
| 106 | GSC 08608-00434 | HD 300683 | 10 19 36.677 | -57 07 38.11 | 9.68-9.94 | 9.68-9.94 | B shell (Bidelman & MacConnell 1982), B3/6 (Loden et al. 1976) | M | l | ObV | GCAS | |
| 107 | GSC 08608-00807 | HD 89990 | 10 21 36.561 | -57 55 52.81 | 9.70-9.78* | 9.70-9.79 | B8 (Loden et al. 1976), B8V (Houk & Cowley 1975) | L | | ObV | GCAS | |
| 108 | GSC 08608-00914 | HD 90187, NSV 18395 | 10 23 07.843 | -57 59 51.23 | 8.67-9.08* | 8.65-9.08 | B1IIne (Turner 1978), B1IIIne (Garrison et al. 1977) | E | l, s, u | SRO | GCAS | |
| 109 | GSC 08612-00331 | HD 302816, CPD-58 2166 | 10 24 26.028 | -59 24 23.88 | 9.81-10.16* | 9.81-10.16 | B2V (Loden et al. 1976) | E | | ObV | GCAS | |
| 110 | GSC 08608-02412 | HD 90551 | 10 25 50.913 | -56 37 15.23 | 8.60-8.73* | 8.55-8.77 | em (Henize 1976), B3Vn(e?) (Houk & Cowley 1975) | E | l, s | LTV | GCAS | |
| 111 | GSC 08612-00380 | HD 302798, ASAS J102559-5859.7 | 10 25 58.956 | -58 59 44.69 | 9.86-10.28* | 9.86-10.32 | B3V(e) (Graham 1970), em (Henize 1976) | E | l | SRO | GCAS | |
| 112 | GSC 08608-00239 | HD 300739, ASAS J102609-5712.1 | 10 26 09.450 | -57 12 06.70 | 10.21-10.46* | 10.21-10.46 | em (MacConnell 1981), O9.5/B0 (Loden et al. 1976) | E | l | SRO | GCAS | |
| 113 | GSC 09397-00615 | HD 91180, ASAS J102758-7613.9 | 10 27 58.238 | -76 13 56.84 | 9.09-9.20* | 9.09-9.20 | em (Henize 1976), B5:e (Houk & Cowley 1975) | M | l | ObV | GCAS | |
| 114 | GSC 08609-01016 | HD 302838, NSV 18417 | 10 28 31.527 | -57 42 18.23 | 9.94-10.31* | 9.87-10.31 | OB:e (Loden et al. 1976), B1Vne (Graham 1970) | E | l | ObV | GCAS | |
| 115 | GSC 08605-01985 | CPD-54 3736, ASAS 102934-5508.2 | 10 29 34.257 | -55 08 10.17 | 10.72-10.84* | 10.72-10.84 | Be (Stephenson & Sanduleak 1977b), B5 (Loden et al. 1976) | M | l | LTV, ObV | GCAS | |
| 116 | GSC 08956-01223 | HD 91597, NSV 18440 | 10 33 01.134 | -60 50 41.77 | 9.54-9.95* | 9.54-9.95 | B1IIIn(e) (Garrison et al. 1977), B7/8IV/V (Houk & Cowley 1975) | M | l | SRO | GCAS | |
| 117 | GSC 08613-01412 | HD 302977, CPD-57 3419 | 10 33 08.321 | -58 26 07.90 | 10.07-10.25 | 10.07-10.25 | B0 (Sundman et al. 1974), em (Wray 1966), A (Nesterov et al. 1995) | E | l | ObV | GCAS | |
| 118 | GSC 08613-00130 | HD 302978, CPD-57 3422 | 10 33 09.422 | -58 28 40.93 | 8.74-8.81* | 8.74-8.81 | B5 (Nesterov et al. 1995) | M | | LTV | BE: | |
| 119 | GSC 08613-01744 | WRAY 15-604, ASAS J103323-5831.8 | 10 33 22.896 | -58 31 50.08 | 11.76-12.41 | 11.76-12.41 | OB+:e (Stephenson & Sanduleak 1977a) | U | l | ObV | GCAS | |
| 120 | GSC 08613-00147 | CPD-58 2312, GDS_J1034088-592853 | 10 34 08.845 | -59 28 54.01 | 10.27-10.84* | 10.27-10.84 | em (Henize 1976), Be (Bidelman & MacConnell 1973) | U | l | SRO | GCAS | |
| 121 | GSC 08957-02248 | HD 92064 | 10 36 19.681 | -60 37 29.81 | 10.03-10.28* | 10.03-10.28 | em (MacConnell 1981), B3/6V: (Houk & Cowley 1975) | M | l | ObV | GCAS | |
| 122 | GSC 08609-00597 | HD 303147, CPD-57 3618 | 10 38 35.398 | -57 53 34.17 | 9.95-10.22* | 9.95-10.22 | B3 (Nesterov et al. 1995), B1/3 (Sundman et al. 1974) | E | | ObV | GCAS | |
| 123 | GSC 08622-01179 | HD 303142, CPD-56 3632 | 10 40 00.582 | -57 32 59.21 | 9.58-9.94* | 9.53-9.94 | em (MacConnell 1981), OB (Stephenson & Sanduleak 1971) | U | l | ObV, NRP | GCAS+LERI | 0.483815(7) |
| 124 | GSC 08622-00429 | HD 92759, ASAS J104121-5739.9 | 10 41 21.316 | -57 39 51.47 | 9.39-9.84* | 9.39-9.84 | em (Schwartz et al. 1990), B2Ve (Graham 1970) | E | l | SRO | GCAS | |
| 125 | GSC 08626-00271 | HD 303190, GDS_J1041231-590918 | 10 41 23.163 | -59 09 18.81 | 10.12-10.23* | 10.12-10.26 | B1V (Fitzgerald & Mehta 1987), B2V (Forte & Orsatti 1981) | E | | LTV | BE | |
| 126 | GSC 08957-03483 | CPD-61 1814 | 10 42 18.409 | -61 45 23.99 | 11.29-11.62* | 11.29-11.62 | | U | | ObV | GCAS | |
| 127 | GSC 08961-01212 | Hen 3-476, 2MASS 10424993-6159377 | 10 42 49.948 | -61 59 37.73 | 11.84-12.22* | 11.79-12.22 | B2:, (Sundman et al. 1974); em (Henize 1976) | E | l | LTV | BE | |
| 128 | GSC 08618-01375 | HD 93051 | 10 43 35.610 | -55 55 42.22 | 9.73-9.83* | 9.73-9.83 | B8/9III (Houk & Cowley 1975) | L | | ObV | GCAS | |

Table A1continued.

| (1) No. | (2) ID GSC | (3) ID alt | (4) $\alpha$ (J2000) | (5) $\delta$ (J2000) | (6) Range($V$) [mag] | (7) Range($V$) lit. [mag] | (8) Spec.type lit. | (9) subtype [E/M/L/U] | (10) emission flag | (11) Var.type [LB17] | (12) Var.type [GCVS/VSX] | (13) Period(s) [d] |
|---|---|---|---|---|---|---|---|---|---|---|---|---|
| 129 | GSC 08622-01758 | HD 93442, NSV 18520 | 10 46 06.001 | -56 45 25.47 | 8.77-8.93* | 8.77-8.93 | B3 (Cannon & Mayall 1949), em (Henize 1976) | E | l | ObV | GCAS | |
| 130 | GSC 08622-01002 | HD 303333, CPD-56 3785 | 10 46 21.227 | -57 37 32.19 | 11.14-11.27* | 11.13-11.28 | A2 (Nesterov et al. 1995) | U | | LTV | GCAS | |
| 131 | GSC 08618-01665 | HD 93561, NSV 18527 | 10 46 59.466 | -54 49 52.60 | 8.86-9.38* | 8.86-9.38 | B1III:nne (Garrison et al. 1977), B1/3e (Houk & Cowley 1975) | E | l, u | SRO | GCAS | |
| 132 | GSC 08622-01017 | HD 93618, ASAS J104722-5710.0 | 10 47 22.142 | -57 10 01.93 | 9.01-9.19* | 9.01-9.19 | OB (Stephenson & Sanduleak 1971), B2Ve (Graham 1970) | E | l, u | LTV, NRP | BE+LERI | 0.88286(2) |
| 133 | GSC 08965-01363 | HD 308023, CPD-63 1673 | 10 48 00.558 | -64 31 49.40 | 9.95-10.10* | 9.94-10.10 | OB-e (Stephenson & Sanduleak 1971), B3IVe (Graham 1970) | E | l | LTV | GCAS | |
| 134 | GSC 08969-00788 | HD 310250, CPD-65 1515 | 10 48 24.521 | -66 23 41.64 | 9.92-10.19* | 9.92-10.19 | em (Henize 1976) | U | l | ObV | GCAS | |
| 135 | GSC 08958-02421 | HD 94288, GDS_J1051408-614136 | 10 51 40.876 | -61 41 36.09 | 9.20-9.54* | 9.20-9.55 | em (Henize 1976), B3/5V (Houk & Cowley 1975) | M | l, u | SRO | GCAS | |
| 136 | GSC 08623-02851 | CPD-55 3970 | 10 54 57.445 | -56 39 43.55 | 11.38-11.52* | 11.38-11.54 | em (MacConnell 1981) | U | l | ObV | GCAS | |
| 137 | GSC 08958-02242 | HD 305829, GDS_J1056461-614635 | 10 56 46.201 | -61 46 35.31 | 10.22-10.48 | 10.15-10.48 | | U | | SRO | GCAS | |
| 138 | GSC 08619-01981 | HD 301197, CPD-54 4219 | 10 59 18.999 | -55 29 34.25 | 9.71-9.90* | 9.63-9.90 | em (Henize 1976) | U | l | LTV, NRP | BE+LERI: | 0.2356186(10) |
| 139 | GSC 08627-00455 | HD 303669, CPD-58 2933 | 11 00 50.878 | -58 59 48.33 | 9.43-9.58* | 9.43-9.58 | OB (Graham & Lynga 1965), B5 (Wallenquist 1931) | M | | LTV, NRP | BE+LERI: | |
| 140 | GSC 08627-01249 | HD 95615, GDS_J1101143-590900 | 11 01 14.361 | -59 09 00.62 | 9.12-9.31* | 9.11-9.40 | B3II/III (Houk & Cowley 1975), B2V (Humphreys 1973) | E | | SRO | GCAS | |
| 141 | GSC 08958-02961 | HD 95826A, ASAS J110223-6030.9 | 11 02 23.915 | -60 30 58.51 | 9.21-9.36* | 9.10-9.36 | B5e (Cannon & Pickering 1919b), em (Henize 1976), Be (Bidelman & MacConnell 1973) | M | l | LTV, IP | BE | 23.0(2) |
| 142 | GSC 08627-02220 | HD 303763, CPD-58 2969 | 11 02 50.868 | -59 19 05.90 | 10.10-10.27* | 10.09-10.27 | em (Henize 1976), Be (Bidelman & MacConnell 1973) | U | l | ObV | GCAS | |
| 143 | GSC 08958-03515 | CPD-59 2978, GDS_J1102595-602911 | 11 02 59.577 | -60 29 11.24 | 10.70-10.95 | 10.63-10.95 | em (Henize 1976) | U | l | LTV | GCAS | |
| 144 | GSC 08627-02146 | HD 303764, GDS_J1103042-592218 | 11 03 04.300 | -59 22 18.72 | 10.82-11.00 | 10.82-11.04 | A (Wallenquist 1931) | L | | ObV | GCAS | |
| 145 | GSC 08958-00887 | HD 95972, ASAS J110310-6146.1 | 11 03 09.683 | -61 46 07.53 | 8.64-9.00* | 8.64-9.02 | B2Vnn (Houk & Cowley 1975), B2Ve (Graham 1970) | E | l, u | ObV | GCAS | |
| 146 | GSC 08958-01376 | HD 96447, CPD-60 2517 | 11 06 00.796 | -61 10 33.95 | 8.94-9.20* | 8.94-9.20 | B3 (Loden 1980), B2/3V: (Houk & Cowley 1975), B9V (Martin 1964) | M | u | LTV | BE | |
| 147 | GSC 08958-03463 | Hen 3-575, 2MASS 11061965-6046352 | 11 06 19.672 | -60 46 35.25 | 11.56-11.74* | 11.56-11.74 | B5e (Stephenson & Sanduleak 1977a) | M | l | LTV | GCAS | |
| 148 | GSC 08958-03384 | HD 306093, GDS_J1108184-604801 | 11 08 18.423 | -60 48 01.10 | 10.71-11.00* | 10.71-11.05 | B8 (Nesterov et al. 1995), B1/3 (Sundman et al. 1974) | U | | ObV | GCAS | |
| 149 | GSC 08959-00482 | HD 306111 | 11 09 19.827 | -61 06 05.36 | 10.09-10.26* | 10.08-10.26 | em (Henize 1976), Be (Bidelman & MacConnell 1973), B2Ve (Graham 1970) | E | l | ObV | GCAS | |
| 150 | GSC 08967-00393 | HD 97136 | 11 09 48.048 | -63 47 39.30 | 9.16-9.30* | 9.16-9.30 | B2III (Garrison et al. 1977) | E | | SRO | GCAS | |
| 151 | GSC 08959-00488 | HD 306082, GDS_J1109573-604622 | 11 09 57.356 | -60 46 23.00 | 9.81-10.09* | 9.81-10.09 | B0 (Johansson 1980), B5e (Stephenson & Sanduleak 1977a) | E | l, u | ObV | GCAS | |
| 152 | GSC 08628-00661 | HD 303887, ASAS J111052-5813.0 | 11 10 52.287 | -58 13 01.86 | 9.27-9.64* | 9.27-9.64 | B5e (Stephenson & Sanduleak 1977a) | M | l | ObV | GCAS | |
| 153 | GSC 08959-00863 | HD 306205 | 11 13 29.502 | -61 15 50.95 | 9.87-10.15* | 9.87-10.15 | Be (Stephenson & Sanduleak 1977a), B1.5Vne (Feast et al. 1961) | E | l | SRO | GCAS | |
| 154 | GSC 08959-00846 | HD 306209, GDS_J1113347-612042 | 11 13 34.727 | -61 20 42.84 | 9.86-10.14* | 9.86-10.16 | B1Ve (Graham 1970) | E | l | ObV | GCAS | |
| 155 | GSC 08620-01856 | HD 97792 | 11 14 20.001 | -56 02 51.23 | 7.89-8.04* | 7.86-8.04 | B7e (Stephenson & Sanduleak 1977a) | L | l, s | ObV | GCAS | |
| 156 | GSC 08959-02476 | HD 306196 | 11 14 29.446 | -61 03 21.45 | 9.52-9.75* | 9.52-9.75 | em (MacConnell 1981), B0:Vn (Feast et al. 1961), O9.5V (Martin 1964) | E | l | ObV | GCAS | |
| 157 | GSC 08625-00369 | HD 304395 | 11 31 08.510 | -57 14 51.12 | 9.30-9.38* | 9.25-9.38 | B7e (Stephenson & Sanduleak 1977a) | L | l | LTV | BE | |
| 158 | GSC 08980-01582 | NSV 18793, ALS 2385 | 11 33 06.654 | -64 42 01.88 | 11.34-11.83* | 11.33-11.83 | em (Henize 1976), OB+e (Stephenson & Sanduleak 1971) | U | l | ObV, LTV | GCAS | |
| 159 | GSC 08972-00064 | HD 306657, NSV 18800 | 11 35 15.172 | -61 41 59.57 | 10.36-10.54* | 10.36-10.54 | em (Henize 1976), OBe (Stephenson & Sanduleak 1971) | U | l, u | ObV | GCAS | |
| 160 | GSC 08972-00932 | HD 101221 | 11 38 21.470 | -60 37 39.76 | 9.20-9.44* | 9.20-9.44 | B9 (Cannon & Pickering 1919b) | L | u | ObV | GCAS | |
| 161 | GSC 08642-01087 | HD 306958, GDS_J1140458-595946 | 11 40 45.061 | -59 59 46.93 | 10.35-10.57* | 10.35-10.57 | em (Henize 1976) | U | l | ObV | GCAS | |
| 162 | GSC 08973-00795 | HD 307007, ASAS J114322-6111.3 | 11 43 21.298 | -61 11 21.18 | 9.91-10.11* | 9.91-10.11 | em (Henize 1976), OB-:e (Stephenson & Sanduleak 1971) | U | l | LTV | BE | |
| 163 | GSC 08973-00729 | HD 309065 | 11 45 45.780 | -61 27 52.77 | 8.72-8.84* | 8.72-8.84 | B:e (Stephenson & Sanduleak 1977a) | U | l, u | LTV | BE | |
| 164 | GSC 08985-01836 | CPD-65 1722, ASAS J114755-6614.7 | 11 47 54.897 | -66 14 41.53 | 11.19-11.48 | 11.19-11.48 | Be (Stephenson & Sanduleak 1977a) | U | l | ObV | GCAS | |
| 165 | GSC 08639-01611 | HD 102564, ASAS J114805-5726.0 | 11 48 05.120 | -57 26 02.30 | 8.66-8.95* | 8.66-8.95 | B2III (Houk & Cowley 1975) | E | s | SRO | GCAS | |
| 166 | GSC 08643-01679 | HD 307293, GDS_J1151597-595906 | 11 51 59.696 | -59 59 07.12 | 9.85-10.10* | 9.85-10.10 | em (Henize 1976), OBe (Stephenson & Sanduleak 1971), B3 (Bourgés et al. 2014) | E | l | ObV | GCAS | |
| 167 | GSC 08973-01406 | HD 307300, ASAS J115305-6023.3 | 11 53 05.073 | -60 23 17.46 | 10.00-10.31 | 10.00-10.31 | OB-e (Stephenson & Sanduleak 1971) | U | l | SRO | GCAS | |
| 168 | GSC 08977-00310 | HD 103574, NSV 19021 | 11 55 21.663 | -63 42 12.81 | 7.90-8.01* | 7.90-8.01 | em (MacConnell 1981), B2V(e) (Houk & Cowley 1975) | E | l, s, u | LTV | BE | |
| 169 | GSC 09234-02316 | HD 103715, NSV 5395 | 11 56 28.268 | -71 39 08.30 | 9.07-9.26* | 9.07-9.26 | em (Henize 1976), B0/3:ne (Houk & Cowley 1975), B2V?ne (Feast et al. 1957) | E | l, u | LTV,ObV | GCAS | |
| 170 | GSC 08973-01861 | HD 103872, GDS_J1157349-613259 | 11 57 34.951 | -61 32 59.12 | 8.83-8.93* | 8.83-8.93 | em (Henize 1976), B5e shell? (Houk & Cowley 1975) | M | l, u | LTV | BE | |
| 171 | GSC 08977-00421 | HD 309397, CPD-61 2828 | 11 57 37.026 | -61 52 54.46 | 10.61-10.70 | 10.61-10.70 | B9 (Nesterov et al. 1995), B7/8 (Sundman et al. 1974) | L | | LTV | GCAS: | |
| 172 | GSC 08978-01510 | HD 309467, GDS_J1202044-631522 | 12 02 04.394 | -63 15 22.05 | 9.94-10.12* | 9.94-10.12 | Be (Stephenson & Sanduleak 1977a) | U | l, u | LTV | BE | |
| 173 | GSC 08974-01031 | HD 104552, NSV 19130 | 12 02 23.398 | -61 27 05.26 | 9.36-9.52* | 9.36-9.53 | em (Henize 1976), B3/5e: (Houk & Cowley 1975) | M | l, u | ObV | GCAS | |
| 174 | GSC 08974-00327 | CD-60 3989, GDS_J1207221-612424 | 12 07 22.003 | -61 24 24.62 | 9.56-9.90* | 9.56-9.90 | OB (Münch 1954), B6-7 (Sundman et al. 1974) | M | | ObV | GCAS | |
| 175 | GSC 08982-00852 | CPD-63 2220 | 12 16 37.856 | -64 17 20.08 | 10.35-10.56 | 10.35-10.56 | em (Wray 1966) | U | l | LTV | GCAS | |
| 176 | GSC 08641-01826 | HD 106793, CPD-55 4964 | 12 16 59.720 | -56 24 24.23 | 9.68-9.89* | 9.66-9.89 | B8/9IVe (Levenhagen & Leister 2006), B7e (Stephenson & Sanduleak 1977a) | L | l | ObV | GCAS | |
| 177 | GSC 08974-00002 | NSV 19349, CPD-59 4156 | 12 17 08.449 | -60 40 29.68 | 10.63-10.89* | 10.63-10.89 | B7e (Stephenson & Sanduleak 1977a) | L | l | ObV | GCAS | |
| 178 | GSC 08979-00623 | CPD -62 2673 | 12 18 09.428 | -63 01 13.42 | 9.37-9.48 | 9.33-9.48 | B7e: (Stephenson & Sanduleak 1977b) | L | l, u | LTV | GCAS | |
| 179 | GSC 08975-03998 | HD 106960, CPD-60 3892 | 12 18 13.632 | -61 29 30.86 | 9.50-9.61* | 9.50-9.61 | em (Henize 1976), B8II/III (Houk & Cowley 1975) | L | l | ObV | GCAS | |
| 180 | GSC 08975-00799 | HD 107208, GDS_J1219489-603018 | 12 19 48.955 | -60 30 18.34 | 8.94-9.10 | 8.94-9.10 | B3II (Houk & Cowley 1975), OB (Münch 1954) | E | u | LTV, EB | EA+BE | 1.68445(5) |
| 181 | GSC 08988-02966 | HD 109474, CPD-60 4167 | 12 35 27.589 | -61 18 08.65 | 8.83-9.02* | 8.83-9.02 | em (Henize 1976), B8Ib/II (Houk & Cowley 1975) | L | l, s, u | ObV | GCAS | |
| 182 | GSC 08992-01008 | HD 111124, CPD-62 2928 | 12 47 50.751 | -62 59 45.28 | 9.27-9.42* | 9.27-9.42 | em (Henize 1976), B1/3I: (Houk & Cowley 1975) | E | l | SRO, LTV | GCAS | |
| 183 | GSC 08988-01196 | HD 111363, CPD-60 4306 | 12 49 27.495 | -60 41 56.82 | 8.80-9.11* | 8.77-9.11 | em (MacConnell 1981), B3III (Houk & Cowley 1975) | E | l, u | ObV | GCAS | |
| 184 | GSC 08989-02518 | HD 312075, CPD-59 4513 | 12 52 36.241 | -60 18 25.34 | 10.30-10.54* | 10.30-10.70 | em (Henize 1976), B0 (Cannon & Mayall 1949) | E | l | LTV | GCAS | |
| 185 | GSC 09001-00465 | HD 111906, CPD-66 1982 | 12 53 41.623 | -67 24 15.84 | 9.42-9.56* | 9.37-9.56 | em (MacConnell 1981), B3/5III/V (Houk & Cowley 1975) | M | l | LTV, NRP | LERI | 0.43626(3) |
| 186 | GSC 09245-01017 | HD 112442, CPD-69 1733 | 12 58 06.815 | -70 32 49.24 | 9.26-9.42* | 9.26-9.42 | B5/6V (Houk & Cowley 1975) | M | | ObV | GCAS | |
| 187 | GSC 08660-01131 | HD 112825, ASAS J130024-5941.1 | 13 00 23.895 | -59 41 07.60 | 9.53-9.78* | 9.53-9.78 | B1.5IVe (Jaschek et al. 1964), em (Henize 1976) | E | l, u | SRO | GCAS | |
| 188 | GSC 08660-00731 | CPD-59 4679 | 13 02 47.858 | -59 54 13.23 | 10.14-10.40* | 10.14-10.42 | Be (Johnston et al. 1992), pulsar companion, Be (Stephenson & Sanduleak 1977a), B5 (Stephenson & Sanduleak 1977a), B8Ib (Kennedy 1996), B2V: (Buscombe 1998) | U | l | ObV | GCAS | |
| 189 | GSC 08652-02075 | HD 113399, ASAS J130418-5610.6 | 13 04 17.639 | -56 10 34.42 | 8.88-9.12 | 8.88-9.12 | em (Henize 1976), B7II (Houk & Cowley 1975) | L | l | ObV | GCAS | |
| 190 | GSC 08998-01018 | HD 114044, CPD-64 2199 | 13 09 09.723 | -64 54 46.42 | 9.40-9.64* | 9.40-9.64 | B5V (Houk & Cowley 1975) | M | | ObV | GCAS | |
| 191 | GSC 09245-00106 | HD 114200, ASAS J131053-7048.5 | 13 10 52.704 | -70 48 31.08 | 8.43-8.82* | 8.43-8.82 | B1IIIne (Garrison et al. 1977), B5Ve (Hill et al. 1974), B0/2V:e (Houk & Cowley 1975) | E | l, s | ObV | GCAS | |
| 192 | GSC 08998-01663 | HD 114516, ASAS J131222-6348.7 | 13 12 22.388 | -63 48 43.94 | 8.22-8.70* | 8.22-8.70 | em (Henize 1976), B0V:e: (Houk & Cowley 1975) | E | l | SRO | GCAS | |

Table A1continued.

| (1) No. | (2) ID GSC | (3) ID alt | (4) α (J2000) | (5) δ (J2000) | (6) Range(V) [mag] | (7) Range(V) lit. [mag] | (8) Spec.type lit. | (9) subtype [E/M/L/U] | (10) emission flag | (11) Var.type [LB17] | (12) Var.type [GCVS/VSX] | (13) Period(s) [d] |
|---|---|---|---|---|---|---|---|---|---|---|---|---|
| 193 | GSC 08990-00217 | HD 115114, NSV 19676 | 13 16 16.119 | -61 45 44.17 | 9.69-9.92* | 9.29-9.92 | B1III:ne (Garrison et al. 1977), B0e (Humphreys 1975) | E | l, u | LTV | GCAS | |
| 194 | GSC 08995-02904 | CPD-61 3736, ASAS J132922-6202.3 | 13 29 22.315 | -62 02 17.42 | 9.49-9.95* | 9.49-9.95 | B2IVe (Garrison et al. 1977) | E | l | ObV | GCAS | |
| 195 | GSC 07793-00222 | CPD-38 5581 | 13 44 45.709 | -39 29 00.84 | 9.82-10.11 | 9.82-10.15 | OB 1995PASP..107..846D, Be (Bidelman & MacConnell 1973) | U | l | LTV | GCAS | |
| 196 | GSC 09016-00519 | HD 119423 | 13 45 18.398 | -66 45 16.78 | 7.35-7.60* | 7.35-7.60 | B3Vne (Levenhagen & Leister 2006), B8e (Stephenson & Sanduleak 1977a) | M | l, s, u | ObV | GCAS+LERI | |
| 197 | GSC 09008-04083 | HD 119763, CPD-62 3579 | 13 47 00.413 | -63 06 12.70 | 9.62-9.86* | 9.55-9.86 | em (MacConnell 1981), B9II Kn (Houk & Cowley 1975) | L | l | LTV | GCAS | |
| 198 | GSC 08676-01771 | HD 120330, ASAS J135026-5944.8 | 13 50 26.092 | -59 44 52.83 | 7.78-8.14* | 7.72-8.14 | B2/3Vnne (Houk & Cowley 1975) | E | l, s | ObV,LTV | GCAS | |
| 199 | GSC 09009-01997 | HD 122669, NSV 20029 | 14 05 22.036 | -62 30 26.63 | 8.94-9.02* | 8.94-9.02 | B1III:e (Garrison et al. 1977), B0.5IIep (Garrison et al. 1975), B0.5Ve (Crampton 1971) | E | l, s, u | LTV | BE | |
| 200 | GSC 09009-02487 | HD 122691, ASAS J140525-6235.4 | 14 05 25.178 | -62 35 18.37 | 9.33-9.81 | 9.20-9.81 | B1:Vnne (Garrison et al. 1977) | E | l | SRO | GCAS | |
| 201 | GSC 09005-03448 | CPD-61 4341, ASAS J140726-6143.2 | 14 07 26.528 | -61 43 13.09 | 10.16-10.40* | 10.16-10.42 | OB- (Lynga 1964) | U | | ObV | GCAS | |
| 202 | GSC 09005-03474 | Hen 3-966, ASAS J140947-6145.0 | 14 09 47.761 | -61 44 58.43 | 11.63-12.07 | 11.63-12.07 | OBe (Stephenson & Sanduleak 1977a) | U | l | IP | VAR | 16.77(1) |
| 203 | GSC 09013-01063 | CPD-63 3152, ASAS J141100-6427.6 | 14 11 00.704 | -64 27 32.82 | 9.88-10.01* | 9.88-10.01 | Be: (Stephenson & Sanduleak 1977b) | U | l | ObV | GCAS | |
| 204 | GSC 09006-04576 | CPD-60 5320, GDS_J1422173-603957 | 14 22 17.323 | -60 39 57.20 | 9.64-9.84* | 9.64-9.84 | Be (Stephenson & Sanduleak 1977a), Be (Bidelman & MacConnell 1973) | U | l | LTV, ObV | GCAS | |
| 205 | GSC 08691-03023 | CPD-59 5640, ASAS J143516-5954.6 | 14 35 16.240 | -59 54 37.65 | 10.56-10.80* | 10.51-10.80 | Be (Stephenson & Sanduleak 1977a) | U | l | SRO | GCAS | |
| 206 | GSC 08688-01283 | HD 129772, CPD-55 6162 | 14 46 36.797 | -56 25 08.11 | 8.57-8.63* | 8.55-8.63 | em (MacConnell 1981), B7/9IIIp: (Houk & Cowley 1975), A2III: (Cannon & Pickering 1920) | L | l, s | LTV | BE | |
| 207 | GSC 09020-02147 | CPD-59 5788, ASAS J150346-6021.9 | 15 03 45.861 | -60 21 55.68 | 10.23-10.52* | 10.23-10.52 | OB- (Stephenson & Sanduleak 1971) | U | | SRO | GCAS | |
| 208 | GSC 07821-02254 | NSV 6905 | 15 04 02.156 | -38 27 18.72 | 11.68-12.02* | 11.68-12.03 | | U | | LTV,ObV | GCAS | |
| 209 | GSC 08305-02320 | HD 133901, CPD-50 7613 | 15 08 52.450 | -51 10 25.57 | 9.21-9.36* | 9.21-9.36 | Be shell (Venn et al. 1998), B8/9Iab (Houk 1978),A5Ibe (Stephenson & Sanduleak 1971) | L | l | LTV | BE | |
| 210 | GSC 09436-00541 | HD 132875 | 15 10 40.749 | -80 22 44.05 | 9.42-9.55* | 9.40-9.55 | B8II/III(e:) (Houk & Cowley 1975) | L | l | ObV | GCAS | |
| 211 | GSC 08702-00469 | ALS 19454, ASAS J151114-5724.8 | 15 11 14.025 | -57 24 47.63 | 11.34-11.97 | 11.19-11.97 | OB+e (Orsatti & Muzzio 1980) | U | l | ObV | GCAS | |
| 212 | GSC 09033-02403 | HD 134401, SAO 253057 | 15 13 12.155 | -65 58 09.02 | 8.89-8.98* | 8.89-9.02 | B2Vne (Levenhagen & Leister 2006), B2IVnnep (shell) (Garrison et al. 1977), B2Vne (Houk & Cowley 1975) | E | l, s, u | LTV, NRP | LERI+BE | 0.423720(2) |
| 213 | GSC 08303-01041 | HD 136556, ASAS J152320-5007.0 | 15 23 20.149 | -50 06 58.43 | 8.99-9.19 | 8.99-9.19 | B2/3V:ne (Houk 1978), B1Vne (Garrison et al. 1977), B:nne (Feast et al. 1961) | E | l, s | LTV,SRO: | GCAS | |
| 214 | GSC 07847-00082 | HD 136935, CPD-43 7060 | 15 24 54.683 | -44 09 14.63 | 7.95-8.09 | 7.95-8.10 | B6II (Houk 1978), B8 (Spencer Jones & Jackson 1939) | U | | ObV | GCAS | |
| 215 | GSC 08307-01059 | HD 137837, CPD-50 8117 | 15 30 23.229 | -51 09 39.72 | 9.00-9.19 | 9.00-9.19 | B5/8II/III (Houk 1978) | M | | ObV | GCAS | |
| 216 | GSC 09022-00605 | HD 138131, ASAS J153311-6054.6 | 15 33 11.287 | -60 54 33.69 | 8.99-9.19* | 8.99-9.19 | B6Vne (Houk & Cowley 1975) | M | l | ObV | GCAS | |
| 217 | GSC 08701-00997 | HD 142237, NSV 20435 | 15 56 05.974 | -54 57 09.55 | 8.75-8.94 | 8.63-8.94 | B1Vne (Levenhagen & Leister 2006), Be (Stephenson & Sanduleak 1977a), B2Vne (Houk & Cowley 1975), B3Ve (Humphreys 1975) | E | l, u, s | ObV | GCAS | |
| 218 | GSC 08719-02158 | HD 146261, CPD-57 7793 | 16 18 24.885 | -57 49 32.97 | 9.13-9.32* | 9.13-9.32 | B8/9II (Houk & Cowley 1975), B5V (Feast 1957) | L | | ObV | GCAS | |
| 219 | GSC 08319-00698 | HD 146444, NSV 20575 | 16 18 39.380 | -49 24 49.94 | 7.47-7.75* | 7.47-7.75 | B2Ve (A&), B2Vne (Houk 1978) | E | l, s | ObV, LTV | GCAS | |
| 220 | GSC 08719-00464 | HD 146324, CPD-57 7816 | 16 18 49.010 | -57 55 51.54 | 7.68-7.99* | 7.68-7.99 | em (Henize 1976), B5V(e) (Feast 1957) | M | l, u, s | LTV | GCAS | |
| 221 | GSC 08711-02092 | HD 146596, GDS_J1619426-524618 | 16 19 42.669 | -52 46 19.00 | 8.03-8.11 | 7.98-8.11 | B5IV/V(e) (Houk 1978), em (Henize 1976) | M | l, s | LTV | BE | |
| 222 | GSC 08723-00042 | HD 146531, CPD-57 7866 | 16 19 55.512 | -58 10 01.86 | 9.72-9.78* | 9.69-9.80 | B3Ve (Levenhagen & Leister 2006), B5/7III (Houk & Cowley 1975) | M | l, u | LTV | BE | |
| 223 | GSC 08715-01941 | HD 147302, CPD-55 7498 | 16 24 01.276 | -55 27 13.35 | 7.65-7.74* | 7.65-7.74 | B2Vn (Garrison et al. 1977), B2III:n(e?) (Houk & Cowley 1975) | E | l, s | NRP | LERI | 0.510717(3) |
| 224 | GSC 08712-02498 | CPD-53 7997, ASAS J162749-5332.1 | 16 27 48.809 | -53 32 05.05 | 9.59-9.89* | 9.59-9.89 | B2.5IVn(e) (Garrison et al. 1977) | E | l, u | SRO | GCAS | |
| 225 | GSC 08325-05810 | HD 148567, ASAS J163103-4628.8 | 16 31 02.812 | -46 28 45.72 | 7.84-8.02* | 7.84-8.11 | B2II:ne shell? (Houk 1978), B1IIIne (Garrison et al. 1977) | E | l, s | ObV, NRP | GCAS+LERI | 0.789476(9) |
| 226 | GSC 09042-01527 | HD 148907, CPD-61 5734 | 16 35 14.063 | -61 53 40.15 | 9.26-9.36* | 9.26-9.36 | B7e (Stephenson & Sanduleak 1977a), B5/7V (Houk & Cowley 1975) | M | l | ObV | GCAS | |
| 227 | GSC 08337-00341 | HD 149814, CPD-51 9804 | 16 39 50.025 | -52 07 15.89 | 9.09-9.16* | 9.05-9.16 | B5/7III/V (Houk 1978), em (Henize 1976) | M | l, s | LTV | BE | |
| 228 | GSC 08325-00916 | HD 328684, ASAS J164052-4639.0 | 16 40 52.280 | -46 39 02.31 | 10.37-10.73* | 10.37-10.73 | em (Vega et al. 1980), A2 (Nesterov et al. 1995) | U | l | ObV | GCAS | |
| 229 | GSC 08330-05153 | HD 150231, CD-47 10953 | 16 41 52.647 | -47 22 49.32 | 9.02-9.41* | 9.02-9.41 | em (MacConnell 1981), B3V (Houk 1978) | E | l | ObV, NRP | GCAS+LERI | 0.580790(7) |
| 230 | GSC 08338-02080 | HD 151083, ASAS J164751-5146.1 | 16 47 51.392 | -51 46 03.99 | 8.78-9.36* | 8.78-9.36 | B2Vn(e) (Houk 1978), B1:IIInne (Garrison et al. 1977) | E | l, u, s | ObV | GCAS | |
| 231 | GSC 08734-02077 | HD 151873, CPD -56 7887 | 16 53 21.134 | -57 01 35.38 | 9.08-9.21 | 9.08-9.21 | em (Henize 1976), A(e)p shell (Houk & Cowley 1975), B shell (Bidelman & MacConnell 1973) | L | l | ObV | GCAS | |
| 232 | GSC 07872-00681 | HD 322282, NSV 20824 | 16 54 29.032 | -40 41 17.37 | 8.90-9.06* | 8.87-9.06 | em (Henize 1976), OBe (Stephenson & Sanduleak 1971), Be (Merrill & Burwell 1949) | U | l, u, s | ObV | GCAS | |
| 233 | GSC 07872-00390 | HD 322447, NSV 20862 | 16 56 18.014 | -40 40 48.78 | 8.77-8.90* | 8.77-8.92 | em (Henize 1976), B1IV (Schild et al. 1969) | E | l | ObV | BE | |
| 234 | GSC 08328-00373 | HD 155280, CPD-46 8449 | 17 12 50.342 | -46 26 19.44 | 8.48-8.76* | 8.48-8.76 | B2/3II (Houk 1978); B3III (Garrison et al. 1977) | E | u, s | SRO | GCAS | |
| 235 | GSC 07878-00246 | HD 155352, ASAS J171300-4238.0 | 17 13 00.391 | -42 38 00.42 | 8.17-8.47* | 8.17-8.47 | B2V (Houk 1978); Be (Bidelman & MacConnell 1973) | E | l, u, s | ObV | GCAS | |
| 236 | GSC 08345-03046 | HD 156008, CPD-47 8152 | 17 17 30.679 | -47 44 22.49 | 9.53-9.63* | 9.52-9.63 | B8IV (Houk 1978), em (Henize 1976) | L | l | LTV | BE | |
| 237 | GSC 07374-00838 | NSV 21486, CPD-35 6938 | 17 20 26.473 | -35 44 06.88 | 10.09-10.27 | 10.09-10.70 | em (Henize 1976), OBe (Stephenson & Sanduleak 1971),B0.5:Ve (Roslund 1966) | E | l | ObV | BE | |
| 238 | GSC 07366-00860 | CPD-32 4502, ASAS J172148-3304.8 | 17 21 47.648 | -33 04 48.74 | 10.66-11.03 | 10.66-11.03 | em (Henize 1976), OB- (Stephenson & Sanduleak 1971) | U | l | SRO | GCAS | |
| 239 | GSC 08341-00889 | HD 157115, ASAS J172336-4526.3 | 17 23 36.284 | -45 26 18.90 | 8.42-8.62* | 8.42-8.62 | B5IV (Houk 1978) | M | | ObV | GCAS | |
| 240 | GSC 07375-00048 | HD 157829, CPD -29 4703 | 17 26 55.205 | -30 05 30.87 | 8.68-8.88* | 8.68-8.88 | B3III/IV (Houk 1982), em (A&) | E | l | ObV, LTV | GCAS | |
| 241 | GSC 08342-00052 | HD 157847, CPD-45 8603 | 17 28 00.368 | -45 45 15.65 | 9.41-9.58* | 9.41-9.58 | B9Ib (Houk 1978), em (MacConnell 1981) | L | l | ObV | GCAS | |
| 242 | GSC 07384-00247 | HD 320103, ASAS J173632-3403.8 | 17 36 31.591 | -34 03 47.71 | 10.40-10.67 | 10.40-10.69 | em (Henize 1976), OBe (Stephenson & Sanduleak 1971) | U | l | ObV | GCAS | |
| 243 | GSC 08342-01635 | HD 159489 | 17 37 13.837 | -45 09 26.63 | 7.98-8.26* | 7.98-8.26 | B1Ve (Levenhagen & Leister 2006), B3V(e)p (shell) (Garrison et al. 1977) | E | l, u | ObV | GCAS | |
| 244 | GSC 06835-00151 | CPD-28 5745, ASAS J173920-2805.7 | 17 39 19.633 | -28 05 39.42 | 11.08-11.30 | 11.08-11.30 | OB-e (Stephenson & Sanduleak 1971) | U | l | ObV | GCAS | |
| 245 | GSC 06839-00611 | CD-29 13831, ASAS J173930-2947.7 | 17 39 29.975 | -29 47 42.08 | 10.59-11.07* | 10.59-11.07 | OB+e (Stephenson & Sanduleak 1971) | U | l | SRO | GCAS | |
| 246 | GSC 07889-01252 | HD 160751, CPD-39 7553 | 17 43 31.964 | -39 49 20.33 | 8.59-8.71* | 8.50-8.71 | B7/8II (Houk 1982) | L | | LTV, ObV | GCAS | |
| 247 | GSC 07385-01338 | HD 161774, CPD-33 4618 | 17 48 51.346 | -33 51 45.05 | 8.64-8.76 | 8.64-8.72 | B8Vnne (Levenhagen & Leister 2006), B5:V:nne (Houk 1982) | L | l, u, s | LTV, NRP | LERI | 0.410651(4) |
| 248 | GSC 07886-02848 | HD 162352 | 17 52 18.591 | -37 45 01.99 | 8.21.8.30* | 8.21-8.31 | B2/3ne: (Houk 1982), B1.5Vn(e) (Garrison et al. 1977) | E | l, s | LTV, NRP | LERI | 0.612158(6) |
| 249 | GSC 06853-01718 | ASAS J175438-2926.8 | 17 54 38.171 | -29 26 46.64 | 11.57-12.42* | 11.57-12.42 | OB-e: (Stephenson & Sanduleak 1977b) | U | l | ObV | GCAS | |
| 250 | GSC 06853-02519 | HD 163453, NSV 24052 | 17 57 25.827 | -28 15 18.41 | 9.23-10.10* | 9.23-10.10 | Bnne (Houk 1982), em (Henize 1976) | U | l | ObV | GCAS | |
| 251 | GSC 06841-01725 | ALS 4506, ASAS 175746-2412.2 | 17 57 46.409 | -24 12 09.47 | 11.11-11.33* | 11.11-11.33 | em (Kohoutek & Wehmeyer 2003), OB (Stephenson & Sanduleak 1971) | U | l | ObV | GCAS | |
| 252 | GSC 06846-01106 | HD 314947, ASAS J180235-2541.7 | 18 02 34.984 | -25 41 43.53 | 10.26-10.44 | 10.17-10.44 | em (Henize 1976), Be (Velghe 1957) | U | l | LTV, ObV | GCAS | |
| 253 | GSC 07399-01124 | HD 165248, CPD-34 7582 | 18 06 31.633 | -34 30 52.00 | 9.30-9.38* | 9.30-9.38 | B3/5V: (Houk 1982) | M | | LTV | BE | |
| 254 | GSC 06263-03157 | HD 165595, CPD-22 6704 | 18 07 38.231 | -22 06 57.46 | 8.45-8.50 | 8.45-8.50 | B3 (Cannon & Mayall 1949) | E | u | LTV, NRP | LERI | 0.540456(6) |
| 255 | GSC 07399-01226 | HD 321289, ASAS J180801-3524.2 | 18 08 00.469 | -35 24 15.29 | 9.90-10.24* | 9.90-10.24 | B5e (Stephenson & Sanduleak 1977a) | M | l | SRO | GCAS | |
| 256 | GSC 06272-02199 | HD 165970, CPD-19 6497 | 18 09 15.914 | -19 43 23.86 | 8.96-9.16* | 8.96-9.16 | B5e (Stephenson & Sanduleak 1977a) | M | l | ObV | GCAS | |

Table A1continued.

| (1) No. | (2) ID GSC | (3) ID alt | (4) α (J2000) | (5) δ (J2000) | (6) Range(V) [mag] | (7) Range(V) lit. [mag] | (8) Spec.type lit. | (9) subtype [E/M/L/U] | (10) emission flag | (11) Var.type [LB17] | (12) Var.type [GCVS/VSX] | (13) Period(s) [d] |
|---|---|---|---|---|---|---|---|---|---|---|---|---|
| 257 | GSC 06268-02490 | HD 166188, ASAS J181018-1811.7 | 18 10 18.319 | -18 11 41.36 | 8.78-9.06* | 8.78-9.06 | B2V:ep (Morgan et al. 1955) | E | l, s | SRO | GCAS | |
| 258 | GSC 06276-00317 | HD 166146 | 18 10 18.747 | -22 23 14.69 | 9.88-9.98* | 9.88-9.98 | B8 (Cannon & Mayall 1949) | L | | LTV | GCAS | |
| 259 | GSC 06851-02063 | HD 166365 | 18 11 27.175 | -27 08 02.62 | 10.14-10.26* | 10.14-10.26 | B9II/III (Houk 1982), B7e: (Stephenson & Sanduleak 1977b) | L | l | ObV | GCAS | |
| 260 | GSC 06847-02930 | HD 315277, ASAS J181208-2508.5 | 18 12 08.042 | -25 08 32.83 | 10.85-11.15 | 10.85-11.15 | OB+e (Stephenson & Sanduleak 1971), Be (Miller & Merrill 1951) | U | l | ObV, LTV | GCAS | |
| 261 | GSC 06851-04189 | HD 166629 | 18 12 38.466 | -27 08 29.26 | 9.10-9.37* | 9.10-9.37 | B5nne (Houk 1982), B9e (Merrill & Burwell 1949) | U | l | ObV | GCAS | |
| 262 | GSC 06268-00943 | HD 312861, ASAS J181307-1821.5 | 18 13 07.086 | -18 21 31.63 | 10.03-10.19* | 10.03-10.20 | B7e (Stephenson & Sanduleak 1977a) | L | l | ObV | GCAS | |
| 263 | GSC 06847-02073 | HD 166967, CPD-25 6403 | 18 14 08.185 | -25 18 39.00 | 8.36-8.48* | 8.35-8.48 | OB (Stephenson & Sanduleak 1971), B5e (Merrill & Burwell 1949) | M | l, u, s | LTV | GCAS | |
| 264 | GSC 06272-00394 | HD 167247, ASAS J181501-1912.5 | 18 15 01.126 | -19 12 27.66 | 9.10-9.20* | 9.10-9.20 | B5III/V (Houk 1978), B9 (Yale 1997) | U | | LTV | GCAS | |
| 265 | GSC 07404-05201 | HD 167233, NSV 24347 | 18 15 50.905 | -36 34 25.55 | 6.57-7.02* | 6.57-7.02 | B3Ve (A&), B3Ve (Buscombe & Kennedy 1969) | E | l, s | ObV | GCAS | |
| 266 | GSC 06269-02592 | ALS 4888, ASAS J181806-1728.9 | 18 18 05.938 | -17 28 54.73 | 10.59-11.00 | 10.59-11.00 | OB (Stephenson & Sanduleak 1971), em (Henize 1976) | U | l | ObV | GCAS | |
| 267 | GSC 06274-00902 | HD 313306, ASAS J182441-1940.4 | 18 24 41.455 | -19 40 25.02 | 9.61-9.84* | 9.59-9.84 | A0 (Cannon & Mayall 1949) | L | | ObV | GCAS | |
| 268 | GSC 07909-02656 | HD 169639 | 18 27 33.814 | -41 35 33.39 | 10.29-10.46 | 10.23-10.46 | em (MacConnell 1982), B8/9Ib/II (Houk 1978) | L | l | LTV | GCAS | |
| 269 | GSC 05703-02553 | HD 170603, BD-15 4995 | 18 30 54.729 | -14 55 39.85 | 9.24-9.54* | 9.24-9.54 | B3V (Hiltner & Iriarte 1955) | E | | ObV | GCAS | |
| 270 | GSC 05124-01543 | ASAS J183442-0638.8 | 18 34 42.525 | -06 38 49.26 | 11.68-11.87 | 11.68-11.87 | B5 (Roslund 1963) | M | | LTV | GCAS | |
| 271 | GSC 05703-01526 | HD 171392, ASAS J183506-1419.8 | 18 35 05.696 | -14 19 50.09 | 9.09-9.38 | 9.09-9.38 | OB (Nassau & Stephenson 1963) | U | | SRO, LTV | GCAS | |
| 272 | GSC 06275-00943 | HD 172122, ASAS J183912-2020.1 | 18 39 12.148 | -20 20 08.58 | 8.62-8.87 | 8.62-8.87 | B (Stephenson & Sanduleak 1977a), B2IVnp shell (Garrison et al. 1977) | E | l, s | LTV | GCAS | |
| 273 | GSC 05692-01642 | HD 172637, ASAS J184139-0803.3 | 18 41 38.802 | -08 03 15.26 | 9.27-9.38* | 9.27-9.38 | OB (Nassau & Stephenson 1963), B3 (Roslund 1963) | E | | ObV | GCAS | |
| 274 | GSC 05125-02006 | ALS 9887, ASAS J184244-0609.2 | 18 42 44.504 | -06 09 13.63 | 10.32-10.66* | 10.32-10.66 | em (MacConnell 1981), B3 (Roslund 1963) | E | l | ObV | GCAS | |
| 275 | GSC 00456-00461 | HD 173530 | 18 45 42.914 | +04 34 42.98 | 8.79-8.89* | 8.79-8.89 | B8II/IIIe (Stephenson & Sanduleak 1977a) | L | l | LTV | BE | |
| 276 | GSC 05693-07523 | TYC 5693-7523-1 | 18 46 31.795 | -07 38 36.51 | 11.77-12.05* | 11.77-12.05 | | U | | ObV | GCAS | |
| 277 | GSC 05126-03377 | ASAS J184740-0630.6 | 18 47 39.672 | -06 30 39.80 | 11.04-11.47* | 11.04-11.47 | em: (Stephenson & Sanduleak 1977b) | U | l | SRO | GCAS | |
| 278 | GSC 05701-00964 | HD 174070, SAO 161859 | 18 49 12.383 | -12 35 43.96 | 9.09-9.22* | 9.09-9.22 | em (MacConnell 1981), B4 (Neubauer 1943) | M | l, u | ObV | GCAS | |
| 279 | GSC 01026-02065 | HD 174571, ASAS J185047+0842.2 | 18 50 47.173 | +08 42 10.05 | 8.67-8.86 | 8.67-8.92 | Be shell? (Stephenson & Sanduleak 1977a), B2.5V:[n]e (Walborn 1971), B2e (Vieira et al. 2003) | E | l, u | ObV | GCAS | |
| 280 | GSC 06289-02980 | HD 174652, CPD-20 7256 | 18 52 16.482 | -20 18 52.91 | 9.03-9.20 | 9.02-9.20 | B9e (Merrill & Burwell 1949) | L | l | ObV | GCAS | |
| 281 | GSC 05123-00145 | HD 175180, ASAS J185427-0522.8 | 18 54 26.577 | -05 22 48.43 | 8.92-9.18 | 8.92-9.18 | B3III (Houk & Swift 1999) | E | | ObV | GCAS | |
| 282 | GSC 00463-02825 | HD 178720 | 19 09 37.901 | +00 55 29.49 | 8.88-9.11* | 8.88-9.11 | B2e (Bidelman 1981) | E | l, u, s | ObV | GCAS | |
| 283 | GSC 05131-01423 | SAO 143395 | 19 26 15.407 | -00 15 16.29 | 9.19-9.29* | 9.19-9.29 | B2 (Kelly & Kilkenny 1986) | E | | ObV | GCAS | |
| 284 | GSC 02129-00864 | HD 338423, ASAS J192626+2431.4 | 19 26 26.254 | +24 31 21.95 | 11.38-11.60 | 11.38-11.60 | em (Coyne et al. 1974), A2 (Cannon & Mayall 1949) | L | l | ObV: | BE: | |
| 285 | GSC 05149-01177 | HD 185092, NSV 24828 | 19 37 36.523 | -02 41 56.43 | 8.56-8.72 | 8.56-8.72 | B8Ib (Anderson & Francis 2012) | L | | ObV | GCAS | |
| 286 | GSC 01645-00281 | HD 347184 | 20 38 30.165 | +21 19 43.96 | 9.08-9.38 | 9.10-9.38 | em (MacConnell 1982) | U | l | ObV | GCAS | |
| 287 | GSC 01124-01184 | BD+08 4699 | 21 35 23.790 | +09 29 18.45 | 10.07-10.37* | 10.07-10.37 | B8 (Wright et al. 2003) | L | | LTV | GCAS | |

## APPENDIX B: LIGHT CURVES

**Figure B1.** The ASAS-3 light curves of all sample stars ($N = 287$). For convenience and to provide an easy identification in the corresponding tables (Tables A1 and D1), all stars were numbered in order of increasing right ascension (No. 1 – No. 287).

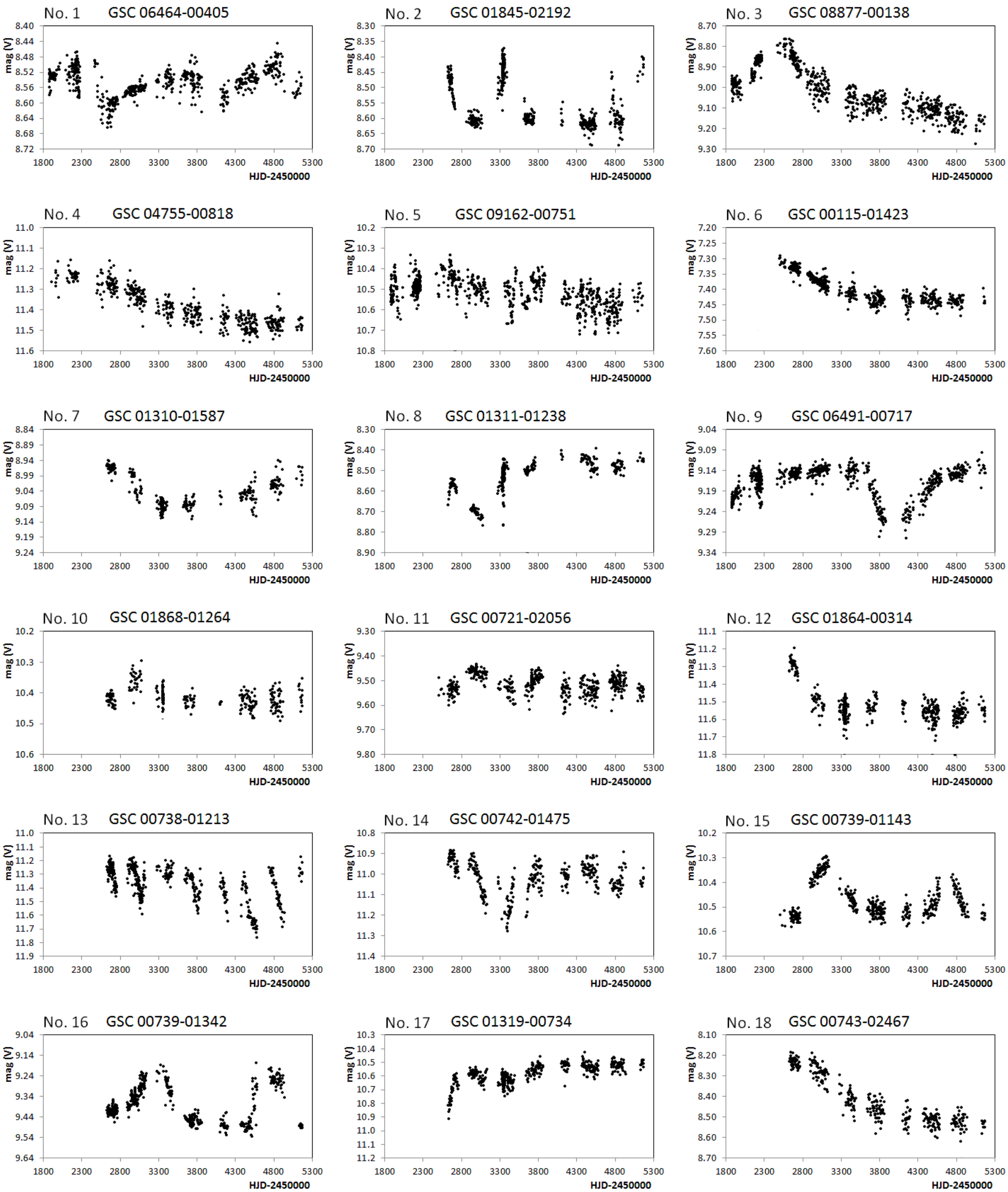

**Figure B1.** continued.

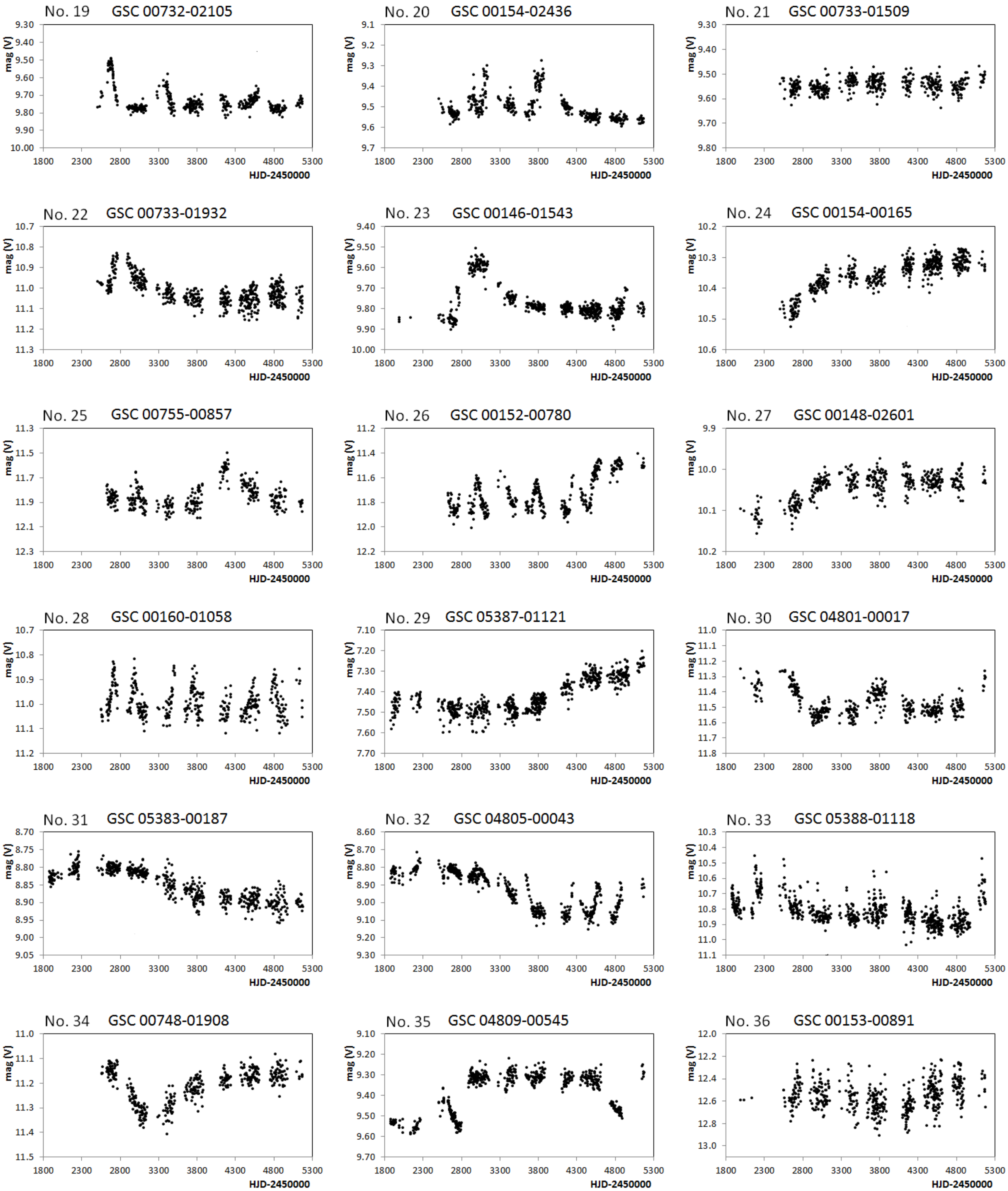

**Figure B1.** continued.

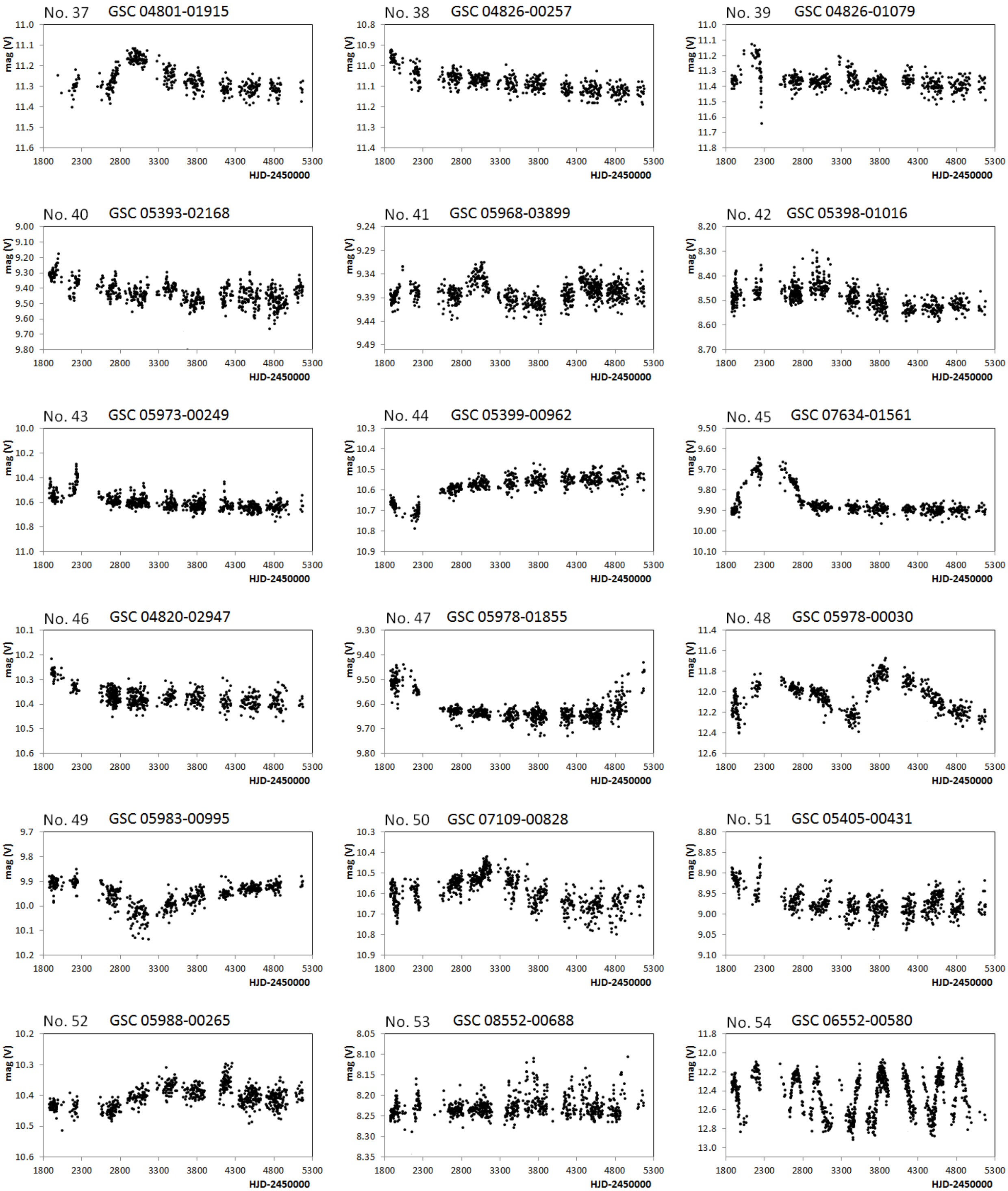

**Figure B1.** continued.

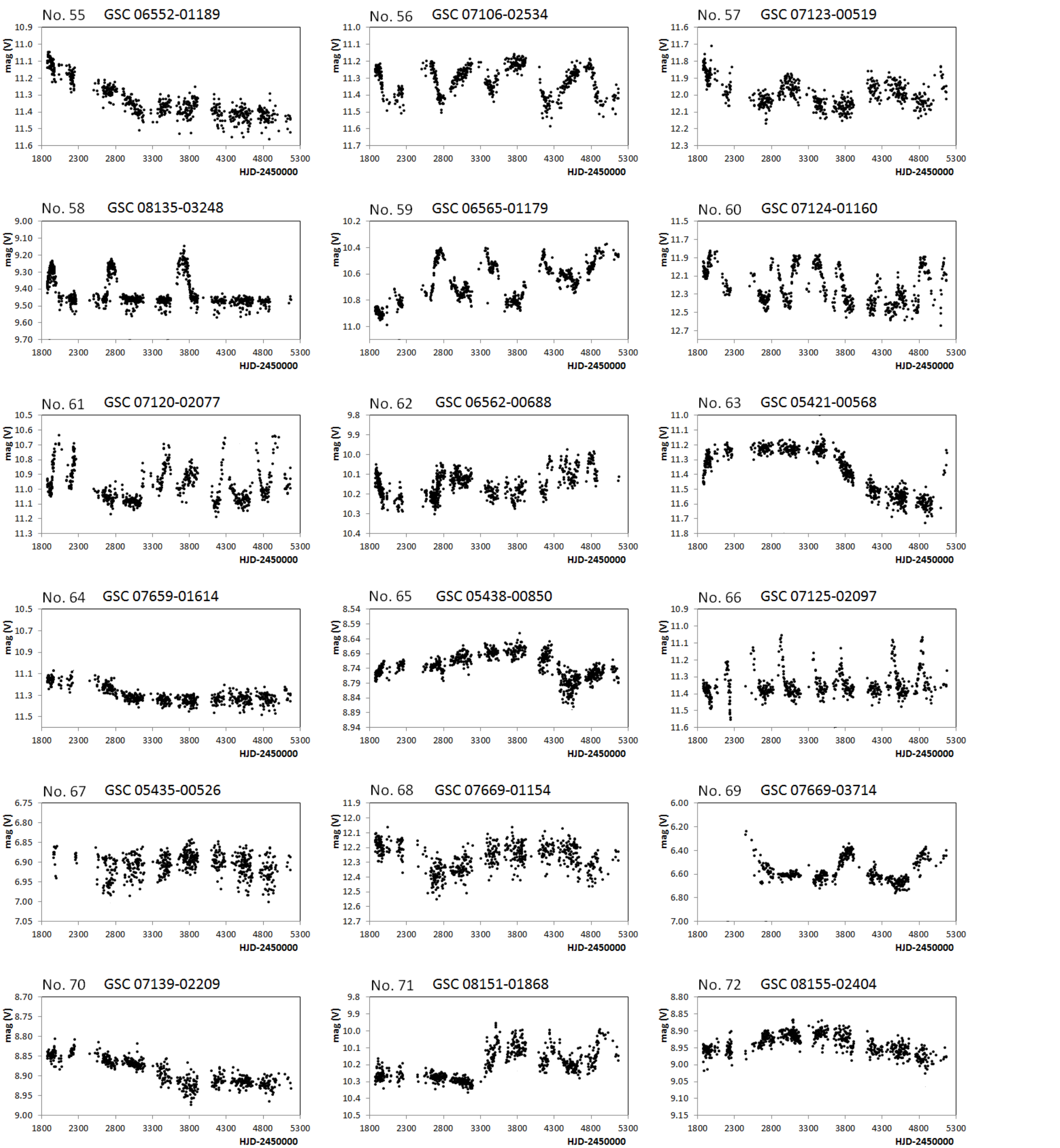

**Figure B1.** continued.

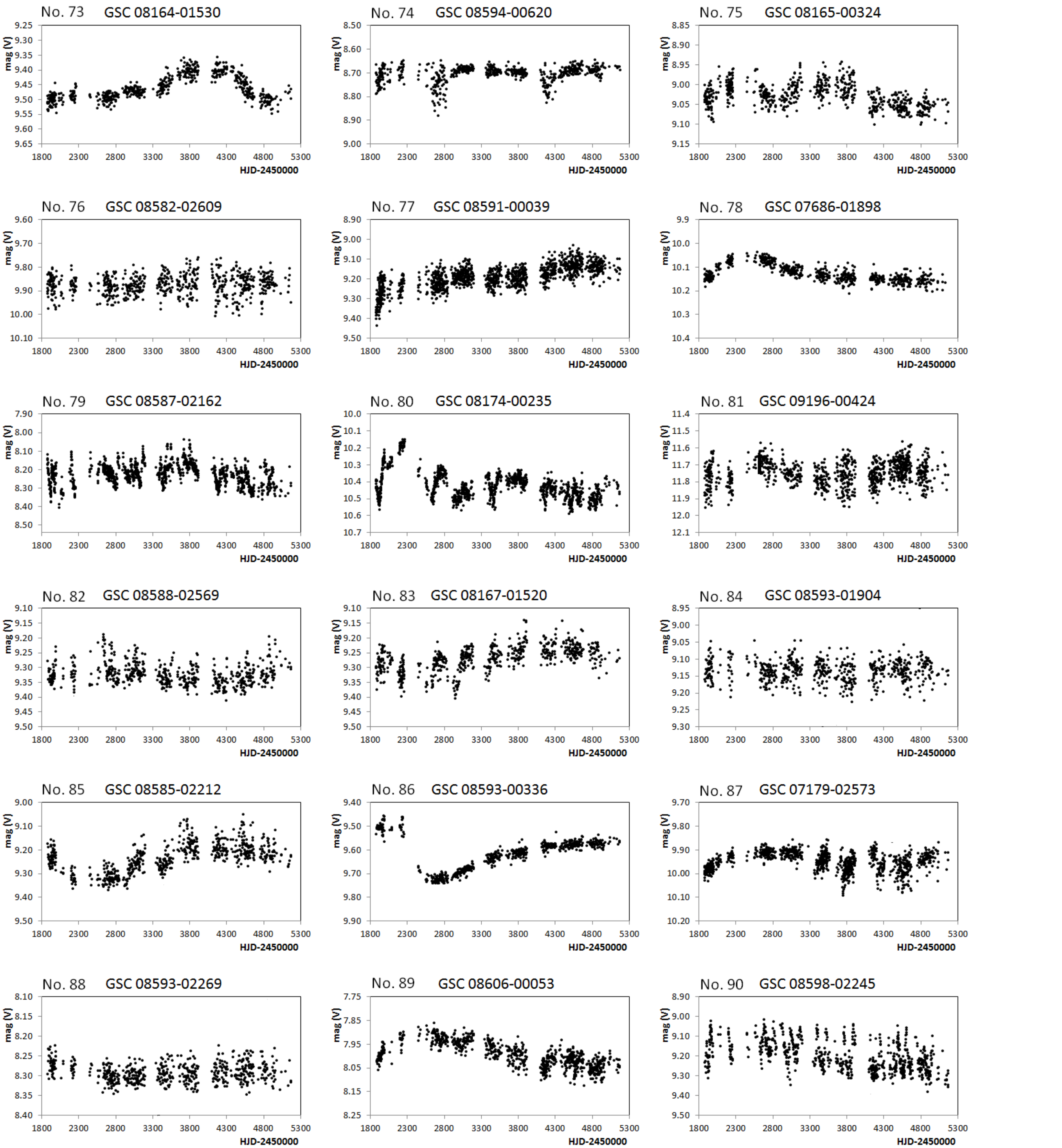

**Figure B1.** continued.

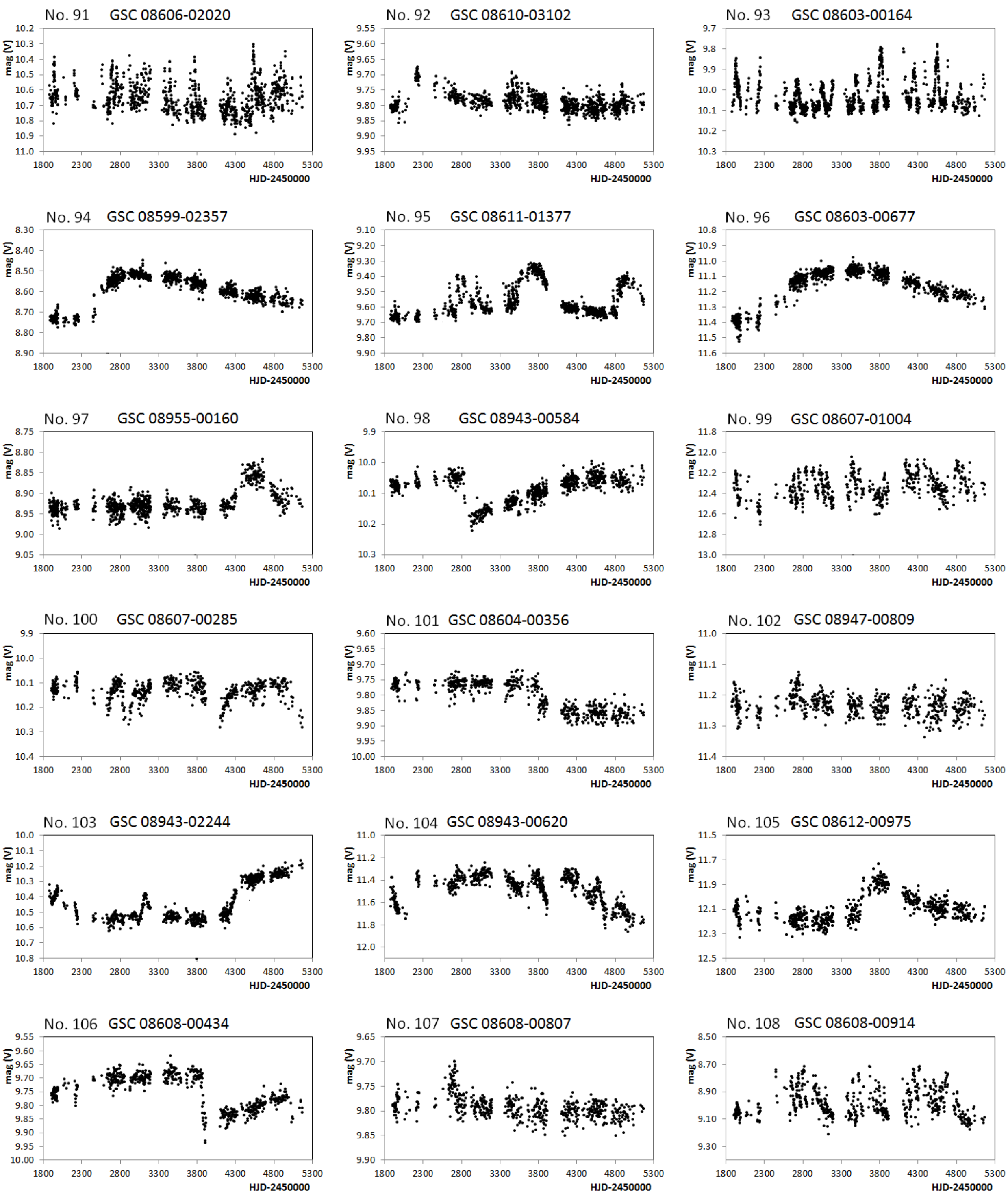

**Figure B1.** continued.

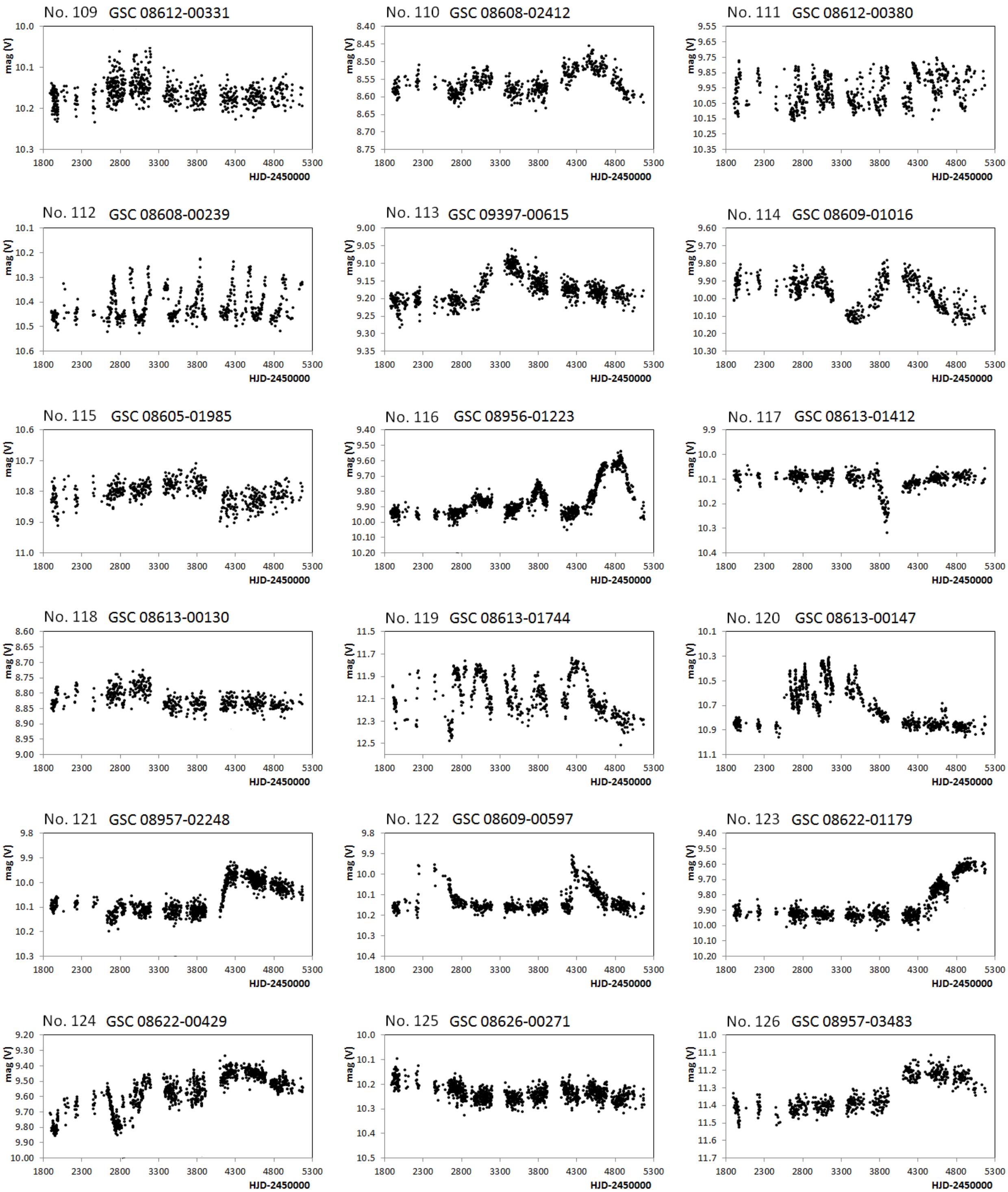

**Figure B1.** continued.

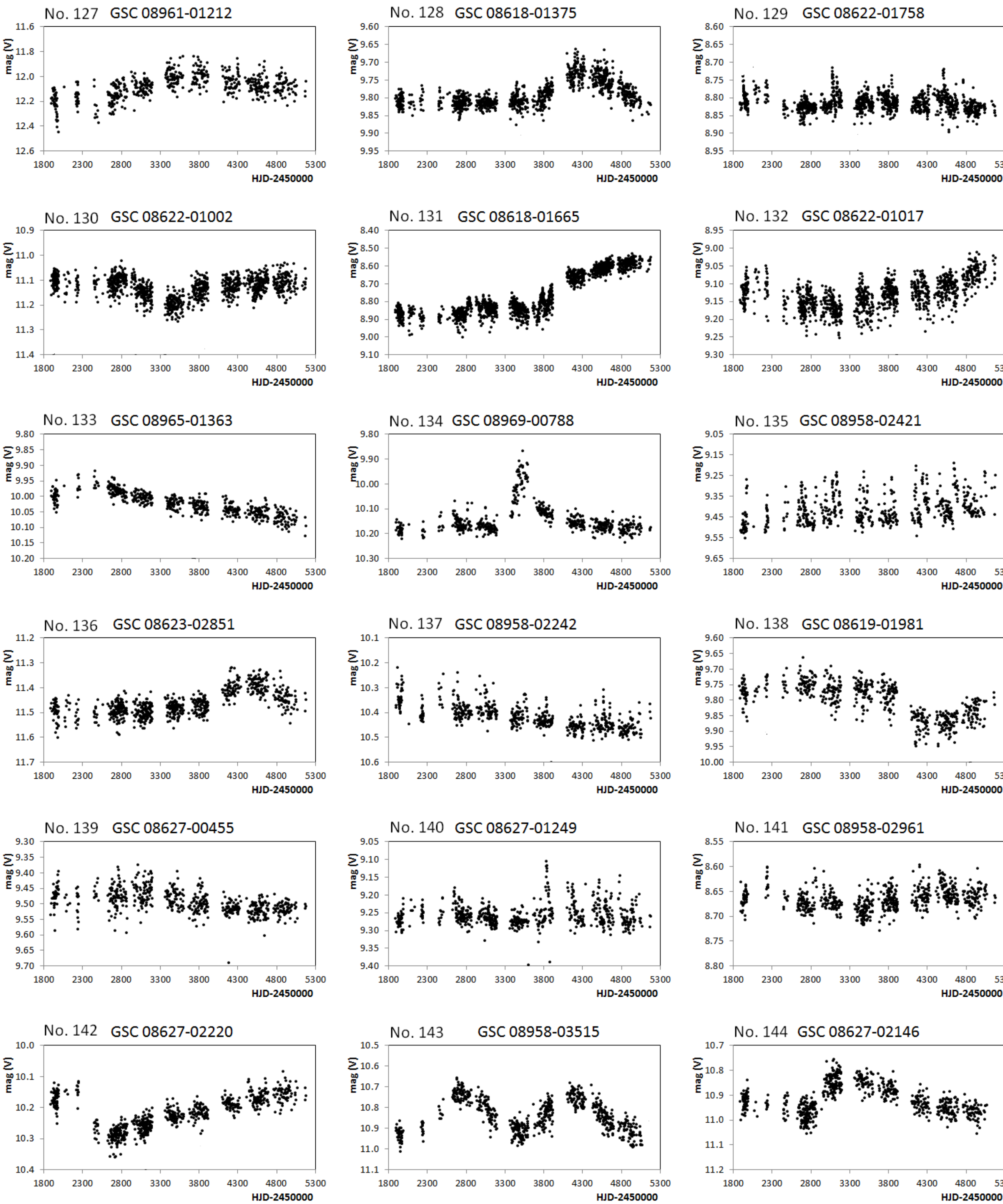

**Figure B1.** continued.

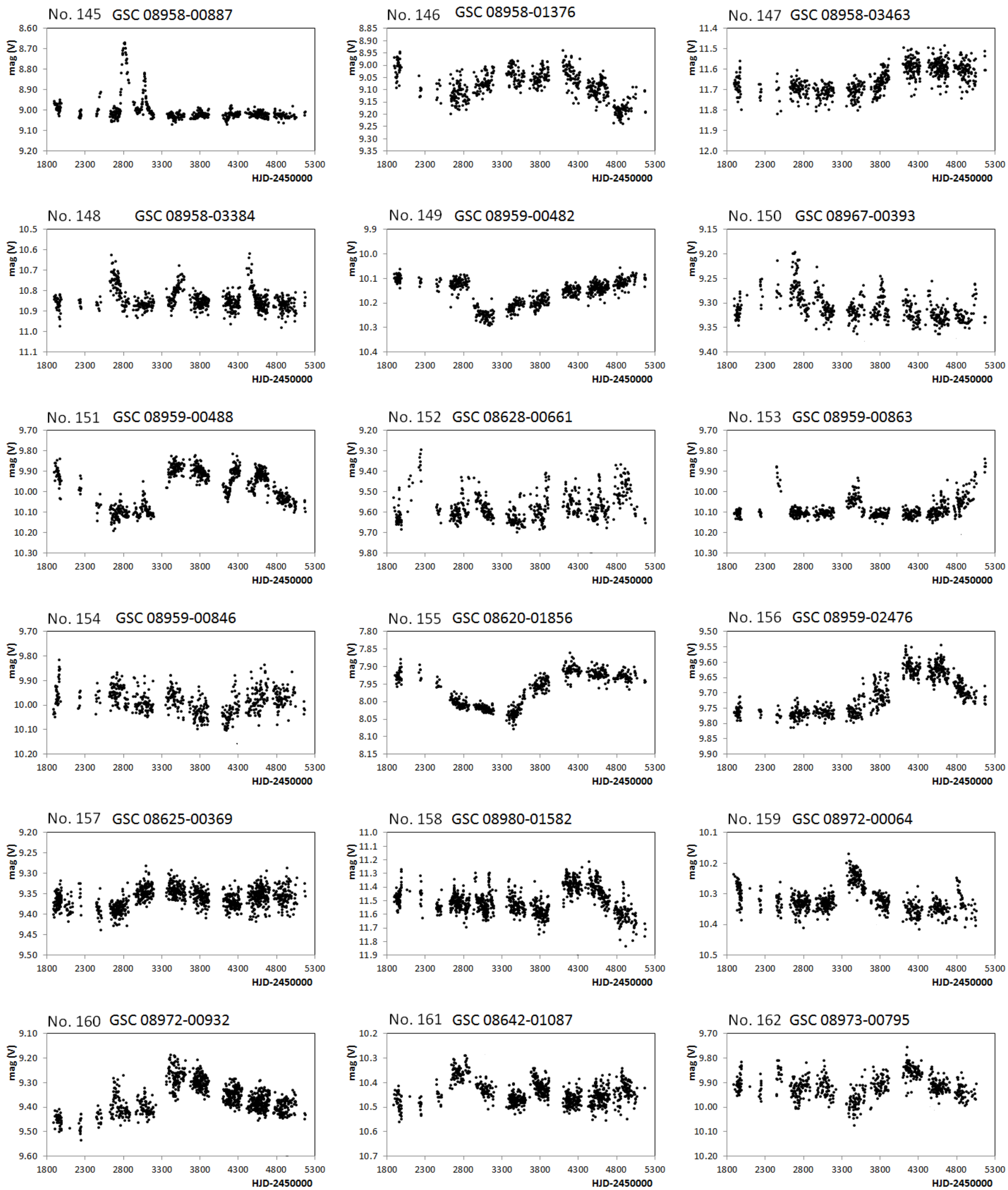

**Figure B1.** continued.

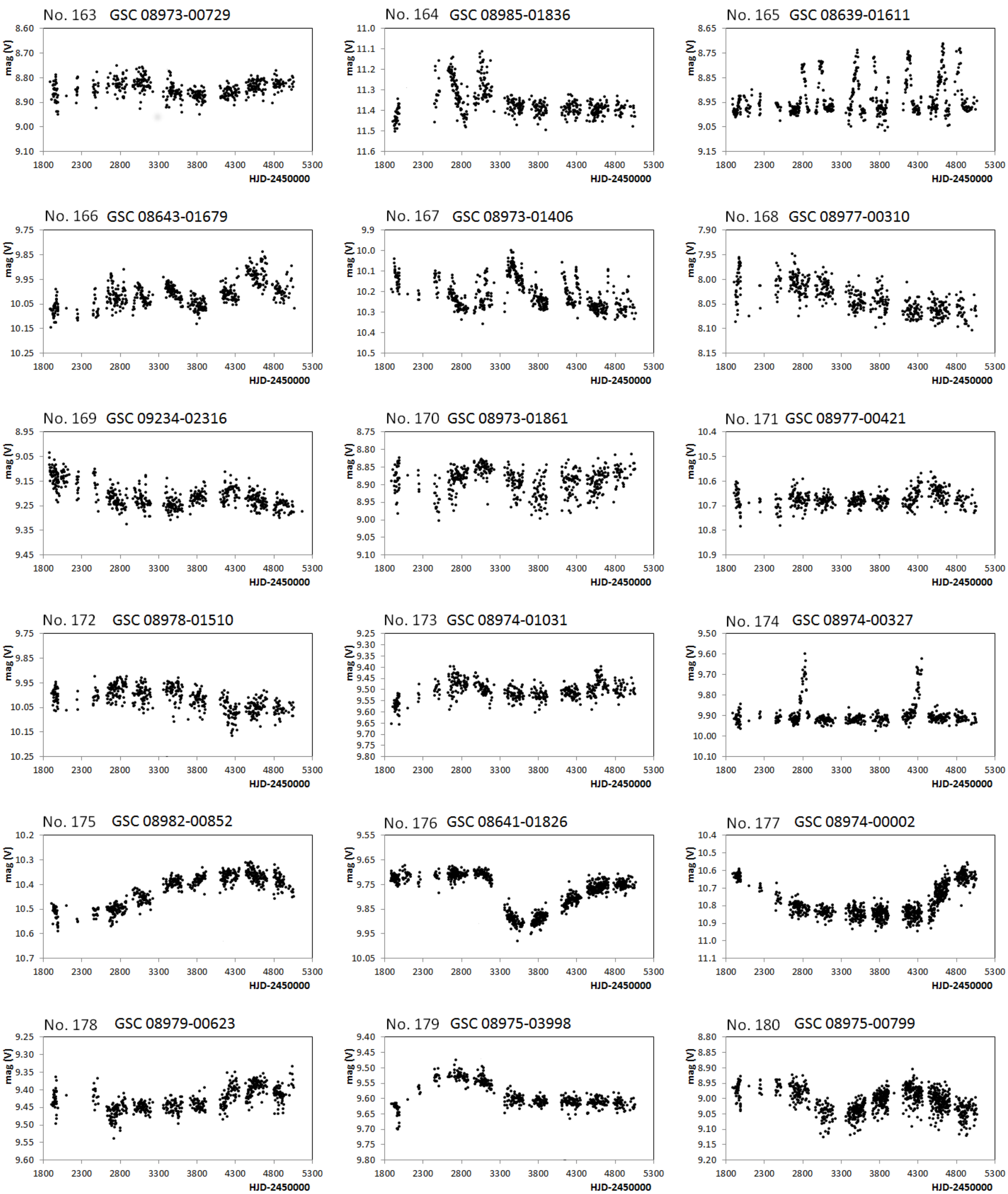

**Figure B1.** continued.

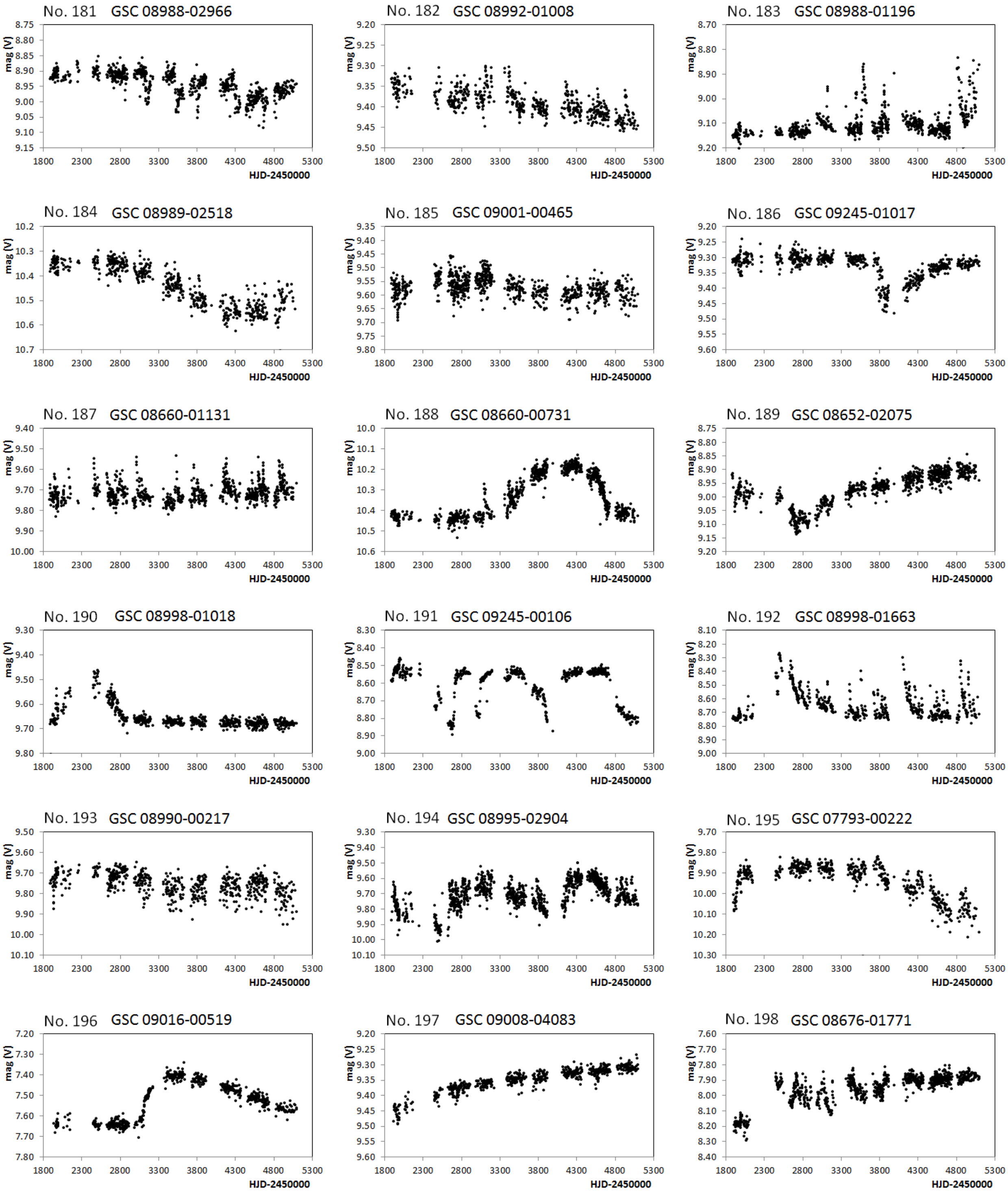

**Figure B1.** continued.

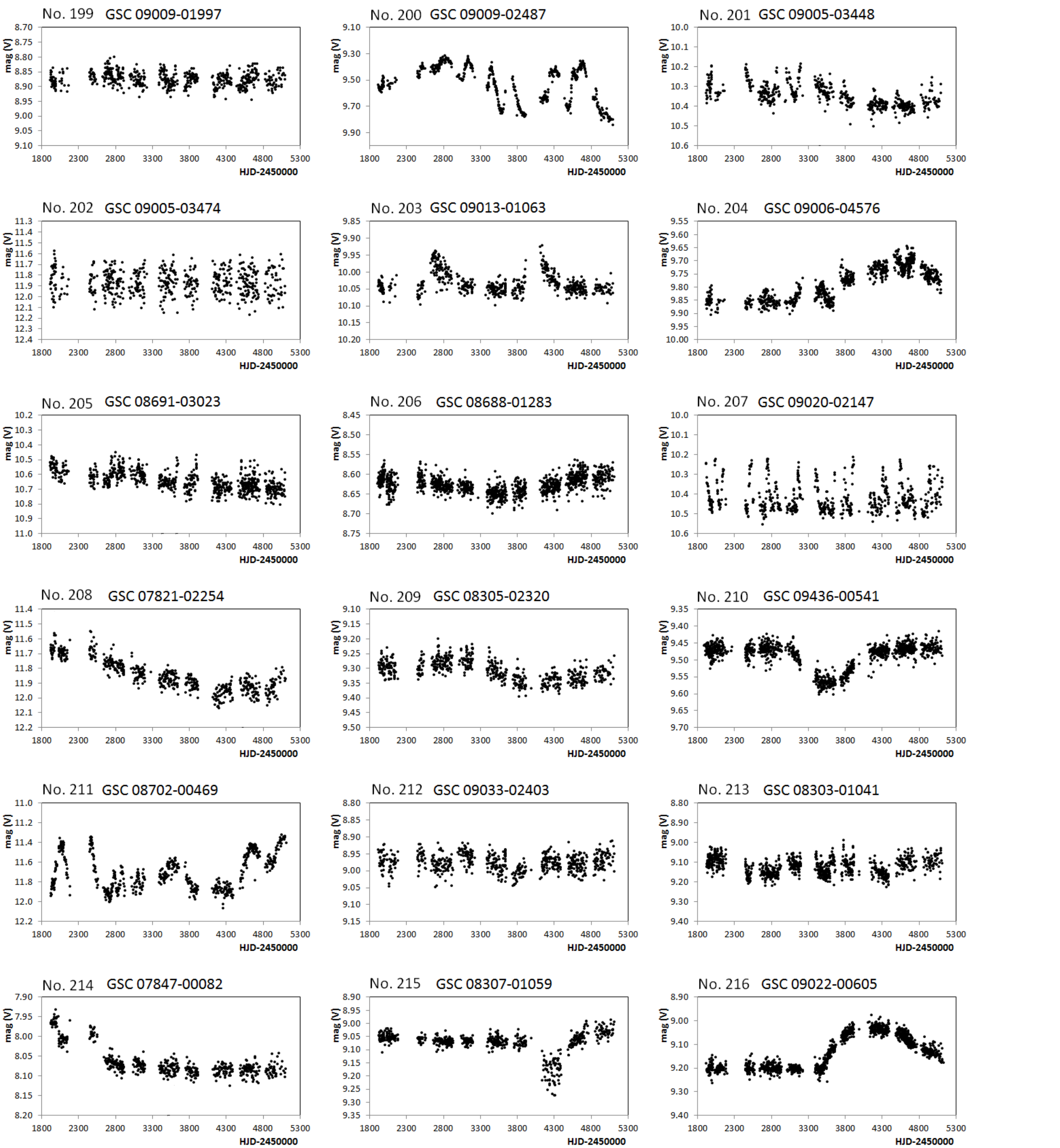

**Figure B1.** continued.

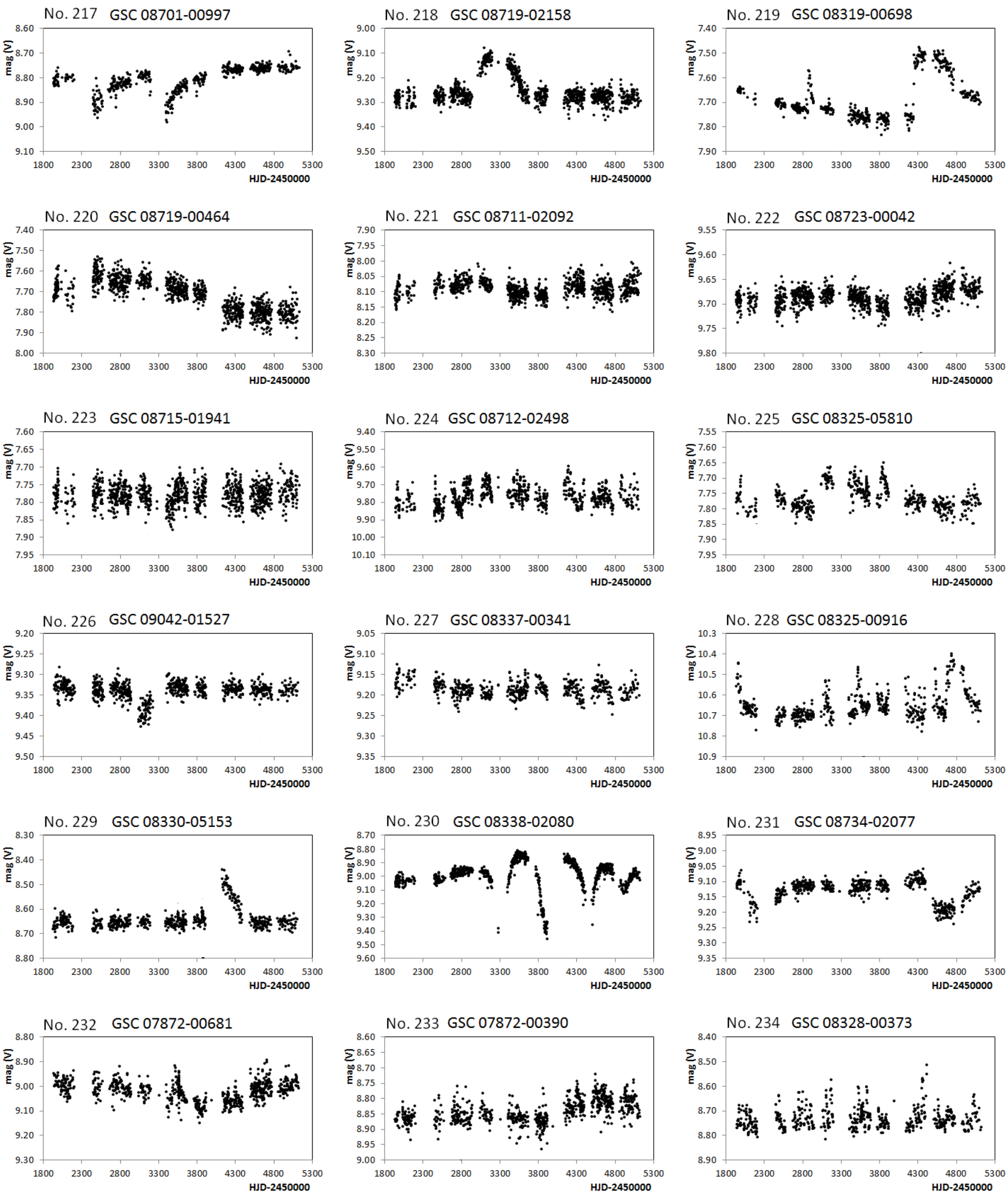

**Figure B1.** continued.

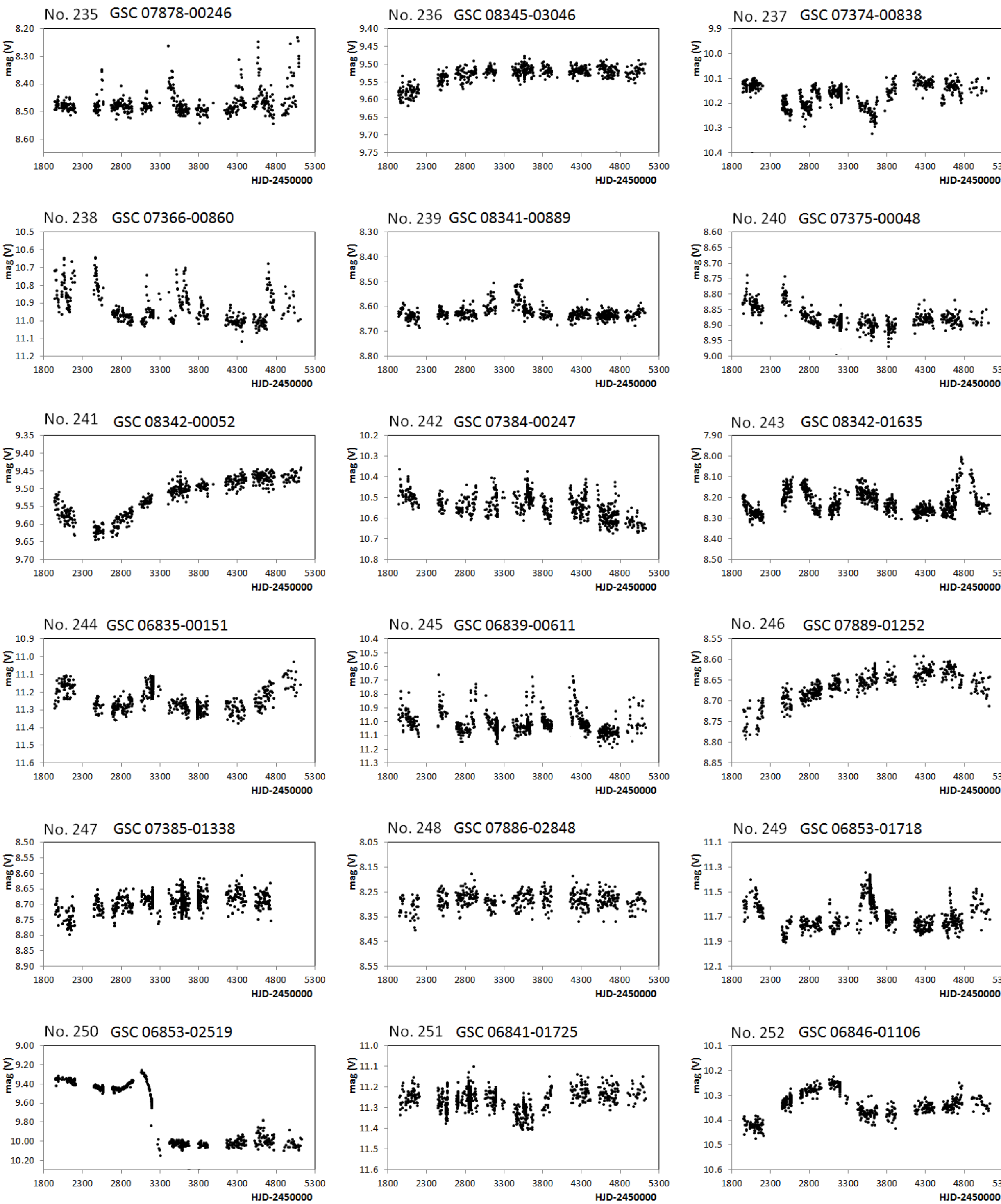

**Figure B1.** continued.

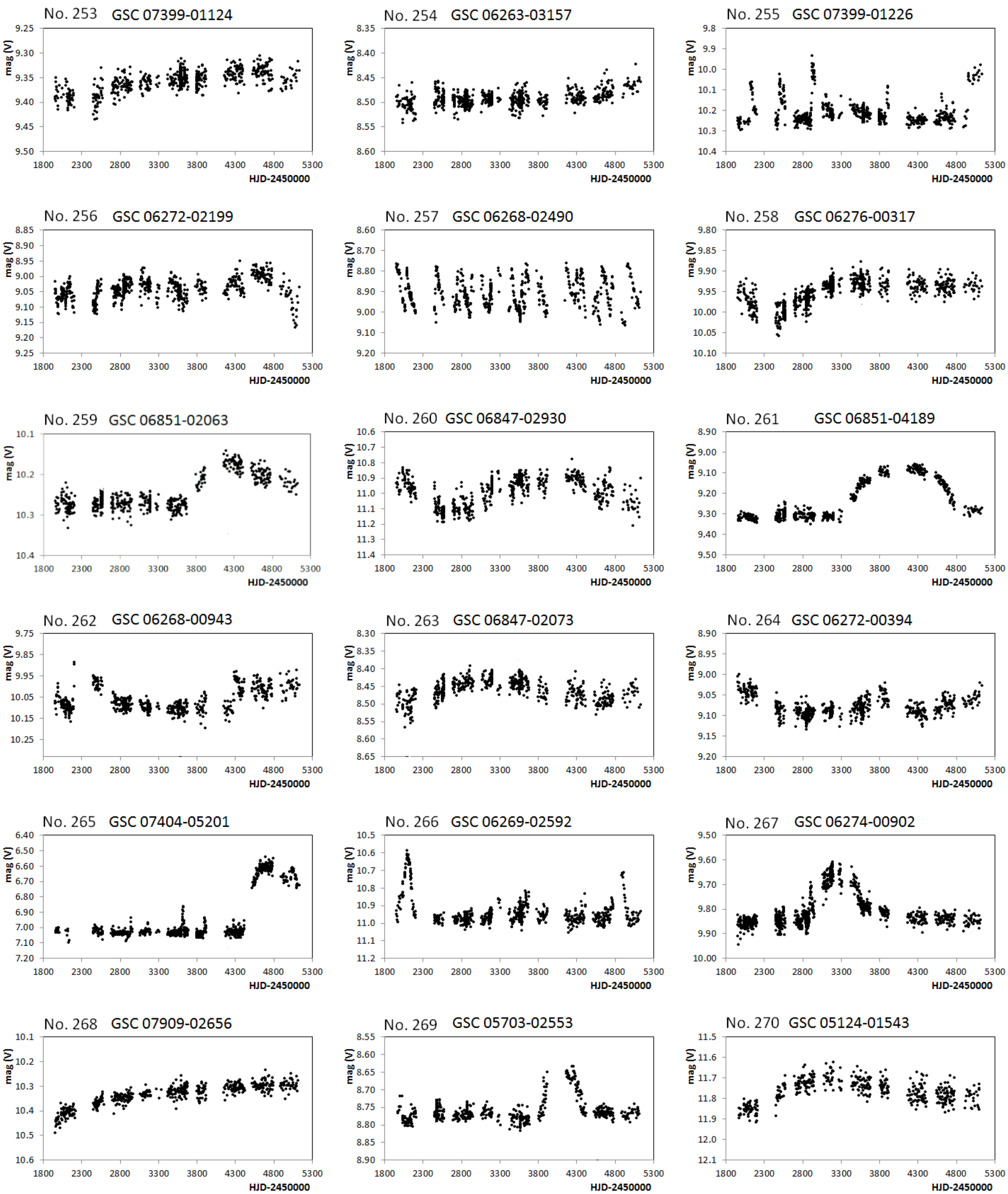

**Figure B1.** continued.

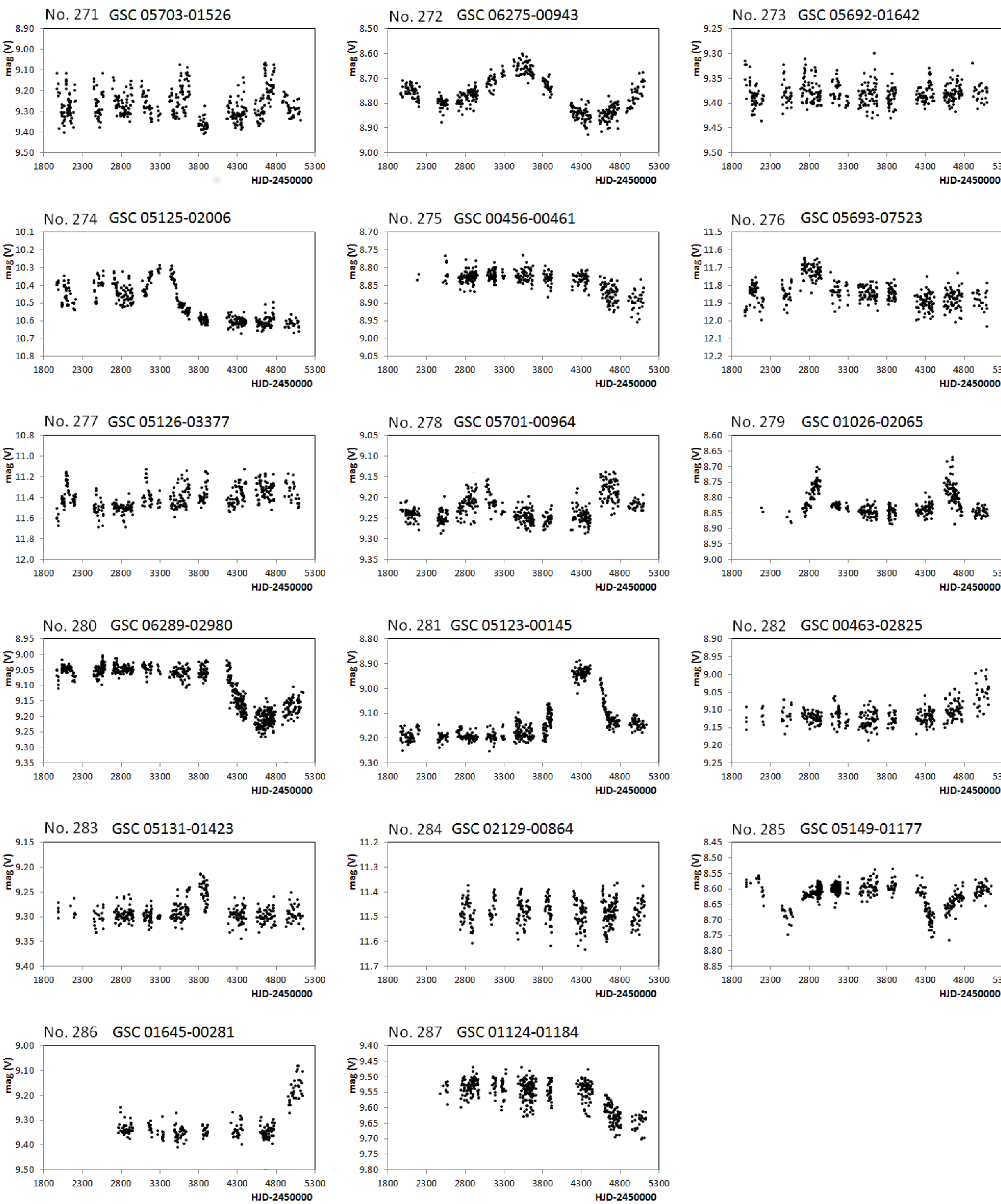

**Figure B2.** Phase plots of the stars showing periodic variability. For the construction of the plots, ASAS-3 data have been folded with the periods listed in Table A1. In order to bring out the periodic variability more clearly in objects that show complex photometric variations, some plots have been based on only part of the ASAS-3 observations, as indicated below the abscissae. All other plots have been based on the full available range of data. To facilitate identification, the internal running numbers are provided in the plots (upper left).

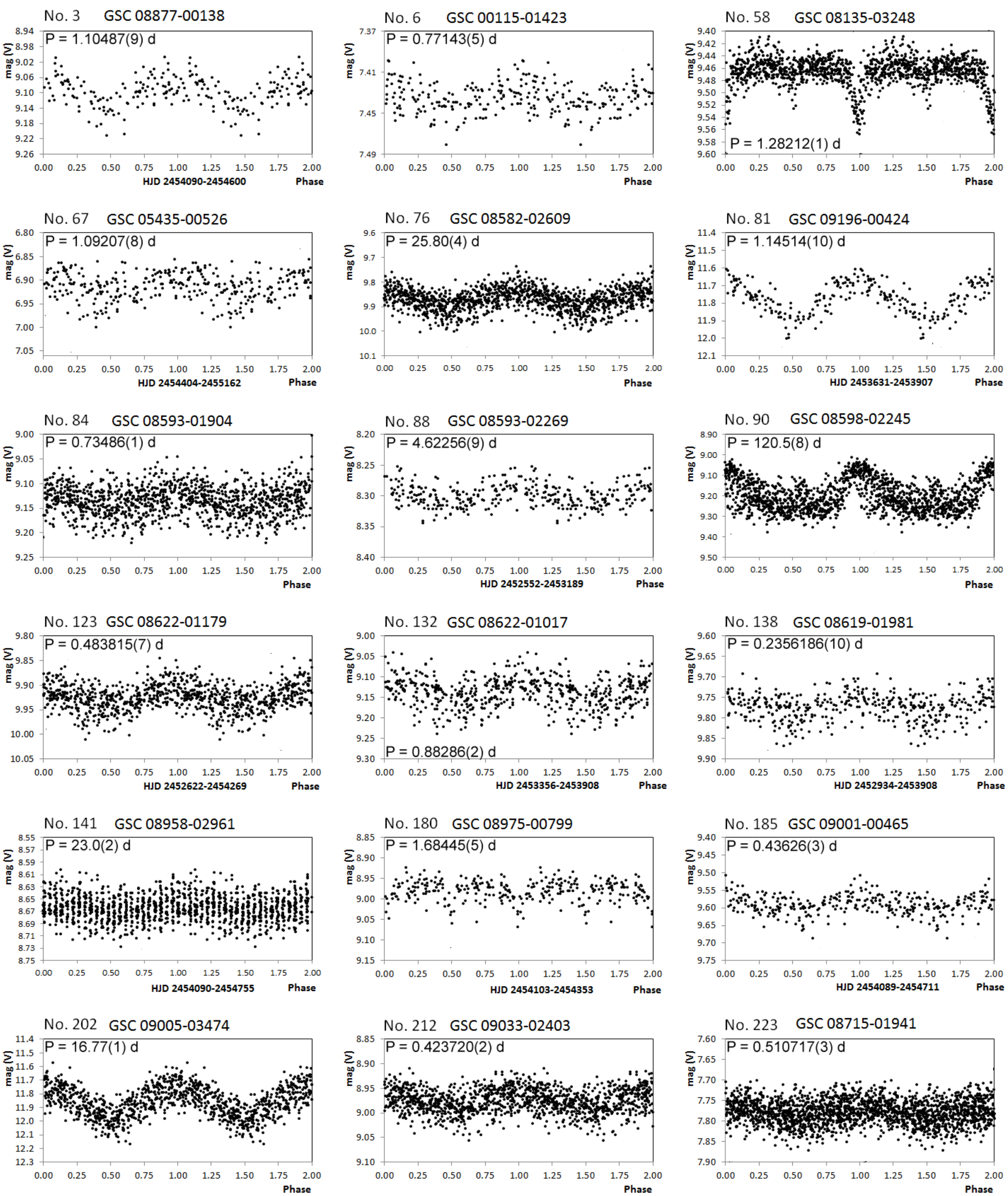

**Figure B2.** continued.

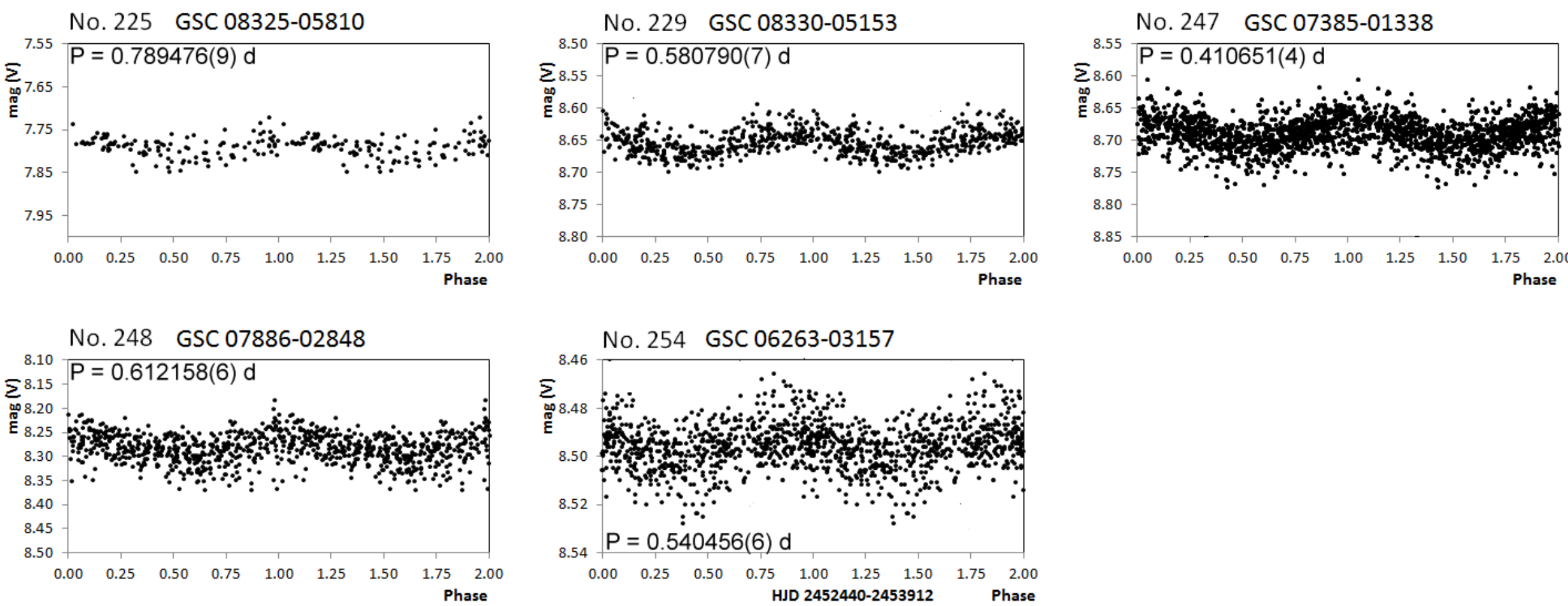

## APPENDIX C: NEWLY-ACQUIRED AND ARCHIVAL SPECTRA

**Figure C1.** The classification resolution spectra obtained at Mirranook Observatory and the available LAMOST spectra of our sample stars, illustrating the region containing the H$\alpha$ and H$\beta$ lines (4700–7100 Å). In order to save space, unit labeling has been omitted from the abscissae and ordinates, which denote, respectively, wavelength (Å) and normalized flux in arbitrary units. Objects have been sorted by increasing right ascension. To facilitate identification, the internal running numbers are provided in the plots.

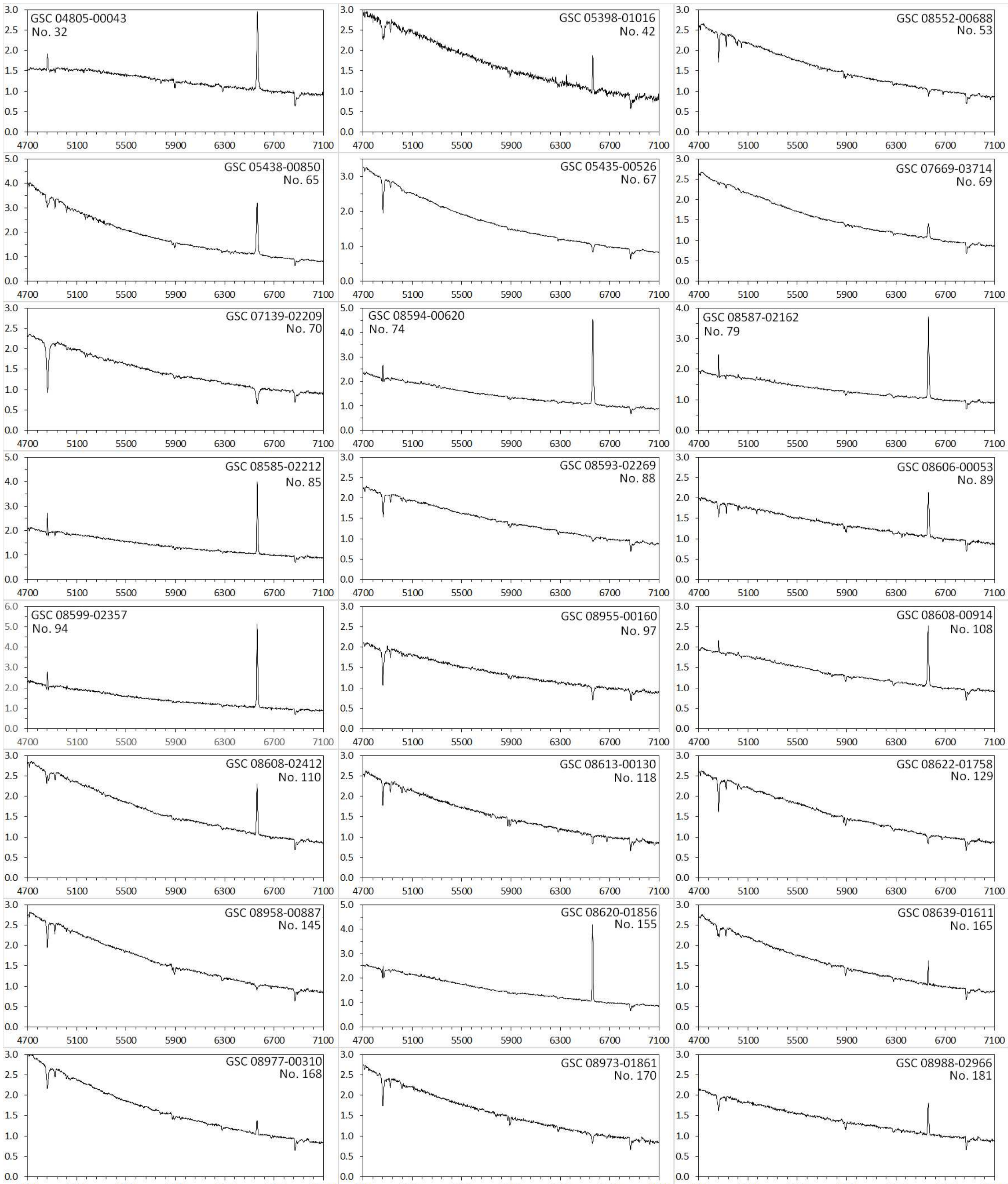

**Figure C1.** continued.

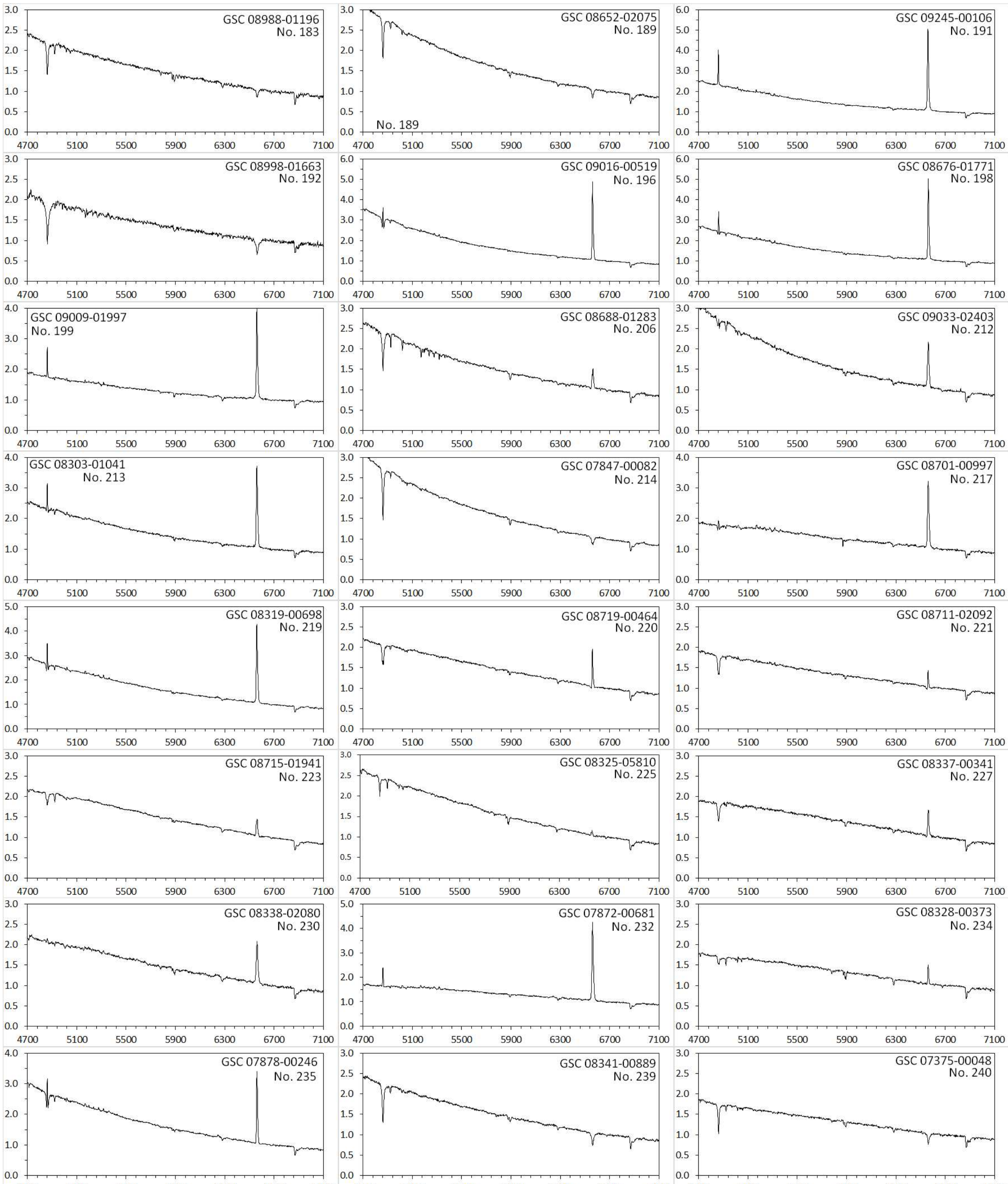

**Figure C1.** continued.

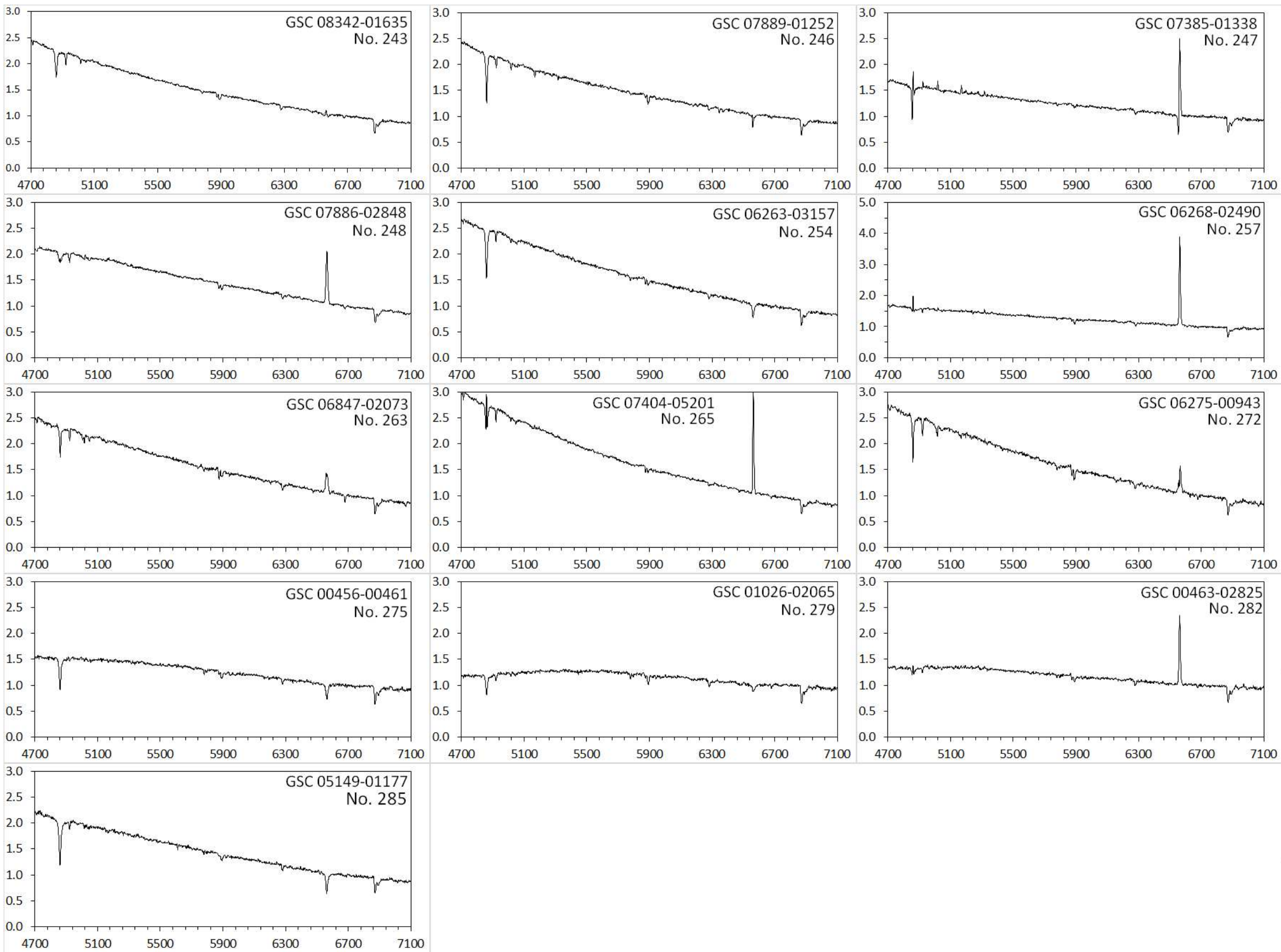

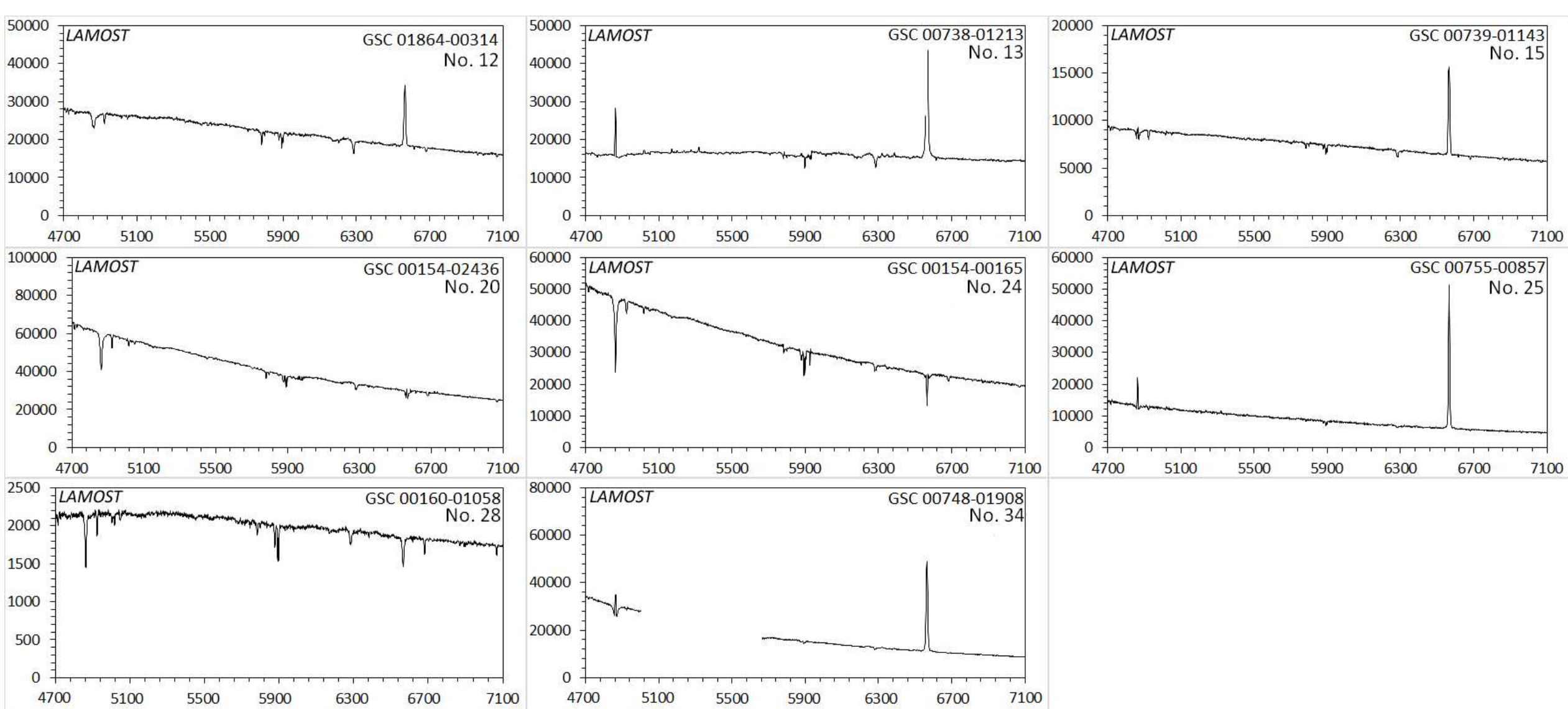

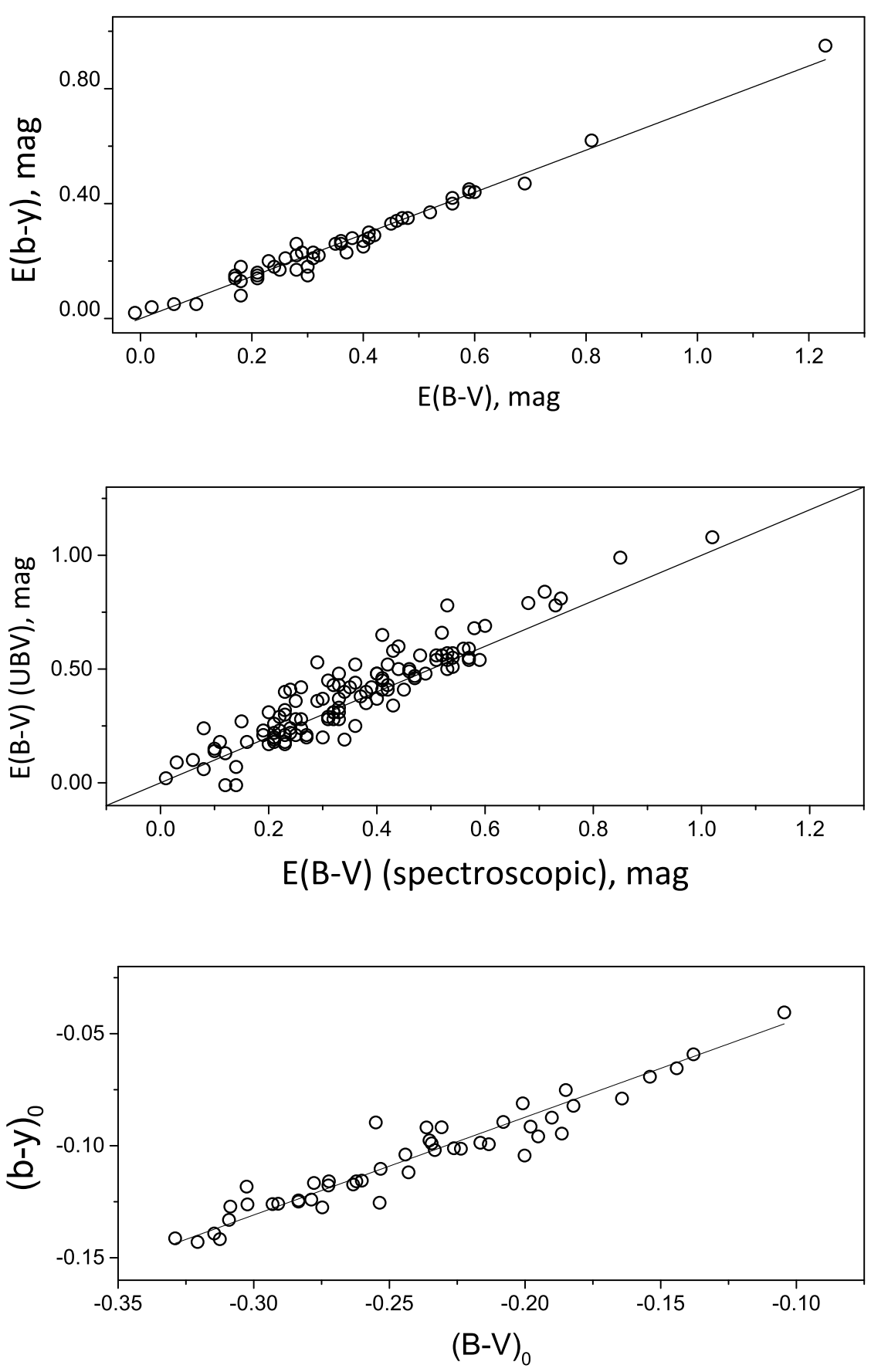

**Figure D1.** The top panel compares the *uvby*-based and *UBV*-based interstellar colour excesses, as obtained following Crawford (1978) and Crawford (1994), respectively. We derive $E(b-y) = 0.733 \pm 0.009\ E(B-V)$, which agrees well with the established coefficient of 0.74 between the two colour excesses. The middle panel illustrates the comparison between the colour excesses derived from *UBV* photometry and spectral classification; the agreement is satisfactory. The bottom panel provides a comparison of the dereddened indices $(B-V)_0$ and $(b-y)_0$. The correlation between these two quantities is very good (standard error ±0.004 mag). The solid lines in the upper and bottom panels represent the best fits to the data; the solid line in the middle panel is the unity line.

## APPENDIX D: INTRINSIC COLOURS AND INTERSTELLAR COLOUR EXCESSES FROM *UBV* AND *UVBY* PHOTOMETRIES

As Be stars enter a mass-losing phase, their colours are prone to change continuously. de Wit et al. (2006) have found that the outflowing material produces a bi-valued colour-magnitude relation in many Small Magellanic Cloud Be stars that leads to loop structures in colour-magnitude diagrams and can be ascribed to optical depth effects. Likewise, complex colour changes have been observed in Galactic Be stars, which – depending on the inclination angle – may become redder or bluer as their brightness increases (e.g. Hirata 1982; Percy & Bakos 2001; Keller et al. 2002). Marr et al. (2018) have investigated variations in linear polarization and *V* and *B* band colour-magnitudes for classical Be star disks and found that, depending on the employed model, the maximum changes in $(B-V)$ do not exceed 0.1 mag.

Be stars, therefore, are not particularly well suited to determine parameters like interstellar colour excess.[7] On the other hand, interstellar colour excesses obtained via *uvby* photometry for classical Be stars have been shown to be consistent with colour excesses based on the Barbier-Chalonge-Divan spectrophotometric system (Gkouvelis et al. 2016), and, given the large sample size and the availability of good-quality optical photometry for a significant part of our sample, we have chosen to investigate the possibilities of colour excess determination with our sample stars.

Of all 287 objects, 64 stars boast *uvby* photometry and 149 stars have complete *UBV* data in either the GCPD (Mermilliod et al. 1997) and/or Paunzen (2015). The *UBV*-based colour excess $E(B-V)$ was obtained following the procedure summarized by Crawford (1994) and also the more recent calibrations derived by Pecaut & Mamajek (2013). Both calibrations provide similar results. The *uvby*-based colour excess $E(b-y)$ was obtained via the calibrations of Crawford (1978). Spectral types are available for 72% of our sample stars. Using this information, we also obtained intrinsic $(B-V)_0$ colours utilizing the calibration of Deutschman et al. (1976). Since uncertainties and variations in spectral type and luminosity class are a general characteristic of Be stars (cf. Section 2.3), we made efforts to adopt the most reliable classification for each star by evaluating the quality of the classification's source. In general, the most recent classification was adopted, and classifications from spectroscopic surveys were favoured.

Figure D1 provides comparisons between the colour excesses as derived by the different methods applied here. The top panel shows a comparison between the *uvby*-based and *UBV*-based colour excesses, the latter obtained following Crawford (1994). From this, we derive $E(b-y) = 0.733 \pm 0.009\ E(B-V)$, which agrees well with the established coefficient of 0.74 between the two colour excesses (cf. e.g. Straižys 1992). This implies a good agreement between these two sets of photometry-based colour excesses, suggesting that the (constant) colour excess due to interstellar absorption is considerably larger than the mean colour changes induced by the variability of our sample stars. This is in line with results from the literature (e.g. de Wit et al. 2006). We also note that, even in the most variable stars of our sample, the mean *V* magnitude difference between two observations at any randomly-chosen epochs during the whole time span of observations amounts to only about 0.1 mag. We therefore expect corresponding changes in $(B-V)$ on the order of 0.025 mag, which is on the upper limit of the calculations of Marr et al. (2018). This is in agreement with the here reported scatter for the intrinsic colours.

The calibration by Pecaut & Mamajek (2013) also provides a good agreement with the *uvby* colour excesses, but a slightly smaller value of the established coefficient. Because of this we give preference to the $E(B-V)$ values based on Crawford (1994). The middle panel shows a comparison between the colour excesses derived from *UBV* photometry and spectral classification; the agreement is satisfactory. The majority of our sample stars have interstellar colour excesses $E(B-V)$ between 0.2 and 0.6 magnitudes.

Figure D1 bottom panel provides a comparison of the dereddened indices $(B-V)_0$ and $(b-y)_0$. Despite the scatter induced by the changing colours of Be stars, the correlation between these two quantities is strong (standard error ±0.004 mag), which makes us confident of the applicability of our approach. We therefore calculated a combined $(B-V)_0$ index based on an average of both indices, where available. To this end, we used the relation $(B-V)_0 = (b-y)_0/0.436$ to convert the $(b-y)_0$ values to $(B-V)_0$, which was obtained based on our comparisons. A comparison between this index and intrinsic $(B-V)_0$ colours derived from the spectral types is provided in Fig. D2. The resulting correlation is mediocre, which we attribute mostly to the lower precision of the spectrophotometric intrinsic colour determinations and the known inaccuracies in the spectral classification of Be stars (cf. Section 2.3).[8]

[7] In this section, the term colour excess always refers to interstellar colour excess.

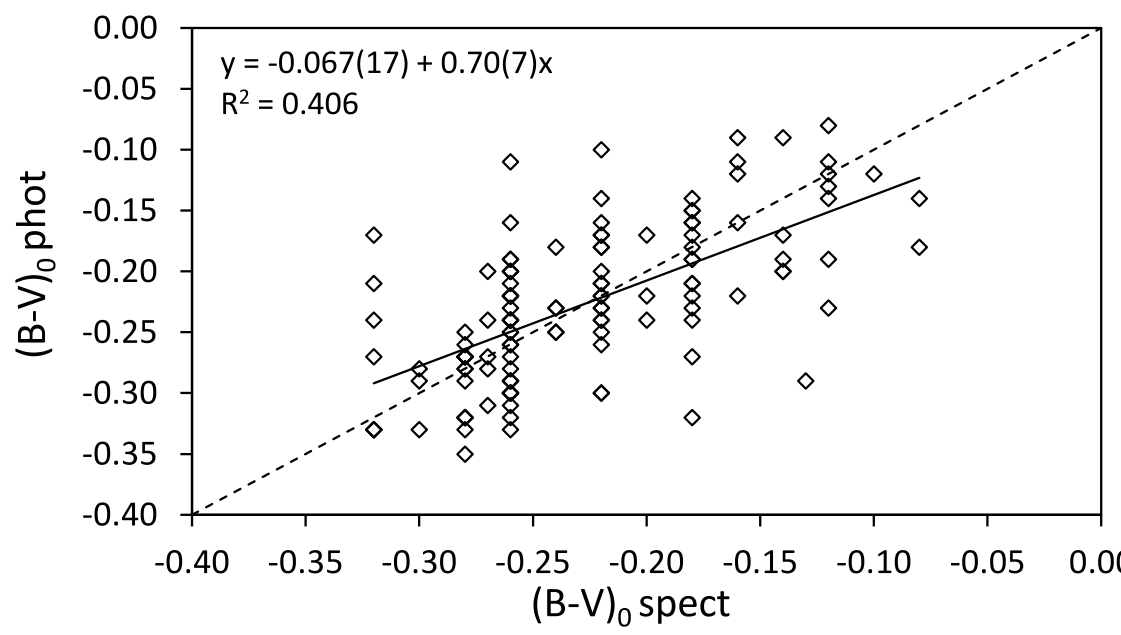

**Figure D2.** A comparison between the combined $(B-V)_0$ index as calculated in the present work and the intrinsic $(B-V)_0$ colours derived from the spectral types. The solid line represents the best fit to the data; the dashed line is the unity line.

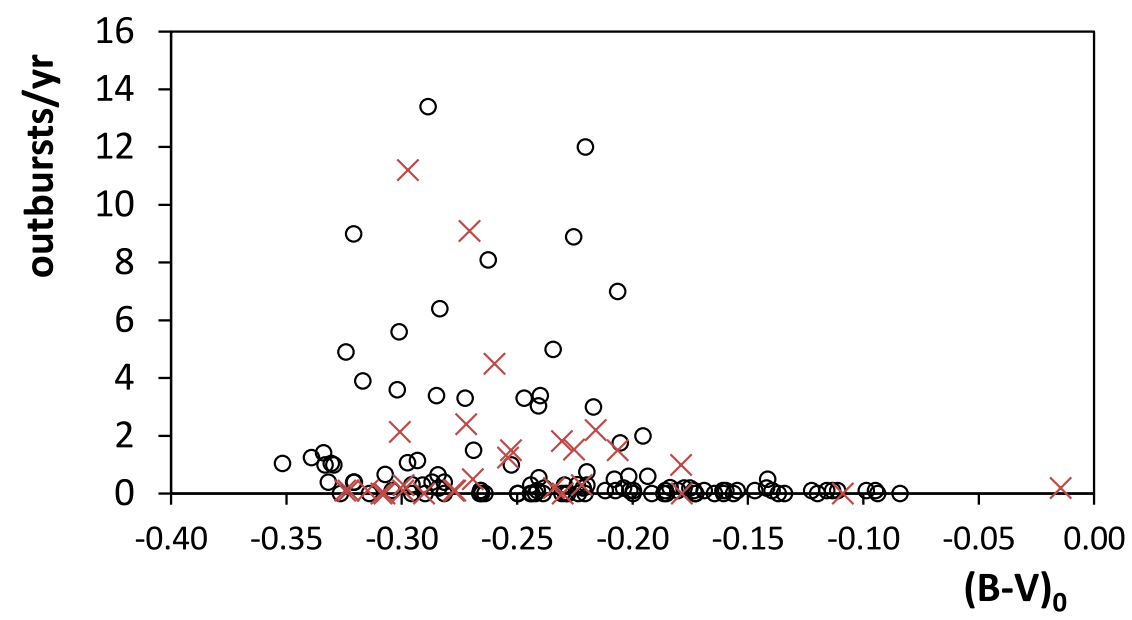

**Figure D3.** Outbursts per year versus average $(B-V)_0$ index, as derived from the available *UBV* and *uvby* data. The 24 stars that boast good photometry, and thus $(B-V)_0$ indices, but only general spectral types such as Be, OBe or em (and hence are not considered in Fig. 14) are indicated by crosses.

Table D1 lists the derived colour indices and colour excesses for our sample stars. We note that for three objects in our sample, the $(B-V)$ index from Kharchenko (2001) deviates from the $(B-V)$ index taken from the GCPD and/or APASS by more than 0.5 mag. These objects are GSC 08702-00469 (#211; difference ~1.0 mag), GSC 00153-00891 (#36; difference ~0.8 mag), and GSC 08623-02851 (#136; difference ~0.7 mag). GSC 00153-00891 (#36) is among the faintest objects in our sample ($V \sim 12.6$ mag); in this magnitude range, the Tycho-2 catalog indicates standard measurement errors of ~0.2 mag, which might (partly) explain the observed discrepant values. In the case of the other two objects, an investigation of the corresponding sky regions using the ALADIN visualization tool (Bonnarel et al. 2000) has revealed the presence of close neighbouring stars that will likely have affected the photometry. Therefore, in all likelihood, instrumental effects will be at the root of the observed discrepancies in colour index. We cannot totally exclude an intrinsic variability with time, though.

As there is evidence that spectral classifications of Be stars suffer from inaccuracies (cf. Section 2.3), we have employed the derived intrinsic colours to check our results concerning the percentage of Be stars exhibiting outbursts over the spectral subtype sequence (cf. Section 4). By doing so, we were able to include another 24 stars into the analysis that boast good photometry, and thus $(B-V)_0$ indices, but only general spectral types such as Be, OBe or em. The results are shown in Fig. D3 and are fully consistent with the results based on spectral type. Stars bluer than $(B-V)_0 \approx -0.18$ show more frequent outbursts. This corresponds to a spectral type of ~B4 (Ducati et al. 2001), which defines the red border of the here employed definition of early-type Be stars (Be stars with spectral types earlier than B4; cf. Section 2.3).

[8] In fact, some of the most outlying points in Fig. D2 relate to objective prism spectra taken more than 60 years ago.

This paper has been typeset from a TEX/LATEX file prepared by the author.

**Table D1.** Colour indices and colour excesses for our sample stars, sorted by increasing right ascension. The columns denote: (1) Internal identification number. Stars were numbered in order of increasing right ascension. (2) Identification from the Guide Star Catalog (GSC), version 1.2. (3) $(B-V)$ index, taken from Kharchenko (2001). (4) $(B-V)$ index, taken from the GCPD and/or APASS. When indices from both sources are available, average values have been given. (5) $(U-B)$ index, taken (with very few exceptions) from the GCPD. (6) Spectral type and luminosity class used as input for obtaining the spectroscopic $(B-V)_0$ and $E(B-V)$. (7) $(B-V)_0$ index, as derived from the spectral type. (8) Average $(B-V)_0$ index, as derived from the available $UBV$ and *uvby* data. (9) Colour excess $E(B-V)$, as derived from spectroscopy. (10) Average colour excess $E(B-V)$, as derived from the available $UBV$ and *uvby* data.

| (1) No | (2) GSC | (3) $(B-V)$ Kh01 | (4) $(B-V)$ GCPD APASS | (5) $(U-B)$ GCPD | (6) SpT&LC input | (7) $(B-V)_0$ spec. | (8) $(B-V)_0$ phot. avg. | (9) $E(B-V)$ spec. | (10) $E(B-V)$ phot. avg. |
|---|---|---|---|---|---|---|---|---|---|
| 1 | GSC 06464-00405 | -0.10 | -0.15 | -0.53 | B3V | -0.22 | -0.14 | 0.12 | -0.01 |
| 2 | GSC 01845-02192 | 0.17 | 0.20 | -0.51 | B3IV | -0.22 | -0.22 | 0.39 | 0.42 |
| 3 | GSC 08877-00138 | -0.17 | -0.18 | -0.71 | B5V | -0.18 | -0.21 | 0.01 | 0.03 |
| 4 | GSC 04755-00818 | 0.11 | 0.03 | – | B9V | -0.08 | – | 0.19 | – |
| 5 | GSC 09162-00751 | 0.20 | 0.09 | -0.67 | B2.5V | -0.24 | -0.25 | 0.43 | 0.34 |
| 6 | GSC 00115-01423 | -0.08 | -0.10 | -0.53 | B6V | -0.16 | -0.16 | 0.08 | 0.06 |
| 7 | GSC 01310-01587 | 0.08 | – | – | – | – | – | – | – |
| 8 | GSC 01311-01238 | 0.17 | 0.14 | -0.69 | B1V | -0.28 | -0.27 | 0.45 | 0.41 |
| 9 | GSC 06491-00717 | -0.09 | – | – | B8V | -0.12 | – | 0.02 | – |
| 10 | GSC 01868-01264 | -0.02 | 0.16 | – | – | – | – | – | – |
| 11 | GSC 00721-02056 | 0.15 | – | – | B2V | -0.26 | – | 0.40 | – |
| 12 | GSC 01864-00314 | 0.26 | 0.52 | -0.45 | – | – | -0.28 | – | 0.80 |
| 13 | GSC 00738-01213 | 0.57 | 0.65 | – | – | – | – | – | – |
| 14 | GSC 00742-01475 | 0.82 | 0.75 | – | – | – | – | – | – |
| 15 | GSC 00739-01143 | 0.27 | 0.30 | – | – | – | – | – | – |
| 16 | GSC 00739-01342 | 0.13 | – | – | B8V | -0.12 | – | 0.25 | – |
| 17 | GSC 01319-00734 | 0.37 | 0.55 | -0.57 | – | – | -0.32 | – | 0.87 |
| 18 | GSC 00743-02467 | 0.04 | 0.09 | -0.49 | B8V | -0.12 | -0.19 | 0.15 | 0.28 |
| 19 | GSC 00732-02105 | 0.14 | 0.13 | -0.75 | B1V | -0.28 | -0.28 | 0.42 | 0.41 |
| 20 | GSC 00154-02436 | 0.28 | 0.30 | -0.50 | B0V | -0.32 | -0.24 | 0.59 | 0.54 |
| 21 | GSC 00733-01509 | 0.06 | 0.13 | -0.69 | B1V | -0.28 | -0.27 | 0.34 | 0.38 |
| 22 | GSC 00733-01932 | 0.24 | 0.43 | – | B3V | -0.22 | – | 0.46 | – |
| 23 | GSC 00146-01543 | 0.17 | 0.27 | – | B5V | -0.18 | -0.18 | 0.35 | 0.35 |
| 24 | GSC 00154-00165 | 0.12 | 0.10 | – | B8V | -0.12 | – | 0.24 | – |
| 25 | GSC 00755-00857 | -0.06 | 0.14 | – | – | – | – | – | – |
| 26 | GSC 00152-00780 | 0.35 | 0.35 | – | – | – | – | – | – |
| 27 | GSC 00148-02601 | 0.17 | 0.20 | -0.18 | – | – | -0.11 | – | 0.31 |
| 28 | GSC 00160-01058 | 0.03 | 0.23 | -0.72 | B2V | -0.26 | -0.30 | 0.29 | 0.53 |
| 29 | GSC 05387-01121 | -0.12 | -0.13 | -0.78 | B5V | -0.18 | -0.22 | 0.06 | 0.08 |
| 30 | GSC 04801-00017 | 0.41 | 0.61 | – | – | – | – | – | – |
| 31 | GSC 05383-00187 | 0.00 | – | – | B8V | -0.12 | -0.13 | 0.12 | 0.07 |
| 32 | GSC 04805-00043 | 0.24 | 0.24 | -0.70 | B0.5V | -0.30 | -0.29 | 0.53 | 0.53 |
| 33 | GSC 05388-01118 | 0.52 | 0.51 | – | – | – | – | – | – |
| 34 | GSC 00748-01908 | -0.05 | 0.01 | – | – | – | – | – | – |
| 35 | GSC 04809-00545 | 0.27 | 0.18 | -0.75 | B3V | -0.22 | -0.30 | 0.49 | 0.48 |
| 36 | GSC 00153-00891 | -0.30 | 0.51 | – | – | – | – | – | – |
| 37 | GSC 04801-01915 | 0.02 | 0.15 | – | B5V | -0.18 | – | 0.20 | – |
| 38 | GSC 04826-00257 | 0.08 | 0.22 | – | – | – | – | – | – |
| 39 | GSC 04826-01079 | 0.18 | 0.25 | – | – | – | – | – | – |
| 40 | GSC 05393-02168 | 0.80 | 0.98 | -0.05 | – | – | -0.27 | – | 1.26 |
| 41 | GSC 05968-03899 | 0.01 | 0.02 | – | B7/8V | -0.13 | – | 0.14 | – |
| 42 | GSC 05398-01016 | 0.07 | 0.09 | -0.57 | B2V | -0.26 | -0.22 | 0.33 | 0.29 |
| 43 | GSC 05973-00249 | 0.10 | 0.22 | -0.74 | B2V | -0.26 | -0.30 | 0.36 | 0.52 |
| 44 | GSC 05399-00962 | 0.01 | 0.01 | -0.33 | B8V | -0.12 | -0.11 | 0.12 | 0.12 |
| 45 | GSC 07634-01561 | 0.12 | – | – | B8V | -0.12 | – | 0.24 | – |
| 46 | GSC 04820-02947 | -0.09 | 0.00 | – | B7V | -0.14 | – | 0.05 | – |
| 47 | GSC 05978-01855 | 0.01 | 0.04 | -0.50 | B2.5V | -0.24 | -0.18 | 0.24 | 0.22 |
| 48 | GSC 05978-00030 | | 0.63 | – | – | – | – | – | – |
| 49 | GSC 05983-00995 | 0.08 | 0.16 | – | B5V | -0.18 | – | 0.26 | – |
| 50 | GSC 07109-00828 | 0.18 | 0.20 | -0.72 | – | – | -0.29 | – | 0.49 |
| 51 | GSC 05405-00431 | -0.02 | – | – | B3V | -0.22 | -0.25 | 0.20 | 0.19 |
| 52 | GSC 05988-00265 | 0.32 | 0.41 | -0.62 | – | – | -0.31 | – | 0.72 |
| 53 | GSC 08552-00688 | -0.07 | -0.05 | -0.81 | B2V | -0.26 | -0.26 | 0.19 | 0.21 |
| 54 | GSC 06552-00580 | | 0.52 | – | – | – | – | – | – |
| 55 | GSC 06552-01189 | 0.31 | 0.36 | -0.66 | – | – | -0.31 | – | 0.67 |
| 56 | GSC 07106-02534 | 0.35 | 0.53 | -0.42 | – | – | -0.27 | – | 0.80 |

**Table D1.** continued.

| (1) No | (2) GSC | (3) $(B-V)$ Kh01 | (4) $(B-V)$ GCPD APASS | (5) $(U-B)$ GCPD | (6) SpT&LC input | (7) $(B-V)_0$ spec. | (8) $(B-V)_0$ phot. avg. | (9) $E(B-V)$ spec. | (10) $E(B-V)$ phot. avg. |
|---|---|---|---|---|---|---|---|---|---|
| 57 | GSC 07123-00519 | 0.57 | 1.03 | -0.15 | – | – | -0.30 | – | 1.33 |
| 58 | GSC 08135-03248 | -0.01 | – | – | B2V | -0.26 | -0.22 | 0.24 | 0.22 |
| 59 | GSC 06565-01179 | 0.42 | 0.45 | -0.67 | B0V | -0.32 | -0.33 | 0.73 | 0.78 |
| 60 | GSC 07124-01160 | | 1.00 | -0.29 | – | – | -0.34 | – | 1.34 |
| 61 | GSC 07120-02077 | 0.40 | 0.44 | -0.73 | B1V | -0.28 | -0.35 | 0.68 | 0.79 |
| 62 | GSC 06562-00688 | 0.03 | 0.11 | – | A3V | 0.08 | – | -0.05 | – |
| 63 | GSC 05421-00568 | 0.06 | -0.05 | – | – | – | – | – | – |
| 64 | GSC 07659-01614 | | 0.32 | – | – | – | – | – | – |
| 65 | GSC 05438-00850 | -0.09 | -0.11 | -0.39 | B3V | -0.22 | -0.10 | 0.14 | 0.01 |
| 66 | GSC 07125-02097 | 0.21 | 0.45 | -0.66 | B0V | -0.32 | -0.33 | 0.53 | 0.78 |
| 67 | GSC 05435-00526 | -0.13 | -0.15 | – | B7IV | -0.14 | – | 0.01 | – |
| 68 | GSC 07669-01154 | | 0.73 | – | – | – | – | – | – |
| 69 | GSC 07669-03714 | -0.17 | -0.07 | -0.97 | B2V | -0.26 | -0.31 | 0.08 | 0.24 |
| 70 | GSC 07139-02209 | 0.31 | – | – | A0IV | -0.02 | – | 0.33 | – |
| 71 | GSC 08151-01868 | 0.20 | 0.23 | -0.60 | B2V | -0.26 | -0.26 | 0.46 | 0.49 |
| 72 | GSC 08155-02404 | 0.17 | 0.27 | -0.74 | B1.5V | -0.27 | -0.31 | 0.43 | 0.58 |
| 73 | GSC 08164-01530 | 0.08 | – | – | B1V | -0.28 | – | 0.36 | – |
| 74 | GSC 08594-00620 | -0.02 | -0.05 | -0.68 | B3V | -0.22 | -0.22 | 0.20 | 0.18 |
| 75 | GSC 08165-00324 | 0.26 | 0.32 | -0.49 | B2V | -0.26 | -0.24 | 0.51 | 0.56 |
| 76 | GSC 08582-02609 | 0.71 | – | – | – | – | – | – | – |
| 77 | GSC 08591-00039 | 0.02 | -0.07 | -0.84 | B1IV | -0.28 | -0.27 | 0.30 | 0.20 |
| 78 | GSC 07686-01898 | 0.15 | 0.19 | – | B7V | -0.14 | – | 0.29 | – |
| 79 | GSC 08587-02162 | 0.09 | 0.13 | -0.75 | B2IV | -0.26 | -0.29 | 0.35 | 0.40 |
| 80 | GSC 08174-00235 | 0.26 | 0.29 | -0.41 | B5V | -0.18 | -0.21 | 0.44 | 0.50 |
| 81 | GSC 09196-00424 | | 0.89 | – | – | – | – | – | – |
| 82 | GSC 08588-02569 | 0.16 | 0.20 | -0.57 | B4V | -0.20 | -0.24 | 0.36 | 0.44 |
| 83 | GSC 08167-01520 | 0.20 | 0.25 | -0.75 | – | – | -0.32 | – | 0.57 |
| 84 | GSC 08593-01904 | 0.07 | 0.09 | -0.66 | B2V | -0.26 | -0.24 | 0.33 | 0.33 |
| 85 | GSC 08585-02212 | 0.11 | 0.08 | -0.76 | B1.5V | -0.27 | -0.28 | 0.38 | 0.35 |
| 86 | GSC 08593-00336 | 0.01 | -0.01 | – | B8/9V | -0.10 | – | 0.11 | – |
| 87 | GSC 07179-02573 | 0.00 | – | – | B8V | -0.12 | – | 0.11 | – |
| 88 | GSC 08593-02269 | -0.05 | -0.01 | -0.73 | B2.5V | -0.24 | -0.25 | 0.19 | 0.25 |
| 89 | GSC 08606-00053 | 0.06 | – | | B4V | -0.20 | – | 0.26 | – |
| 90 | GSC 08598-02245 | 0.11 | 0.19 | -0.58 | B3IV | -0.22 | -0.24 | 0.33 | 0.43 |
| 91 | GSC 08606-02020 | 0.64 | 0.66 | -0.41 | – | – | -0.30 | – | 0.96 |
| 92 | GSC 08610-03102 | -0.05 | – | – | B5V | -0.18 | – | 0.14 | – |
| 93 | GSC 08603-00164 | 0.33 | 0.25 | -0.63 | – | – | -0.27 | – | 0.52 |
| 94 | GSC 08599-02357 | 0.07 | 0.12 | -0.64 | B5V | -0.18 | -0.24 | 0.25 | 0.36 |
| 95 | GSC 08611-01377 | -0.09 | -0.09 | – | B2V | -0.26 | -0.21 | 0.17 | 0.09 |
| 96 | GSC 08603-00677 | 0.59 | 0.61 | – | B7V | -0.14 | – | 0.73 | – |
| 97 | GSC 08955-00160 | 0.04 | – | – | B4V | -0.20 | – | 0.24 | – |
| 98 | GSC 08943-00584 | -0.02 | 0.02 | – | B5V | -0.18 | – | 0.16 | – |
| 99 | GSC 08607-01004 | | 1.51 | – | – | – | – | – | – |
| 100 | GSC 08607-00285 | 0.43 | 0.52 | -0.58 | B1V | -0.28 | -0.32 | 0.71 | 0.84 |
| 101 | GSC 08604-00356 | 0.15 | – | – | B6.5V | -0.15 | – | 0.30 | – |
| 102 | GSC 08947-00809 | 0.14 | 0.09 | -0.90 | – | – | -0.32 | – | 0.41 |
| 103 | GSC 08943-02244 | -0.05 | 0.08 | -0.93 | B0V | -0.32 | -0.33 | 0.26 | 0.41 |
| 104 | GSC 08943-00620 | 0.16 | 0.32 | -0.75 | B2V | -0.26 | -0.33 | 0.41 | 0.65 |
| 105 | GSC 08612-00975 | | 0.64 | – | – | – | – | – | – |
| 106 | GSC 08608-00434 | -0.09 | – | – | B4V | -0.20 | – | 0.12 | – |
| 107 | GSC 08608-00807 | 0.05 | 0.04 | – | B8V | -0.12 | – | 0.16 | – |
| 108 | GSC 08608-00914 | 0.16 | 0.27 | -0.76 | B1V | -0.28 | -0.32 | 0.44 | 0.59 |
| 109 | GSC 08612-00331 | 0.14 | 0.09 | – | B2V | -0.26 | – | 0.40 | – |
| 110 | GSC 08608-02412 | -0.12 | -0.05 | -0.73 | B3V | -0.22 | -0.23 | 0.11 | 0.15 |
| 111 | GSC 08612-00380 | 0.33 | – | – | B3V | -0.22 | – | 0.55 | – |
| 112 | GSC 08608-00239 | 0.31 | 0.37 | – | B0V | -0.32 | – | 0.63 | – |

**Table D1.** continued.

| (1)<br>No | (2)<br>GSC | (3)<br>$(B-V)$<br>Kh01 | (4)<br>$(B-V)$<br>GCPD<br>APASS | (5)<br>$(U-B)$<br>GCPD | (6)<br>SpT&LC<br>input | (7)<br>$(B-V)_0$<br>spec. | (8)<br>$(B-V)_0$<br>phot. avg. | (9)<br>$E(B-V)$<br>spec. | (10)<br>$E(B-V)$<br>phot. avg. |
|---|---|---|---|---|---|---|---|---|---|
| 113 | GSC 09397-00615 | 0.20 | – | – | B5V | -0.18 | – | 0.38 | – |
| 114 | GSC 08609-01016 | 0.29 | 0.28 | -0.59 | B1V | -0.28 | -0.27 | 0.57 | 0.55 |
| 115 | GSC 08605-01985 | 0.04 | 0.05 | -0.48 | B5V | -0.18 | -0.17 | 0.22 | 0.22 |
| 116 | GSC 08956-01223 | 0.11 | 0.10 | -0.80 | B7/8V | -0.13 | -0.29 | 0.23 | 0.37 |
| 117 | GSC 08613-01412 | -0.03 | -0.14 | – | B0V | -0.32 | – | 0.29 | – |
| 118 | GSC 08613-00130 | 0.08 | – | – | B5V | -0.18 | – | 0.26 | – |
| 119 | GSC 08613-01744 | | 0.70 | – | – | – | – | – | – |
| 120 | GSC 08613-00147 | 0.57 | 0.48 | -0.55 | – | – | -0.30 | – | 0.78 |
| 121 | GSC 08957-02248 | 0.01 | 0.21 | – | B4V | -0.20 | – | 0.21 | – |
| 122 | GSC 08609-00597 | 0.02 | 0.08 | – | B3V | -0.22 | – | 0.24 | – |
| 123 | GSC 08622-01179 | 0.19 | 0.28 | -0.76 | – | – | -0.32 | – | 0.60 |
| 124 | GSC 08622-00429 | 0.26 | 0.34 | -0.71 | B2V | -0.26 | -0.32 | 0.52 | 0.66 |
| 125 | GSC 08626-00271 | -0.09 | 0.06 | -0.70 | B1V | -0.28 | -0.25 | 0.20 | 0.31 |
| 126 | GSC 08957-03483 | 0.09 | 0.10 | – | – | – | – | – | – |
| 127 | GSC 08961-01212 | -0.03 | 0.10 | – | B2V | -0.26 | – | 0.22 | – |
| 128 | GSC 08618-01375 | 0.09 | 0.11 | – | B8/9V | -0.10 | – | 0.19 | – |
| 129 | GSC 08622-01758 | 0.08 | – | – | B3V | -0.22 | – | 0.30 | – |
| 130 | GSC 08622-01002 | 0.17 | 0.29 | – | A2V | 0.05 | – | 0.12 | – |
| 131 | GSC 08618-01665 | 0.01 | 0.05 | -0.90 | B1V | -0.28 | -0.32 | 0.29 | 0.36 |
| 132 | GSC 08622-01017 | 0.12 | 0.13 | -0.66 | B2V | -0.26 | -0.25 | 0.37 | 0.38 |
| 133 | GSC 08965-01363 | 0.24 | 0.26 | -0.54 | B3IV | -0.22 | -0.24 | 0.46 | 0.50 |
| 134 | GSC 08969-00788 | 0.13 | 0.13 | – | – | – | – | – | – |
| 135 | GSC 08958-02421 | -0.01 | – | – | B4V | -0.20 | – | 0.19 | 0.16 |
| 136 | GSC 08623-02851 | -0.19 | 0.48 | – | – | – | – | – | – |
| 137 | GSC 08958-02242 | -0.04 | 0.12 | – | – | – | – | – | – |
| 138 | GSC 08619-01981 | -0.04 | -0.01 | – | – | – | – | – | – |
| 139 | GSC 08627-00455 | 0.22 | – | – | B5V | -0.18 | – | 0.40 | – |
| 140 | GSC 08627-01249 | 0.15 | 0.23 | -0.48 | B2V | -0.26 | -0.22 | 0.41 | 0.45 |
| 141 | GSC 08958-02961 | -0.02 | – | – | B5V | -0.18 | -0.27 | 0.16 | 0.32 |
| 142 | GSC 08627-02220 | 0.17 | 0.19 | -0.80 | – | – | -0.31 | – | 0.50 |
| 143 | GSC 08958-03515 | 0.10 | 0.16 | – | – | – | – | – | – |
| 144 | GSC 08627-02146 | 0.29 | 0.15 | – | A1V | 0.01 | – | 0.28 | – |
| 145 | GSC 08958-00887 | 0.08 | 0.07 | -0.85 | B2V | -0.26 | -0.29 | 0.33 | 0.34 |
| 146 | GSC 08958-01376 | -0.01 | 0.00 | -0.78 | B3V | -0.22 | -0.26 | 0.21 | 0.27 |
| 147 | GSC 08958-03463 | 0.05 | 0.30 | – | B5V | -0.18 | – | 0.23 | – |
| 148 | GSC 08958-03384 | 0.28 | 0.23 | – | B8V | -0.12 | – | 0.40 | – |
| 149 | GSC 08959-00482 | 0.21 | 0.16 | -0.79 | B2V | -0.26 | -0.30 | 0.47 | 0.46 |
| 150 | GSC 08967-00393 | 0.01 | 0.03 | -0.73 | B2V | -0.26 | -0.25 | 0.26 | 0.28 |
| 151 | GSC 08959-00488 | 0.24 | 0.26 | -0.47 | B5V | -0.16 | -0.22 | 0.40 | 0.48 |
| 152 | GSC 08628-00661 | 0.00 | – | – | B5V | -0.18 | – | 0.18 | – |
| 153 | GSC 08959-00863 | 0.03 | 0.12 | -0.64 | B1.5V | -0.27 | -0.24 | 0.30 | 0.36 |
| 154 | GSC 08959-00846 | 0.04 | 0.15 | -0.73 | B1V | -0.28 | -0.28 | 0.32 | 0.43 |
| 155 | GSC 08620-01856 | -0.05 | -0.05 | -0.63 | B7V | -0.14 | -0.20 | 0.10 | 0.15 |
| 156 | GSC 08959-02476 | 0.20 | 0.33 | -0.39 | O9.5V | -0.32 | -0.21 | 0.51 | 0.54 |
| 157 | GSC 08625-00369 | -0.03 | – | – | B7V | -0.14 | – | 0.11 | – |
| 158 | GSC 08980-01582 | 0.87 | 0.70 | -0.27 | – | – | -0.26 | – | 0.96 |
| 159 | GSC 08972-00064 | 0.02 | 0.09 | – | – | – | -0.23 | – | 0.27 |
| 160 | GSC 08972-00932 | -0.01 | – | – | B9V | -0.08 | -0.18 | 0.08 | 0.16 |
| 161 | GSC 08642-01087 | 0.12 | 0.24 | – | – | – | – | – | – |
| 162 | GSC 08973-00795 | -0.01 | 0.06 | -0.59 | – | – | -0.21 | – | 0.27 |
| 163 | GSC 08973-00729 | 0.01 | 0.04 | -0.73 | – | – | -0.23 | – | 0.25 |
| 164 | GSC 08985-01836 | 0.15 | 0.22 | – | – | – | – | – | – |
| 165 | GSC 08639-01611 | -0.03 | -0.01 | – | B2V | -0.26 | – | 0.23 | – |
| 166 | GSC 08643-01679 | 0.10 | 0.08 | -0.64 | B3V | -0.22 | -0.23 | 0.32 | 0.31 |
| 167 | GSC 08973-01406 | 0.06 | 0.07 | -0.62 | – | – | -0.23 | – | 0.30 |
| 168 | GSC 08977-00310 | -0.03 | -0.02 | -0.58 | B2V | -0.26 | -0.20 | 0.23 | 0.19 |

**Table D1.** continued.

| (1) No | (2) GSC | (3) $(B-V)$ Kh01 | (4) $(B-V)$ GCPD APASS | (5) $(U-B)$ GCPD | (6) SpT&LC input | (7) $(B-V)_0$ spec. | (8) $(B-V)_0$ phot. avg. | (9) $E(B-V)$ spec. | (10) $E(B-V)$ phot. avg. |
|---|---|---|---|---|---|---|---|---|---|
| 169 | GSC 09234-02316 | 0.08 | 0.18 | -0.77 | B2V | -0.26 | -0.30 | 0.33 | 0.48 |
| 170 | GSC 08973-01861 | 0.07 | 0.10 | -0.47 | B5V | -0.18 | -0.19 | 0.25 | 0.29 |
| 171 | GSC 08977-00421 | 0.12 | 0.18 | – | B8V | -0.12 | – | 0.23 | – |
| 172 | GSC 08978-01510 | 0.08 | 0.07 | -0.50 | – | – | -0.18 | – | 0.24 |
| 173 | GSC 08974-01031 | 0.03 | 0.06 | -0.66 | B4V | -0.20 | -0.22 | 0.23 | 0.27 |
| 174 | GSC 08974-00327 | 0.12 | 0.15 | – | B6V | -0.16 | – | 0.28 | – |
| 175 | GSC 08982-00852 | 0.13 | 0.16 | – | – | – | – | – | – |
| 176 | GSC 08641-01826 | 0.00 | 0.02 | -0.35 | B8/9IV | -0.10 | -0.12 | 0.10 | 0.14 |
| 177 | GSC 08974-00002 | 0.26 | 0.29 | -0.39 | B7V | -0.14 | -0.20 | 0.40 | 0.49 |
| 178 | GSC 08979-00623 | 0.10 | – | – | B7V | -0.14 | -0.19 | 0.24 | 0.26 |
| 179 | GSC 08975-03998 | 0.02 | – | – | B8V | -0.12 | – | 0.13 | – |
| 180 | GSC 08975-00799 | 0.19 | – | – | B3V | -0.22 | -0.17 | 0.41 | 0.42 |
| 181 | GSC 08988-02966 | 0.12 | – | – | B8V | -0.12 | -0.14 | 0.24 | 0.35 |
| 182 | GSC 08992-01008 | 0.63 | 0.69 | -0.40 | B3V | -0.22 | -0.30 | 0.85 | 0.99 |
| 183 | GSC 08988-01196 | 0.11 | 0.09 | -0.49 | B3V | -0.22 | -0.20 | 0.33 | 0.31 |
| 184 | GSC 08989-02518 | 0.23 | 0.34 | -0.27 | B0V | -0.32 | -0.17 | 0.54 | 0.51 |
| 185 | GSC 09001-00465 | 0.08 | – | – | B4V | -0.20 | – | 0.28 | – |
| 186 | GSC 09245-01017 | 0.01 | – | – | B6V | -0.16 | – | 0.17 | – |
| 187 | GSC 08660-01131 | 0.29 | 0.32 | -0.58 | B1.5IV | -0.27 | -0.27 | 0.56 | 0.59 |
| 188 | GSC 08660-00731 | 0.27 | 0.31 | -0.33 | B2V | -0.26 | -0.19 | 0.53 | 0.50 |
| 189 | GSC 08652-02075 | 0.07 | 0.10 | -0.21 | B7V | -0.14 | -0.09 | 0.21 | 0.19 |
| 190 | GSC 08998-01018 | 0.03 | 0.04 | -0.45 | B5V | -0.18 | -0.16 | 0.21 | 0.20 |
| 191 | GSC 09245-00106 | 0.06 | 0.09 | -0.89 | B5V | -0.18 | -0.32 | 0.24 | 0.41 |
| 192 | GSC 08998-01663 | 0.16 | 0.19 | -0.69 | B0V | -0.32 | -0.27 | 0.47 | 0.47 |
| 193 | GSC 08990-00217 | 0.29 | 0.30 | -0.65 | B1V | -0.28 | -0.29 | 0.57 | 0.60 |
| 194 | GSC 08995-02904 | 0.17 | 0.25 | -0.63 | B2IV | -0.26 | -0.27 | 0.42 | 0.52 |
| 195 | GSC 07793-00222 | -0.02 | -0.08 | – | – | – | – | – | – |
| 196 | GSC 09016-00519 | -0.03 | -0.03 | – | B3V | -0.22 | -0.21 | 0.19 | 0.15 |
| 197 | GSC 09008-04083 | 0.12 | – | – | B9V | -0.08 | – | 0.20 | – |
| 198 | GSC 08676-01771 | 0.03 | 0.10 | – | B2.5V | -0.24 | – | 0.26 | – |
| 199 | GSC 09009-01997 | 0.30 | 0.36 | -0.72 | B0.5V | -0.30 | -0.33 | 0.60 | 0.66 |
| 200 | GSC 09009-02487 | 0.30 | 0.35 | -0.74 | B1V | -0.28 | -0.33 | 0.58 | 0.68 |
| 201 | GSC 09005-03448 | 0.27 | 0.24 | -0.36 | – | – | -0.18 | – | 0.42 |
| 202 | GSC 09005-03474 | | 1.59 | – | – | – | – | – | – |
| 203 | GSC 09013-01063 | 0.04 | 0.13 | – | – | – | – | – | – |
| 204 | GSC 09006-04576 | 0.43 | 0.39 | – | – | – | – | – | – |
| 205 | GSC 08691-03023 | 0.23 | 0.37 | -0.42 | – | – | -0.23 | – | 0.60 |
| 206 | GSC 08688-01283 | 0.04 | 0.11 | – | B9V | -0.08 | – | 0.12 | – |
| 207 | GSC 09020-02147 | 0.58 | 0.56 | -0.24 | – | – | -0.22 | – | 0.78 |
| 208 | GSC 07821-02254 | | 1.08 | – | – | – | – | – | – |
| 209 | GSC 08305-02320 | 0.27 | 0.32 | -0.02 | B8V | -0.12 | -0.08 | 0.38 | 0.40 |
| 210 | GSC 09436-00541 | 0.08 | 0.13 | – | B8V | -0.12 | – | 0.20 | – |
| 211 | GSC 08702-00469 | -0.02 | 1.03 | – | – | – | – | – | – |
| 212 | GSC 09033-02403 | 0.06 | 0.06 | -0.63 | B2V | -0.26 | -0.23 | 0.31 | 0.30 |
| 213 | GSC 08303-01041 | 0.28 | 0.29 | -0.65 | B2V | -0.26 | -0.29 | 0.53 | 0.58 |
| 214 | GSC 07847-00082 | 0.05 | 0.07 | -0.28 | B6V | -0.16 | -0.11 | 0.21 | 0.18 |
| 215 | GSC 08307-01059 | 0.01 | 0.07 | – | B7V | -0.14 | – | 0.15 | – |
| 216 | GSC 09022-00605 | 0.11 | 0.09 | -0.28 | B6V | -0.16 | -0.12 | 0.27 | 0.21 |
| 217 | GSC 08701-00997 | 0.07 | 0.11 | -0.51 | B2V | -0.26 | -0.20 | 0.32 | 0.31 |
| 218 | GSC 08719-02158 | -0.02 | 0.04 | -0.40 | B5V | -0.18 | -0.15 | 0.16 | 0.18 |
| 219 | GSC 08319-00698 | 0.07 | 0.04 | -0.68 | B2V | -0.26 | -0.24 | 0.32 | 0.28 |
| 220 | GSC 08719-00464 | 0.07 | 0.07 | -0.36 | B5V | -0.18 | -0.14 | 0.25 | 0.21 |
| 221 | GSC 08711-02092 | 0.08 | 0.08 | -0.42 | B5IV/V | -0.18 | -0.16 | 0.26 | 0.24 |
| 222 | GSC 08723-00042 | 0.01 | 0.04 | -0.46 | B3V | -0.22 | -0.17 | 0.23 | 0.20 |
| 223 | GSC 08715-01941 | -0.02 | 0.00 | -0.71 | B2V | -0.26 | -0.24 | 0.24 | 0.24 |
| 224 | GSC 08712-02498 | 0.18 | 0.21 | -0.52 | B2.5IV | -0.24 | -0.23 | 0.42 | 0.44 |

**Table D1.** continued.

| (1) No | (2) GSC | (3) $(B-V)$ Kh01 | (4) $(B-V)$ GCPD APASS | (5) $(U-B)$ GCPD | (6) SpT&LC input | (7) $(B-V)_0$ spec. | (8) $(B-V)_0$ phot. avg. | (9) $E(B-V)$ spec. | (10) $E(B-V)$ phot. avg. |
|---|---|---|---|---|---|---|---|---|---|
| 225 | GSC 08325-05810 | 0.26 | 0.29 | -0.56 | B1V | -0.28 | -0.26 | 0.54 | 0.55 |
| 226 | GSC 09042-01527 | -0.11 | -0.08 | -0.56 | B7V | -0.14 | -0.17 | 0.03 | 0.09 |
| 227 | GSC 08337-00341 | 0.05 | 0.11 | -0.20 | B6V | -0.16 | -0.09 | 0.21 | 0.20 |
| 228 | GSC 08325-00916 | 0.30 | 0.30 | – | A2V | 0.05 | – | 0.25 | – |
| 229 | GSC 08330-05153 | -0.01 | – | – | B3V | -0.22 | – | 0.21 | – |
| 230 | GSC 08338-02080 | 0.16 | 0.18 | -0.70 | B2V | -0.26 | -0.28 | 0.41 | 0.46 |
| 231 | GSC 08734-02077 | 0.02 | 0.06 | 0.00 | – | – | -0.01 | – | 0.07 |
| 232 | GSC 07872-00681 | 0.17 | 0.22 | -0.76 | – | – | -0.30 | – | 0.51 |
| 233 | GSC 07872-00390 | 0.13 | 0.14 | -0.71 | B1IV | -0.28 | -0.27 | 0.41 | 0.40 |
| 234 | GSC 08328-00373 | 0.30 | 0.36 | -0.36 | B3V | -0.22 | -0.21 | 0.52 | 0.55 |
| 235 | GSC 07878-00246 | 0.05 | 0.08 | -0.54 | B2V | -0.26 | -0.19 | 0.31 | 0.26 |
| 236 | GSC 08345-03046 | 0.05 | – | – | B8V | -0.12 | – | 0.17 | – |
| 237 | GSC 07374-00838 | 0.72 | 0.80 | -0.27 | B0.5V | -0.30 | -0.28 | 1.02 | 1.08 |
| 238 | GSC 07366-00860 | 0.72 | 0.91 | 0.04 | – | – | -0.21 | – | 1.12 |
| 239 | GSC 08341-00889 | 0.03 | 0.07 | -0.41 | B5V | -0.18 | -0.15 | 0.21 | 0.22 |
| 240 | GSC 07375-00048 | 0.20 | – | – | B3IV | -0.22 | – | 0.42 | – |
| 241 | GSC 08342-00052 | 0.02 | 0.09 | – | B9V | -0.08 | – | 0.10 | – |
| 242 | GSC 07384-00247 | 0.67 | 0.62 | -0.31 | – | – | -0.25 | – | 0.87 |
| 243 | GSC 08342-01635 | -0.02 | -0.02 | -0.61 | B2V | -0.26 | -0.22 | 0.23 | 0.21 |
| 244 | GSC 06835-00151 | 0.30 | 0.57 | – | – | – | – | – | – |
| 245 | GSC 06839-00611 | 0.56 | 0.74 | -0.22 | – | – | -0.25 | – | 0.99 |
| 246 | GSC 07889-01252 | 0.13 | – | – | B7/8V | -0.13 | – | 0.26 | – |
| 247 | GSC 07385-01338 | 0.20 | 0.23 | -0.50 | B8V | -0.12 | -0.23 | 0.31 | 0.45 |
| 248 | GSC 07886-02848 | 0.00 | 0.02 | -0.58 | B1.5V | -0.27 | -0.20 | 0.27 | 0.22 |
| 249 | GSC 06853-01718 |  | – | – | – | – | – | – | – |
| 250 | GSC 06853-02519 | 0.49 | 0.53 | -0.51 | – | – | -0.30 | – | 0.83 |
| 251 | GSC 06841-01725 | 0.57 | 0.97 | – | – | – | – | – | – |
| 252 | GSC 06846-01106 | 0.75 | 0.87 | -0.20 | – | – | -0.28 | – | 1.15 |
| 253 | GSC 07399-01124 | 0.06 | – | – | B4V | -0.20 | – | 0.26 | – |
| 254 | GSC 06263-03157 | 0.08 | – | – | B3V | -0.22 | -0.16 | 0.30 | 0.23 |
| 255 | GSC 07399-01226 | 0.01 | – | – | B5V | -0.18 | – | 0.19 | – |
| 256 | GSC 06272-02199 | 0.36 | 0.38 | -0.29 | B5V | -0.18 | -0.19 | 0.54 | 0.57 |
| 257 | GSC 06268-02490 | 0.31 | 0.35 | -0.33 | B2V | -0.26 | -0.20 | 0.57 | 0.55 |
| 258 | GSC 06276-00317 | -0.03 | – | – | B8V | -0.12 | – | 0.09 | – |
| 259 | GSC 06851-02063 | 0.11 | – | – | B7V | -0.14 | – | 0.25 | – |
| 260 | GSC 06847-02930 | 0.47 | 0.64 | -0.46 | – | – | -0.31 | – | 0.95 |
| 261 | GSC 06851-04189 | -0.02 | – | – | B5V | -0.18 | – | 0.16 | – |
| 262 | GSC 06268-00943 | 0.20 | 0.26 | – | B7V | -0.14 | – | 0.34 | – |
| 263 | GSC 06847-02073 | 0.05 | 0.08 | -0.64 | B5V | -0.18 | -0.23 | 0.23 | 0.31 |
| 264 | GSC 06272-00394 | 0.16 | 0.17 | – | B5IV | -0.18 | – | 0.34 | – |
| 265 | GSC 07404-05201 | -0.08 | -0.11 | -0.60 | B3V | -0.22 | -0.18 | 0.14 | 0.07 |
| 266 | GSC 06269-02592 | 0.77 | 0.78 | -0.10 | – | – | -0.22 | – | 1.00 |
| 267 | GSC 06274-00902 | 0.34 | 0.37 | – | A0V | -0.02 | – | 0.36 | – |
| 268 | GSC 07909-02656 | -0.10 | 0.00 | – | B8V | -0.12 | – | 0.02 | – |
| 269 | GSC 05703-02553 | 0.18 | 0.19 | -0.40 | B3V | -0.22 | -0.18 | 0.40 | 0.37 |
| 270 | GSC 05124-01543 | 1.18 | 0.77 | – | B5V | -0.18 | – | 1.36 | – |
| 271 | GSC 05703-01526 | 0.29 | 0.32 | – | – | – | – | – | – |
| 272 | GSC 06275-00943 | 0.08 | 0.08 | -0.28 | B2IV | -0.26 | -0.11 | 0.34 | 0.19 |
| 273 | GSC 05692-01642 | 0.37 | 0.39 | – | B3V | -0.22 | – | 0.59 | – |
| 274 | GSC 05125-02006 | 0.34 | 0.26 | – | B3V | -0.22 | – | 0.56 | – |
| 275 | GSC 00456-00461 | 0.37 | 0.44 | -0.04 | B8V | -0.12 | -0.12 | 0.48 | 0.56 |
| 276 | GSC 05693-07523 | 0.10 | – | – | – | – | – | – | – |
| 277 | GSC 05126-03377 | 0.13 | 0.35 | – | – | – | – | – | – |
| 278 | GSC 05701-00964 | 0.27 | 0.25 | – | B4V | -0.20 | -0.17 | 0.47 | 0.46 |
| 279 | GSC 01026-02065 | 0.51 | 0.57 | -0.29 | B2.5V | -0.24 | -0.23 | 0.74 | 0.82 |
| 280 | GSC 06289-02980 | 0.14 | 0.16 | -0.30 | B9V | -0.08 | -0.14 | 0.22 | 0.30 |

**Table D1.** continued.

| (1) No | (2) GSC | (3) $(B-V)$ Kh01 | (4) $(B-V)$ GCPD APASS | (5) $(U-B)$ GCPD | (6) SpT&LC input | (7) $(B-V)_0$ spec. | (8) $(B-V)_0$ phot. avg. | (9) $E(B-V)$ spec. | (10) $E(B-V)$ phot. avg. |
|---|---|---|---|---|---|---|---|---|---|
| 281 | GSC 05123-00145 | 0.43 | 0.47 | – | B3V | -0.22 | – | 0.65 | – |
| 282 | GSC 00463-02825 | 0.45 | 0.61 | – | B2V | -0.26 | – | 0.70 | 0.65 |
| 283 | GSC 05131-01423 | 0.10 | 0.09 | -0.41 | B2V | -0.26 | -0.16 | 0.36 | 0.25 |
| 284 | GSC 02129-00864 | 0.57 | 0.54 | – | – | – | – | – | – |
| 285 | GSC 05149-01177 | 0.13 | – | – | B8V | -0.12 | – | 0.24 | – |
| 286 | GSC 01645-00281 | -0.12 | – | – | – | – | – | – | – |
| 287 | GSC 01124-01184 | -0.05 | -0.03 | – | B8V | -0.12 | – | 0.07 | – |